\documentclass[12pt]{article}

\usepackage[left=2cm,top=2cm,right=2cm,bottom=2cm]{geometry}

\usepackage[english]{babel}

\usepackage{grfext}

\usepackage{xcolor}
\usepackage{tikz-cd}
\usepackage{amsfonts,epsfig,epstopdf}
\usepackage{amssymb,amsmath,enumerate,float}
\makeatother
\usepackage{graphicx,bbm,xcolor,latexsym,epsfig,hyperref,tikz}
\usepackage[title]{appendix}
\usepackage{amsthm}
\usepackage{placeins}
\usepackage{amscd}

\newtheorem{theorem}{Theorem}[section]
\newtheorem{lemma}{Lemma}[section]
\newtheorem{definition}{Definition}[section]
\newtheorem{proposition}{Proposition}[section]
\newtheorem{corollary}
{Corollary}[section]

\newtheorem{example}{Example}[section]

\newenvironment{remark}{{\vskip 6pt \noindent \em Remark.}}{$\diamond$ \vskip 6pt}

\definecolor{myred}{RGB}{160,0,0}
\definecolor{mygreen}{RGB}{0,160,0}
\hypersetup{
    colorlinks = true,
    linkcolor = {myred},
    citecolor = {mygreen},
}

\addto\captionsenglish{\renewcommand{\figurename}{Fig.}}

\renewcommand{\i}{i}
\newcommand{\ptl}{\partial}
\newcommand{\vph}{\varphi}

\newcommand{\cX}{\mathcal{X}}
\newcommand{\rectangle}{{%
		\ooalign{$\sqsubset\mkern3mu$\cr$\mkern3mu\sqsupset$\cr}%
}}
\newcommand{\be}{\begin{equation}}
\newcommand{\ee}{\end{equation}}
\newcommand{\beq}{\begin{equation}}
\newcommand{\eeq}{\end{equation}}

\renewcommand{\Im}{\mathop{\rm Im}}

\newcommand{\cL}{{\mathcal L}}

\newcommand{\cM}{\mathcal{M}}
\newcommand{\ppsi}{\varpi}

\newcommand{\PI}{\mathcal{P}}
\newcommand{\cR}{\mathcal{R}}
\newcommand{\cE}{\mathcal{E}}

\newcommand{\cB}{\mathcal{B}}

\newcommand{\bB}{\mathbb{B}}
\newcommand{\tpB}{\tilde{\partial \bB}}

\newcommand{\pp}{\mathfrak{p}}

\newcommand{\bL}{\mathcal{L}}
\newcommand{\NR}{N\mathbb{R}}

\newcommand{\mesNik}[1]{\color{black}#1\color{black}}
\newcommand{\purple}[1]{{\color{black}{#1}}} 

\begin{document}

\title{Matrix representation  of Picard--Lefschetz--Pham theory near the real plane in 
$\mathbb{C}^2$}

\author{A.V.Shanin, A.I.Korolkov, N.M.Artemov, R.C.Assier}

\maketitle

\tableofcontents

\newpage

 \section*{Notations throughout the paper}
\begin{tabular}{r|l|l}
 {\bf notation} & {\bf comment} &  {\bf page}   \\
 \hline
  $\NR^n$  & small neighborhood of $\mathbb{R}^n$ in $\mathbb{C}^n$ & \pageref{e:r2_def} \\
  $\bB$     & a sufficiently large ball in $\mathbb{C}^2$  &  \pageref{e:r2_def} \\ 
  $\sigma = \cup \sigma_j$ & singularity and its irreducible components &  \pageref{e:i0002}  \\
  $g_j(z; t)$ & defining functions of singularities &  \pageref{e:i0002}  \\ 
  $\cL$ & The Landau set & \pageref{e:i0002a}, \pageref{def:Landau}\\
   $\sigma = \cup \sigma_j$ & singularity in the $z$-space and its irreducible components &  \pageref{e:i0002}  \\
  $\sigma' = \cup \sigma'_j$ & intersections of $\sigma$ and $\sigma_j$ with $\mathbb{R}^2$ &  \pageref{e:i0002}  \\
    $\sigma^{(1)}$, $\sigma^{(0)}$ & strata of $\sigma$ &  \pageref{sec:universal}   \\

  $\pi_1 (X)$ & fundamental group of $X$ &     \\
  $\Pi(z^* , z)$ & set of paths from $z^*$ to $z$ in certain space &  \pageref{sec:universal}   \\
  $\Pi_D(z^s , z^e)$ & set of local paths from $z^s$ to $z^e$ in a small ball $D$ minus $\sigma$ &  \pageref{f:01003i8z}   \\
    $\cX$  &  $\NR^2$ or $\mathbb{C}^2$ &  \pageref{sec:universal} \\
  $\PI$  & $\pi_1 (\NR^2 \setminus \sigma)$ & \pageref{e:def_PI} \\
  $\tilde U$ & The universal Riemann domain of $\cX \setminus \sigma$  & \pageref{def:URD} \\
  $\tilde U_2$, $\tilde U_1$, $\tilde U_0$ & strata of $\tilde U$  & \pageref{def:URD} \\
   $\tilde U'$ & $\tilde U_1 \cup \tilde U_0$ singularities of $\tilde U$  & \pageref{e:def_Up} \\
   $\tilde{\partial \bB}$ & $\tilde U|z\in\ptl \bB$ & \pageref{e:def_tilde-dB}   \\
   $\pp$ & projection $\tilde U \to \cX$ & \pageref{def:p_projection}   \\
  $\hat U$   & Riemann domain of function $F$ & \pageref{pr:single-valued} \\
  $\tilde U^*$   &  $\tilde U|z\in\mathbb{R}^2$ & \pageref{e:U_retract} \\
  $\cR$  & retraction $\tilde U \to \tilde U^*$ &  \pageref{pr:retract} \\
  $\ptl$   & ``boundary'' homomorphism $H_2(\tilde U, \tilde U' \cup\tilde{\partial \bB}) \to H_1 (\tilde U' , \tpB)$  & \pageref{e:cond_H2U} \\
  $\Omega$  & group ring of $\PI$ over $\mathbb{Z}$ & \pageref{e:01318} \\
  $\cM_k$  &  space of vectors-rows of $k$ elements from $\Omega$ & \\
    & \qquad \qquad \qquad \qquad \qquad (right module of rank $k$ over $\Omega$)& \pageref{def:module} \\
  $\cE$  & inflation operator $H_2(\tilde U^* , \tpB) \to H_2(\tilde U_2, \tpB)$  &  \pageref{e:inflation_def} \\
  $\Sigma$ & { Auxiliary surface used to define base paths }& \pageref{pr:surface_Sigma} \\
  $\cB$  & space of regular points of parameter $t$ & \pageref{e:def_B}  \\
   $\sigma_j^{t}$ & singularities in the $t$-space & \pageref{def:Landau} \\
  $\psi_\lambda$ & action of the bypass $\lambda$ on different objects & \pageref{e:variation} \\
  ${\rm var}_\lambda (w)$ & variation of homology for path $\lambda$& \pageref{e:def_var} 
\\
$\langle w^1 \, | \, w^2 \rangle$  & intersection index between homologies & \pageref{e:intersection_index}
\end{tabular} 

\newpage

\begin{abstract}
A matrix formalism is proposed for computations based on Picard--Lefschetz theory in a 2D case. The formalism is essentially  equivalent to the computation of the intersection indices necessary for the Picard--Lefschetz formula and enables one to prove non-trivial topological identities for integrals depending on parameters. 

We introduce the universal Riemann domain $\tilde U$, i.e.\ a sort of ``compactification'' of the universal covering space $\tilde U_2$ over a small tubular neighborhood $\NR^2$ of $\mathbb{R}^2\backslash\sigma$ in $\bB \setminus \sigma$,
where $\bB\subset\mathbb{C}^2$ is a big ball, and $\sigma$ is a one-dimensional complex analytic set (the set of singularities). We compute the Picard-Lefschetz monodromy of the relative homology group of the space $\tilde U$ modulo the singularities and the boundary for the standard local degenerations of type $P_1 ,P_2,P_3$ in Pham's \cite{Pham1967IntroductionAL}
notations and for more complicated configurations in~$\mathbb{C}^2$. 
We consider this homology group as a module over the group ring of the $\pi_1((\NR^2 \cap \bB)\backslash\sigma)$ over $\mathbb{Z}$. The results of the computations are presented in the form of a matrix of the monodromy operator calculated in a certain natural basis. 
We prove an ``inflation'' theorem, which states that the integration surfaces of interest (i.e.\ the elements of the homology group $H_2(\tilde U_2,\tpB)$) (the surfaces in the branched space possibly passing through singularities) are injectively mapped to the group $H_2(\tilde U,\tilde U'\cup\tpB)$ (the surfaces avoiding the singularities). The matrix formalism obtained describes the behaviour of integrals depending on parameters and can be applied to the study of Wiener-Hopf
method in two complex variables.


\end{abstract}

\section{Introduction}

\subsection{Problem under consideration}

We study integrals of the form 
\begin{equation}
I(t) = \int_\Gamma F(z ; t) \, dz_1 \wedge dz_2. 
\label{e:i0001}
\end{equation}
Here $z = (z_1 , z_2) \in \mathbb{C}^2$, $t = (t_1 , t_2, \dots t_M) \in \mathbb{C}^M$, 
$\Gamma$ is some oriented surface of integration avoiding the singularities of~$F$.

\purple{The} function $F(z ; t)$ is holomorphic with
respect to all $M + 2$ complex variables almost everywhere. 
In particular, for any fixed $t$, 
$F$ is holomorphic for
$z \in \NR^2 \setminus \sigma$, where $\NR^2$ is a small tubular neighbourhood of the real plane and the set $\sigma= \sigma(t)$ is the singularity set of $F$ in the $z$-space:
\[
\sigma = \cup_j \sigma_j,
\]   
where $\sigma_j = \sigma_j(t)$ are irreducible singularity components (analytic sets of complex codimension~1).  
In general case, \purple{we} denote a small tubular neighbourhood of~$\mathbb{R}^M\subset\mathbb{C}^M$ by $\NR^M$: 
\begin{equation}
\NR^M = \{ 
z \in \mathbb{C}^M \, : \, \, 
\sum_{j=1}^M{\Im}[z_{j}]^2 \le \delta
\}
\label{e:r2_def}
\end{equation} 
for some small positive $\delta$.

In applications, the function $F(z; t)$ may contain exponential factors (e.g.\ for Fourier integrals), so we cannot assume 
that $F$ has an \purple{algebraic} behaviour in~$z$. This circumstance has an important consequence below: we will not use a projective
compactification of~$\mathbb{C}^2$.

The sets $\sigma_j$ are assumed to be polar or branch sets of $F(z;t)$ (for fixed $t$).
The branching can be logarithmic.  The function $F$ can be continued analytically everywhere in 
$\NR^2 \times \NR^M$ except at the singularities. 

Some of the singularity components are immovable: for them
\purple{
\[
\sigma_j (t) = \sigma_j (z).
\] 
}
The other 
components are movable.

An example of \purple{an integral of a type}  (\ref{e:i0001}) is
\begin{equation}
I(t) = \int_\Gamma 
\frac{1}{(z_1 - t_1) (z_2 - t_2)\sqrt{z_1^2 + z_2^2 -1} }
 \, dz_1 \wedge dz_2. 
\label{e:i0001a}
\end{equation}
with $\sigma = \sigma_1 \cup \sigma_2 \cup \sigma_3$,
\begin{equation}
\sigma_1:\,\, z_1^2 + z_2^2 - 1 = 0, 
\qquad 
\sigma_2(t):\,\, z_1 - t_1 = 0, 
\qquad 
\sigma_3(t):\,\, z_2 - t_2= 0. 
\label{e:i0002a}
\end{equation}

Our global aim is to study the ramification of the integral (\ref{e:i0001}) as a function of~$t$. This ramification can be imagined as a continuous deformation of the singularities 
in the $z$-space with a simultaneous appropriate deformation of the surface of  integration~$\Gamma$. 
The current work mainly  studies the ramification of $\Gamma$ under
loops in the $t$-space. 


\begin{remark}
There are two complex spaces considered: $z$ and $t$. \purple{The} function $F$ (as a function of $z$ for fixed $t$) may have {\em branching\/} at some \purple{singularity} set. At the same time, the integral (\ref{e:i0001}) can be {\em ramified\/} as a function of~$t$. We use the synonyms ``branching'' and ``ramification'' to indicate  which space ($z$ or $t$) is under consideration. We do not consider the space of all variables $z$ and $t$ as it is done sometimes. 
\end{remark}

As it is known, the multidimensional Cauchy theorem \cite{Shabat2} can be applied to the integral 
(\ref{e:i0001}): one can deform $\Gamma$ in $\NR^2 \setminus \sigma$ without the 
integral being changed.      
Thus, the integral $I(t)$ has a singularity only when one cannot deform $\Gamma$
in such a way that it avoids the singularity; in other words, when the surface $\Gamma$
becomes pinched by the singularities. 

The possible values of $t$ for which such a pinching can happen form \purple{what is known as} a {\em Landau set\/}. 
The formal definition of a Landau set is not elementary \cite{Pham2011,Berghoff2022},    
but we use a simple concept: the Landau set \purple{associated to (\ref{e:i0001}), and} denoted $\bL$ is composed of \purple{the} points $t$ \purple{for which}
the topological properties of $\NR^2 \setminus \sigma(t)$
differ from the ``general case''. 
As the topological property, 
we take the relative
homology group~$H_2 (\NR^2 , \sigma(t))$:
some elements of it vanish for $t \in \bL$.  
 
The integration surface $\Gamma$ can be treated as an element of 
a certain 2D relative homology group of the space $\NR^2 \setminus \sigma$ modulo the boundary of a sufficiently large ball $\bB\subset\mathbb{C}^2$, taking into account the branching of $F$ on~$\sigma$. 
This group is introduced below and denoted $H_2 (\tilde U_2,\tpB)$.
The ramification of the integral $I(t)$ as the point $t$ bypasses $\cL$ is caused by the ramification of 
$H_2 (\tilde U_2,\tpB)$, i.e.\ such a bypass, generally, does not map the elements of $H_2 (\tilde U_2,\tpB)$ to themselves. 

The Picard--Lefschetz theory (its ``twisted'' version, see \cite{Vassiliev2002}) describes the ramification of   
$H_2 (\tilde U_2,\tpB)$ about the Landau set in slightly different terms.
To apply it,  one should compute the intersection numbers of $\Gamma$
and the so-called {\em vanishing cells}. 
These intersection 
numbers are not easy to compute.
In this paper, we propose a convenient alternative approach 
to the calculations in \purple{complex} dimension~2 for this problem. Namely, the integration surfaces are represented as 
``vectors'' (more rigorously, elements of a module over a group ring, see Section~\ref{LR}), and the ramification is described 
as multiplication by matrices. The corresponding matrices are calculated in Section~\ref{Elt} for typical Pham degenerations $P_1,P_2,P_3$ (see \cite{Pham1967IntroductionAL}) and in Section~\ref{BMT} for more complicated cases. 

\mesNik{In Appendix~\ref{app:G}, we verify our results by computing twisted intersection indices and applying the Picard–Lefschetz–Pham formula. Our answers can also be verified by applying Lemma 3.2 from \cite{Kita1994}.}

Integrals \purple{of the type} (\ref{e:i0001}) naturally emerge when one tries to apply the Wiener--Hopf method in several complex variables \cite{Assier2019}. We expect that the formalism developed below will be useful in moving the Wiener--Hopf studies forward.

\subsection{Basic assumptions}

We assume that the integration surface $\Gamma$ belongs to $\NR^2$,
which is is a four-dimensional manifold with a boundary. 
We assume everywhere that 
$\sigma$ is transversal to the boundary of~$N\mathbb{R}^2$. 
We also assume that $\Gamma$ can ``go to infinity''; 
to formalize this, introduce a sufficiently large ball $\bB\subset\mathbb{C}^2$ centred at the origin. The radius $R$ of the ball is chosen in such a way that $\sigma$ behaves ``in a simple way'' outside the ball (there are no crossings of the singularity components, etc); in particular, 
$R \gg \delta$, and $\sigma$ is transversal to~$\partial\bB$. We consider the integral over $\Gamma \cap \bB$
and study the limit $R\to \infty$. Assuming that $F(z; t)$ decays fast enough as $|z| \to \infty$, we can state that the integral over $\Gamma \cap \bB$ tends to (\ref{e:i0001}). Introduction of the ball $\bB$ enables us to keep the considered spaces compact and use the standard methods of the homology theory.  

 Note that $N\mathbb{R}^2\cap\partial \bB$ is homeomorphic to 
 a product of a one-dimensional real sphere and a two-dimensional real disc
 $S^1\times D^2$. 


Also, 
we are particularly interested in the values of $t$ belonging to the small tubular neighborhood $\NR^M$. 

Each component of singularity is defined by an equation 
\begin{equation}
\sigma_j = \{ z \in \mathbb{C}^2 \,\, : \,\,
g_j (z ; t) = 0
 \} ,
\label{e:i0002}
\end{equation}  
where $g_j$ is a holomorphic function of all $2+M$ complex variables everywhere in $\NR^2 \times \NR^M$. 

We assume that for any fixed $t$ the gradient $\nabla g_j(z ; t)$ taken with respect to $z$ is non-zero in~$\NR^2$. Thus, each $\sigma_j \cap \mathbb{R}^2$ has a structure of a smooth manifold. It will be important for us that $\sigma_j$ does not have self-crossings. 

\begin{definition}
	\label{def:realproperty}
  An immovable irreducible singularity $\sigma_j$ has the {\em real property\/} if
  its defining function $g_j$ is real whenever its arguments are real:
  \begin{equation}
   g_j (z) \in \mathbb{R}, 
    \hspace{1em} \text{ if } \hspace{1em} 
    z \in \mathbb{R}^2   .
    \label{e:def_deal}
  \end{equation}

For a movable singularity component,  we say that it has the real property 
if $\sigma_j(t)$ has the real property for real~$t$.

\end{definition}

As a consequence of the real property, the intersection between $\mathbb{R}^2$ and an irreducible singularity 
$\sigma_j (t)$, $t \in \mathbb{R}^M$, is a one-dimensional curve. Note that this is not generally the case. 
If the real property is not valid, the intersection of a
manifold $g_j(z)= 0$ with the real plane of $z$ is a set of discrete points or empty.

    Let $t$ be real. The one-dimensional curve  $\sigma_j \cap \mathbb{R}^2$ is
	called the {\em real trace} of $\sigma_j$ and denoted by $\sigma'_j$:
	\[
	\sigma'_j = \sigma_j \cap \mathbb{R}^2.
	\]
	Introduce also $\sigma'=\cup_j \sigma'_j$.


The main reason to study the singularities having the real property is that the intersection of such a singularity and the real plane is unstable, i.e.\ there exists $\Gamma$, that is a slightly deformed $\mathbb{R}^2$, not crossing $\sigma_j$. We also should note that singularities having the real properties  naturally emerge in applications \cite{Assier2019,Assier2022,Assier2024,Mironov2021,Shanin2024}.

For the singularities having the real property, we assume that 
the neighborhood $\NR^2$ is so small that
all crossings of singularity components $\sigma_j$ 
in $\NR^2$ belong to~$\mathbb{R}^2$.





\subsection{Prior work}

Loosely speaking, the Picard-Lefschetz formula describes how homology group transform under analytic continuation around a singularity. First, it was derived in $\mathbb{C}^2$ by Picard in \cite{Picard1897}, and later extended to higher dimensions by Lefschetz in \cite{lefschetz1924analysis}. Next comprehensive study of Picard-Lefshetz theory was conducted by Pham in \cite{Pham1965}, culminating in the handbook \cite{Pham1967IntroductionAL}. We mainly follow \cite{Pham1967IntroductionAL} and a more recent Pham's book \cite{Pham2011}. It is important to note that neither of these books contains a proof of the main theorem; for that, one should refer to \cite{Pham1965}. In his monographs, Pham applied Picard-Lefschetz theory to the study of the ramification of integrals. He briefly explored the case where the integrand itself exhibits branching behaviour. 

A significant advancement in this area was made by Vassiliev in \cite{Vassiliev2002,Vassiliev2012-tu}, who systematically developed a ``twisted'' version of Picard--Lefschetz theory for analysing branching integrals. He  introduced a stratified version of Picard--Lefschetz formula which describe ramification of homology groups associated  with singular algebraic varieties. The ``inflation theorem'', formulated and proved in this work, is closely related to \cite[VI]{Vassiliev2002}.

More recently, in \cite{Berghoff2022} a { detailed study of  the Picard--Lefschetz formula for relative homologies } 
was held. The authors applied their results to Feynman integrals, which exhibit only polar singularities. They focused on integration surfaces with boundaries (relative homologies), allowing these surfaces to intersect the singularities of the integrand. Their analysis demonstrated that so-called linear pinches lead to non-zero variations in this case.

Ramification of integral functions arise in many physical problems. Some of them are the Archimedes--Newton problem on integrable bodies \cite{Vassiliev2002}, Feynman integrals which arise in quantum electrodynamics \cite{Hwa1966}, and study of wave behaviour in the neighboorhood of caustics \cite{Arnold2012-vj}.  In this work, we are  motivated by the problem of wave diffraction by a canonical object --- a quarter-plane.  Physical formulation of this problem, along with the derivation of the 2D Wiener-Hopf equation and relevant integrals, can be found in \cite{Assier2019}. Asymptotic estimation of corresponding integrals have been developed in \cite{Assier2022,Assier2024}.





\subsection{The structure of the paper}

In Section~\ref{sec:2} we introduce a universal Riemann domain for a given set of singularities. We do this in a slightly non-standard way (comparatively to the universal covering of the space minus the singularities), since our aim is to include the singularities into the consideration. As a result, we get a stratified universal Riemann domain $\tilde U$, whose main stratum $\tilde U_2$ is the standard universal covering, and the union of the lower strata $\tilde U'$ is built over the singularities. Any Riemann domain of a function having the same singularities can be reduced to the universal Riemann domain in an easy algebraic way.  

In Section~\ref{sec:3} we introduce algebraic and visual notations for the objects under consideration. Namely, we introduce homology groups $H_2(\tilde U_\purple{2},\tpB)$, whose elements are eligible surfaces of integration, $H_2(\tilde U, \tilde U' \cup \tpB)$, whose elements can have boundaries on the singularities, and $H_2(\tilde U,\tpB)$, whose elements cannot have boundaries but can pass through the singularities. The group $H_2 (\tilde U_2 ,\tpB)$ is what we would like to study,  $H_2(\tilde U, \tilde U' \cup \tpB)$ is what is easy to study, and $H_2 (\tilde U ,\tpB)$ has an intermediate position. We demonstrate that a) $H_2(\tilde U_2,\tpB)$ is isomorphic to 
$H_2 (\tilde U,\tpB)$ (this is the ``inflation theorem''), b) that $H_2 (\tilde U,\tpB)$ can be identified with the set of  elements $w \in H_2 (\tilde U, \tilde U' \cup \tpB)$ such that $\ptl w = 0$. The ``boundary'' operator  $\ptl $ here stands for the ``boundary''  homomorphism
\[
\ptl : \quad H_2 (\tilde U, \tilde U' \cup \tpB) \rightarrow H_1 (\tilde U', \tpB)   
\]
from the exact sequence of homology for the triple
\[
(\tilde U, \, \tilde U'\cup \tpB, \, \tpB). 
\]
{This is the formal definition using the language of algebraic topology; 
on the intuitive level, the boundary operator just takes boundaries of (curvilinear) polygons
from $H_2 (\tilde U,\tilde U' \cup \tpB)$.}

In Section~\ref{sec:4} we describe the ramification of $H_2 (\tilde U , \tilde U' \cup \tpB)$ as the parameter $t$ is carried along some path~$\lambda$. Such a ramification is described as follows: an element of $H_2 (\tilde U , \tilde U' \cup \tpB)$ corresponds to a vector-row, and the transformation 
induced by $\lambda$ is a multiplication of this vector by some square matrix. The elements of the vectors and the matrices belong to the group ring $\Omega$ of the fundamental group $\PI$ 
of $\NR^2 \setminus \sigma$ over~$\mathbb{Z}$.
We derive the transformation matrices for \purple{some specific} elementary cases and demonstrate on several examples that the outlined procedure is enough to build transformation matrices in a quite general case.

In Section~\ref{sec:5} we demonstrate some benefits of the new approach. In particular, we show that the matrix approach a) simplifies computation of the intersection indices that is crucial 
for the Picard--Lefschetz theory, b) contains a lot of topological information about the space of parameters, c) can provide rather sophisticated formulas such as the additive crossing identities.



\section{Riemann domains}
\label{sec:2}
\subsection{Universal Riemann domain}
\label{sec:universal}

\subsubsection{Definition of the universal Riemann domain}

Let $\cX$ be the complex space $\mathbb{C}^2$ or the neighborhood of the 
real plane $\NR^2$ 
(the consideration of this section is rather general, so it can be applied not only to~$\NR^2$).
Here we
introduce \purple{the} {\em universal Riemann domain\/} over 
$\cX \setminus \sigma$, where $\sigma$ is the singularity set described by functions~$g_j$.  
This universal Riemann domain will be denoted~$\tilde U$.

Note that the universal Riemann domain is not linked to any 
function $F(z)$ singular on $\sigma$; 
instead, the universal Riemann domain is defined only by the singularity set, 
which is in this sense the set that should be avoided by certain homologies (the ``integration surfaces'').
By construction, $\tilde U$
is the widest Riemann domain for a function 
having the given singular set in $\cX$.
What is important, we include the singularities into the universal Riemann 
domain.

Everywhere in this section, we assume the parameter $t$ to be fixed and do not write it explicitly.

The singularity set $\sigma$ is stratified in the usual sense.
The main stratum is $\sigma$ minus the union of all intersections 
$\sigma_j \cap \sigma_k$. The real dimension of this stratum is~$2$.
The next stratum is the union of all binary intersections
$\sigma_j \cap \sigma_k$. 
The real dimension of this stratum is~0.
We denote these strata by $\sigma^{(1)}$ and~$\sigma^{(0)}$,
respectively. 
 
Fix a reference point $z^* \in \cX \setminus \sigma$.
Let $\Pi'(z^* , z)$ be the set of continuous paths going from $z^*$ to some point 
$z \in \cX$ ($z$ can belong to $\sigma$ or not), such that all points of such a path, maybe except the 
end point $z$, belong to $\cX \setminus \sigma$. 

Let $\sim$ be the equivalence relation provided by homotopy of paths in $\cX \setminus \sigma$ with fixed ends. 
We define the set of classes 
\[
\Pi(z^* , z) = \Pi'(z^* , z) / \sim
\]
(still, we will refer to elements of $\Pi(z^* , z)$ as paths if this causes no confusion).



\begin{definition}
\label{def:URD}
The universal Riemann domain $\tilde U$ is a set of pairs $(z , \gamma)$, 
where $z\in \cX\cap \bB$,    $\gamma \in \Pi(z^* , z)$, equipped with the natural topology.
\end{definition}

{In our notation, we do not explicitly indicate the dependence of $\tilde U$ on the choice of~$z^*$.}

Let us also \purple{define} the spaces 

\begin{equation}
   \tilde U_2 \equiv \{(z , \gamma)\in\tilde U  \, \, : \, \,     z\in\cX \backslash \sigma\},
\label{e:def_U2}
\end{equation}
\begin{equation}
       \tilde U'\equiv\{(z , \gamma)\in\tilde U \, \, : \, \, z \in \sigma \},
       \label{e:def_Up}
   \end{equation}
   \begin{equation}
     \tpB \equiv\{(z , \gamma)\in\tilde U   \, \, : \, \,     z\in\partial \bB\} .
     \label{e:def_tilde-dB}
   \end{equation}

The element $z$ of the pair is the {\em affix\/} of the point, 
$\gamma$ is the {\em path-index}.
We will say that all points of $\tilde U$ having affix $z$ are the points
over~$z$ on different sheets of~$\tilde U$. 

The universal Riemann domain $\tilde U$ is a ``stratified'' topological space 
\begin{equation}
\tilde U = \tilde U_2 \cup \tilde U_1 \cup \tilde U_0 
\label{e:U_stratification}
\end{equation}



The stratum to which the point $(z, \gamma)$ belongs 
is defined by the affix~$z$. The main stratum 
$\tilde U_2$ corresponds to $z \in \cX \setminus \sigma$ (see (\ref{e:def_U2})). 
The next stratum $\tilde U_{1}$ is built for $z \in \sigma^{(1)}$, 
the stratum $\tilde U_0$ is built for $z \in \sigma^{(0)}$. 
According to (\ref{e:def_Up}),
\begin{equation}
\tilde U' = \tilde U_1 \cup \tilde U_0.
\label{e:singularity_set}
\end{equation}
Each stratum $\tilde U_j$ is a manifold of real dimension~$2j$, moreover, it is a complex manifold of complex dimension~$j$.


The main stratum $\tilde U_2$ is  \purple{the} {\em universal covering\/}  
over $\cX \setminus \sigma$, \purple{which is} a well-known object. Our idea here is to add singularities to~$\tilde U_2$.

The universal Riemann domain $\tilde U$ may be considered as a Riemann domain of some function $F(z)$
having logarithmic branching on all $\sigma_j$ and obeying no additional restrictions. 
Later on, we describe a reduction of the universal Riemann domain to some smaller Riemann domain of a particular function of~$z$.

\begin{remark}
We consider only the singularities in the finite part of $\mathbb{C}^2$ (in the large ball $\bB$). The infinity requires a separate 
study.   
\end{remark}

\begin{remark}
The ``stratification'' (\ref{e:U_stratification}) is not a regular stratification in the Whitney sense, since the external boundary of, say, 
$\tilde U_2$ is not removed from~$\tilde U_2$. Such a stratification is needed for a correct application of the Thom's first isotopy lemma; 
we discuss it below. 
\end{remark}


\subsubsection{Structure of $\tilde U_2$}

First note that
the set $\Pi(z^* , z^*)$ 
can be turned into 
the fundamental group 
$\pi_1 (\cX \setminus \sigma)$ by 
equipping it with the operations of multiplication and taking the 
inverse. Indeed, the multiplication is a concatenation. 
A notation 
\[
\gamma = \gamma_1 \gamma_2
\]
means that the path $\gamma_2$ goes after~$\gamma_1$.
Taking the inverse of $\gamma$ is passing it in the opposite direction.
The trivial bypass is denoted by~$e$. 

Let \purple{us fix} $z \in \cX \setminus \sigma$. 
It is well-known that path-indices for the points of $\tilde U$ over $z$ \mesNik{(i.e.\ all points of $\tilde U_2$) } are linked with $\pi_1 (\cX \setminus \sigma)$. Namely, the following proposition is valid: 

\begin{proposition}
\label{pr:U2_str}
Let $\gamma'$ be an arbitrarily chosen element of $\Pi(z^*, z)$,
$z \in \cX \setminus \sigma$.
Then any element of $\gamma \in \Pi(z^*, z)$ can be uniquely represented as
\begin{equation}
\gamma = \gamma_1 \gamma',  
\label{e:U2_str}
\end{equation}
where $\gamma_1 \in \pi_1 (\cX \setminus \sigma)$.
\end{proposition}

Thus, one can assign a group structure to the points of $\tilde U_2$ over~$z$.

Let 
\begin{equation}\label{def:p_projection}
\pp:\quad \tilde U\rightarrow\cX 
\end{equation}
denote the projection of the branched covering
(taking the affix of the point).  
Note that for any point $z\in\cX\backslash\sigma$ the full preimage $\pp^{-1}$ of a small \mesNik{ball $D(z)\subset\cX$ around $z$ } is homeomorphic to an infinite number of disjoint copies of \mesNik{$D(z)$}. Each copy corresponds bijectively to an element of $\pi_1(\cX\backslash\sigma)$. 
\mesNik{Locally, this set of copies of $D(z)$ forms~$\tilde U_2$.}


\subsubsection{Structure of $\tilde U_1$. Locality lemma for $\sigma^{(1)}$}

Let be $z \in \sigma^{(1)}$, i.e.\ $z$ belongs to a singularity but does not belong to a crossing of two or more singularity components.  
The points of $\tilde U$ over $z$ (i.e.\ the points of $\tilde U_1$)  are $(z, \gamma)$, where $\gamma \in \Pi(z^* , z)$.
It seems useful to describe the singularity using  
$\Pi(z^*,z')$ for some $z' \in \cX \setminus \sigma$ located near $z$, 
since such $\Pi(z^*,z')$ is a more regular object (it is linked to the fundamental group). 

\begin{figure}[h]
\centering
  \includegraphics[width=0.4\textwidth]{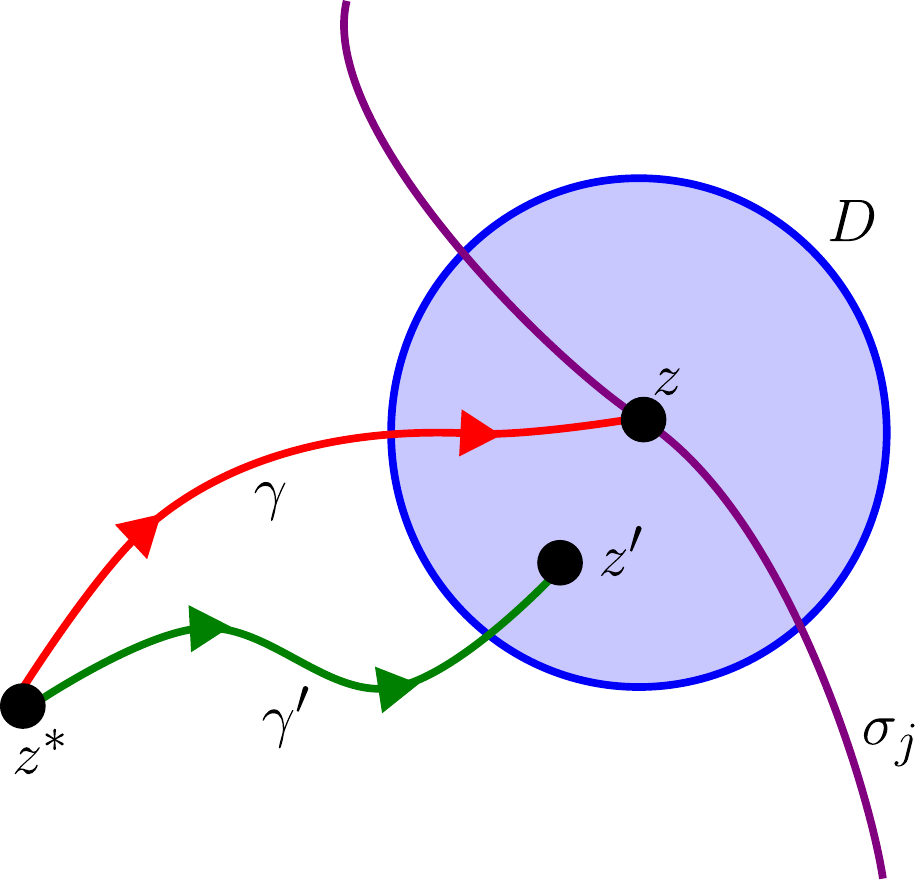}
  \caption{Small ball $D$ and  $z \in \sigma$}
  \label{f:01003i8z}
\end{figure}

Introduce an open ball \mesNik{$D(z)\subset \cX$ } centred at~$z$.
Let the ball be small enough not to cross any other singularity components.  
Take a point $z'$ in $D$ close to $z$ (see \figurename~\ref{f:01003i8z}).

Introduce
a set of {\em local paths\/}  $\Pi_D(z' , z') \subset \Pi(z', z')$
as the set of paths going 
from $z'$ to $z'$ entirely in $D \setminus \sigma$.
The paths are defined up to homotopies 
in $\cX \setminus \sigma$, i.e.\ formally speaking they are classes of paths.                                                                                                                                                                                             
The set of paths $\Pi_D (z', z')$ forms a group. Let us make it a subgroup of $\pi_1 (\cX \setminus \sigma)$. For this, introduce 
some path $\gamma' \in \Pi(z^*,z')$. 
The map 
\[
\ppsi: \qquad \gamma'' \to \gamma' \gamma'' (\gamma')^{-1},  \quad \gamma'' \in \Pi_D (z', z')
\]
maps $\Pi_D (z', z')$ \purple{to} $\pi_1 (\cX \setminus \sigma)$. Denote $\ppsi (\Pi_D (z', z'))$ by $\pi_1^{loc} [z]$ and refer to this group as to the {\em local subgroup\/} of $\pi_1 (\cX \setminus \sigma)$ for the point $z$.
The definition of $\pi_1^{loc} [z]$ depends on $\gamma'$, but any two such subgroups with different $\gamma'$ are conjugate in
$\pi_1 (\cX \setminus \sigma)$.

The following lemma is valid:

\begin{lemma}[\purple{L}ocality for $\sigma^{(1)}$]
\label{le:locality}
\purple{Let us fix} $z \in \sigma^{(1)}$. Then 

a) The local subgroup $\pi_1^{loc} [z]$ is isomorphic to $\mathbb{Z}$. The generator for the group is $\ppsi( \gamma^{el})$,
where $\gamma^{el}$ is a simple loop about $\sigma$ in $\Pi_D (z',z')$. 

b) The set of all possible $\gamma$ that are path-indices of $(z, \gamma) \in \tilde U_1$ is the set of left cosets of 
the subgroup $\pi_1^{loc} [z]$ in $\pi_1 (\cX \setminus \sigma)$. In other words, 
\[
\gamma = \gamma_0 \Pi(z', z'), \qquad \gamma_0 \in \Pi (z^* , z').
\]
\end{lemma}

In the formulation of the lemma we used the following definition: 

\begin{definition}
A simple loop about \purple{a} singularity $\sigma_j$ is a path 
$\gamma^\dag \gamma^\ddag (\gamma^\dag)^{-1}$, where $\gamma^\ddag$ is a small loop about 
$\sigma_j$ encircling it one time in the positive direction. 
\end{definition}

The proof of the lemma is given in Appendix~\ref{app:A}.

Since $\pi_1^{loc} [z]$ is generally not a normal subgroup of $\pi_1 (\cX \setminus \sigma)$,
the path-indices for $\tilde U_1$ do not have the group structure.

\begin{figure}[h]
\centering
  \includegraphics[width=0.5\textwidth]{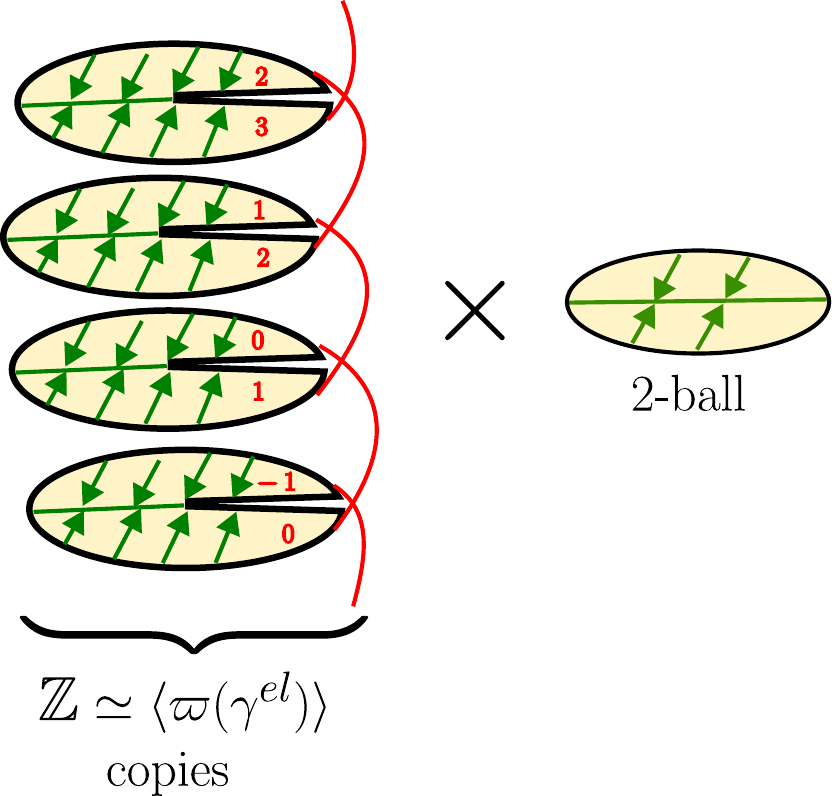}
  \caption{Topology of $W$ for  $z\in\sigma_j\backslash\cup_{n\neq j}\sigma_n$.
  } 
  \label{f:N1}
\end{figure}

\mesNik{For any point $z\in\sigma_j\backslash\cup_{n\neq j}\sigma_n$ the space $\pp^{-1}(D(z))$ (see \eqref{def:p_projection}) is homeomorphic to an infinite number of copies of a space $W$ that is homeomorphic to ($\mathbb{Z}\simeq\langle\varpi(\gamma^{el})\rangle$ copies of cut 2-ball)$\times$(2-ball), glued together as shown in \figurename~\ref{f:N1}. Here 
$\langle \varpi(\gamma^{el}) \rangle$ is a subgroup of $\pi_1(\cX\backslash\sigma)$
generated by $\varpi(\gamma^{el})$. Each copy of the space $W$ corresponds to an element of $\pi_1(\cX\backslash\sigma)/\langle\varpi(\gamma^{el})\rangle$. The assembly of countably many (one-to-one with $\mathbb{Z}$) copies of 2-balls can be considered as a fragment of a (usual) Riemann surface with a 
logarithmic branch point.} 

\subsubsection{Structure of $\tilde U_0$. Locality lemma for $\sigma^{(0)}$}
Let be $z \in \sigma^{(0)}$. To describe the structure of path-indices for the points of $\tilde U$ over~$z$ (i.e.\ the points of $\tilde U_0$), we follow the scheme developed above.  Consider a small ball $D \subset \cX$
centred at $z$, select an arbitrary point $z'$ in $\cX \setminus \sigma$ and introduce a local fundamental group $\pi_1^{loc} [z]$ and the map~$\ppsi$. One can formulate the following lemma, whose conditions are slightly more restrictive than the conditions of Lemma~\ref{le:locality}:

\begin{lemma}
\label{le:locality_x}
Let $z$ be a crossing of exactly two singularities, say $\sigma_j$ and $\sigma_k$. Assume further that
the point $z$ is regular on the singularities (i.e.\ gradients $\nabla g_j$ and $\nabla g_k$ 
at $z$ are non-zero), and 
that the crossing is transversal (i.e.\ the gradients $\nabla g_j$ and $\nabla g_k$ at $z$ are not proportional).  Then 

a) The local subgroup $\pi_1^{loc} [z]$ is isomorphic to $\mathbb{Z}^2$. The generators for the group are $\ppsi( \gamma_1)$, $\ppsi( \gamma_2)$,
where $\gamma_1$ and $\gamma_2$ are simple loops about $\sigma_j$ and $\sigma_k$ in $\Pi_D (z',z')$. 

b) The set of all possible $\gamma$ that are path-indices of $(z, \gamma) \in \tilde U$ is the set of left cosets of 
the subgroup $\pi_1^{loc} [z]$ in $\pi_1 (\cX \setminus \sigma)$.

\end{lemma}

The proof of point a) is different from Lemma~\ref{le:locality}, so we should comment on it here. 
Introduce  local coordinates $w=(w_1, w_2)$ in $D$ by 
\[
w_1 = g_j (z), \qquad w_2 = g_k(z). 
\]
The singularities in the new coordinates are $w_1 = 0$ and $w_2 = 0$. There exists a strict deformation retraction of $D \setminus \sigma$ on a real 2D torus given by 
\[
(w_1 , w_2) \to \left( \frac{w_1}{|w_1|} , \frac{w_2}{|w_2|} \right).
\]
As it is known, the fundamental group of a topological space is the same as that of its retract. 
\mesNik{As it is well-known, the fundamental group of a torus is~$\mathbb{Z}^2$. Let us demonstrate this in a simple way.}

\begin{figure}[h]
\centering
  \includegraphics[width=0.8\textwidth]{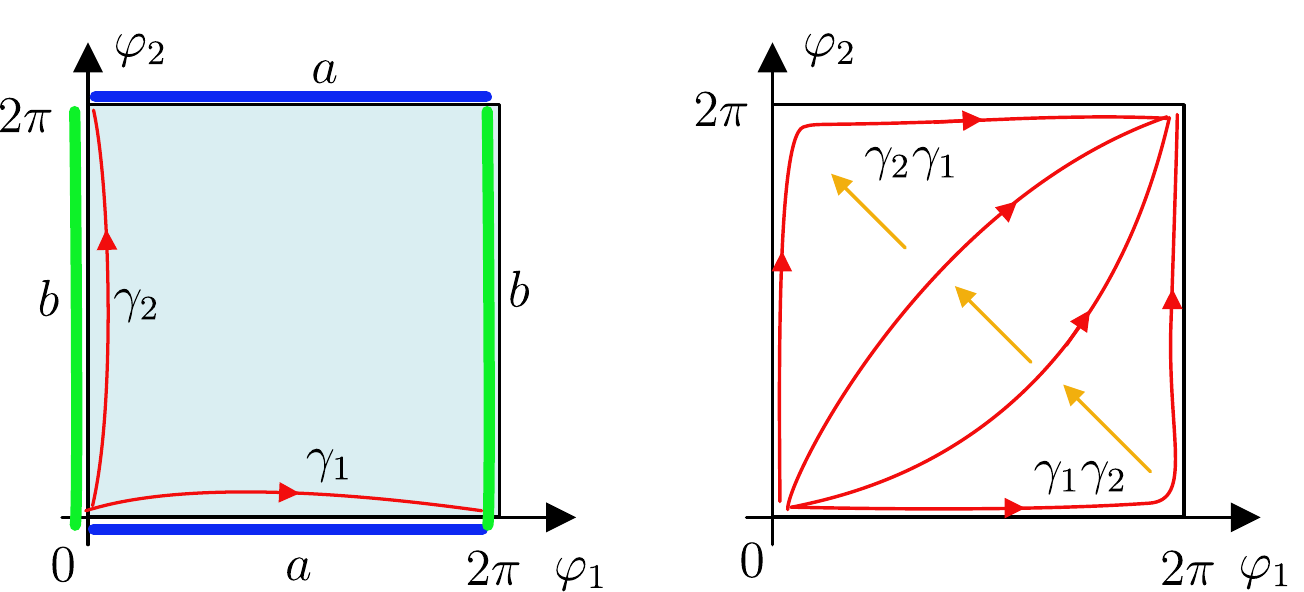}
  \caption{Computation of $\pi_1$ of a torus}
  \label{f:01013}
\end{figure}

The simple loop $\gamma_1$ about $\sigma_j$ is as follows: the variable $w_2$ is fixed, and the variable $w_1$ makes a single loop about zero in the positive direction. The bypass $\gamma_2$
is constructed similarly. 
To prove pt.~a), we should show that the bypasses $\gamma_1$ and $\gamma_2$ commute, i.e.\ that the 
paths $\gamma_1 \gamma_2$ and $\gamma_2 \gamma_1$ are homotopical.  For this, introduce the coordinates 
on the torus: 
\[
\vph_1 = {\rm Arg}[w_1], 
\qquad 
\vph_2 = {\rm Arg}[w_2] .
\]
The torus is  the square $[0,2\pi] \times [0,2\pi]$ in the real plane $(\vph_1, \vph_2)$
with the opposite sides identified (see \figurename~\ref{f:01013}, left, the sides are $a$ and $b$).
The required homotopy is shown in the right part of the figure.

To finish the proof of a) 
we should demonstrate  that no nontrivial paths in $\pi_1 (D\setminus \sigma)$ become trivial 
in $\pi_1 (\cX \setminus \sigma)$. For this, one should consider the function 
\[
f(z) = \log(g_j (z)) + \beta \log(g_k(z))
\]
with arbitrary irrational $\beta$. This function takes different values on different sheets of $\tilde U$ in $D$, so no external path can trivialize this. 

Note that in this proof it is essential that there are {\em two\/} different functions $g_j$ and $g_k$, so this reasoning breaks for a point of self-crossing of a single singularity component.  

The proof of b) is similar to that of Lemma~\ref{le:locality}.

\begin{figure}[h]
\centering
  \includegraphics[width=0.6\textwidth]{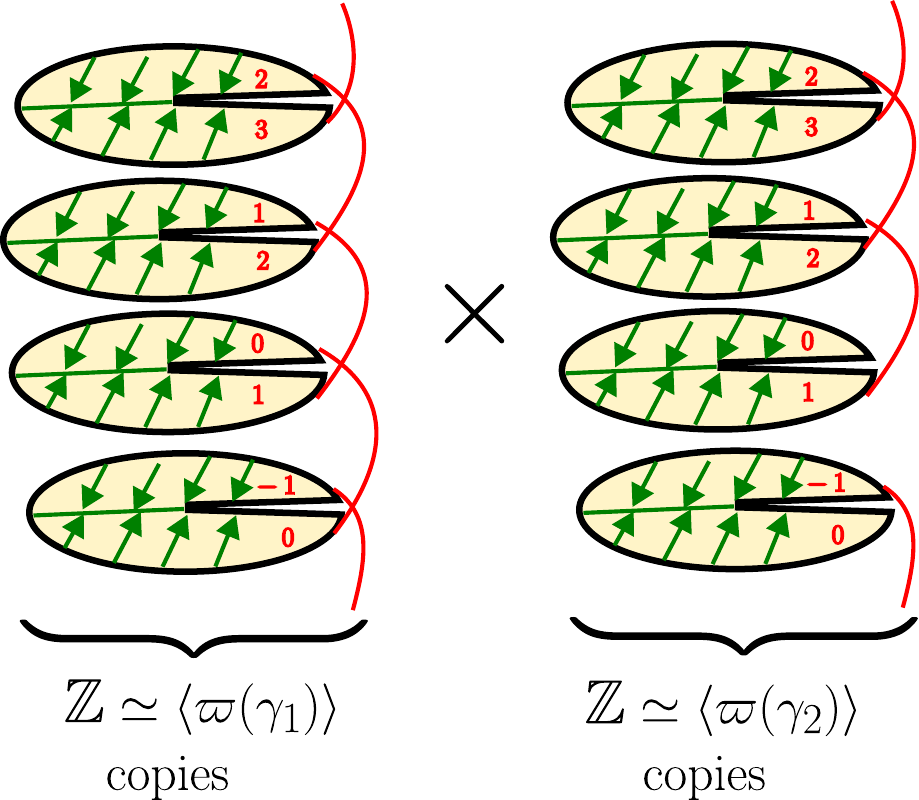}
  \caption{Topology of $W$ for $z\in\sigma_j\cap\sigma_k$.}
  \label{f:N2}
\end{figure}

\mesNik{For any point $z\in\sigma_j\cap\sigma_k$, $j\neq k$, the space $\pp^{-1}(D(z))$ (see \eqref{def:p_projection}) is homeomorphic to an infinite number of copies of some space $W$. Each such copy corresponds to an element of $\pi_1(\cX\backslash\sigma)/(\langle\varpi(\gamma_1)\rangle\times\langle\varpi(\gamma_2)\rangle)$, and $W$ is homeomorphic to ($\mathbb{Z}\simeq\langle\varpi(\gamma_1)\rangle$ copies of $2$-ball)$\times(\mathbb{Z}\simeq\langle\varpi(\gamma_2)\rangle$ copies of $2$-ball), glued together as \purple{shown in} \figurename~\ref{f:N2}.}

\FloatBarrier

\subsubsection{Structure of $\tilde{\partial\bB}$}

\begin{figure}[h]
\centering
  \includegraphics[width=0.6\textwidth]{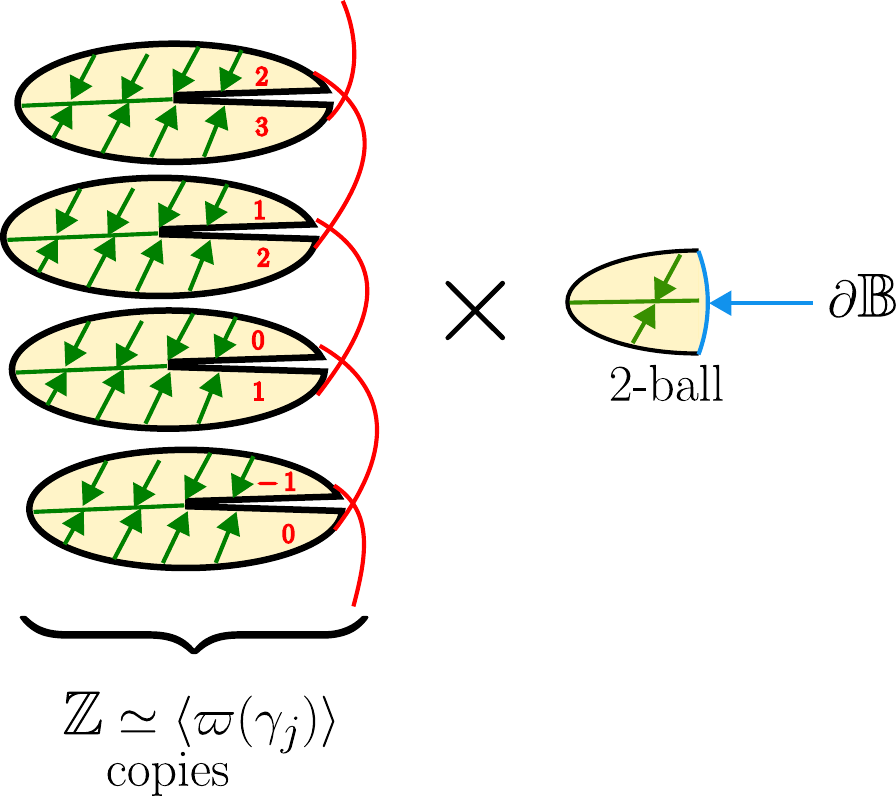}
  \caption{Topology of $W$ for $z\in\partial\bB\cap\sigma_j$ }
  \label{f:N3}
\end{figure}

\mesNik{If $z\in\partial\bB\backslash\sigma$, then the space $\pp^{-1}(D(z))$ (see \eqref{def:p_projection}) is homeomorphic to an  infinite number of copies (one-to-one with $\pi_1(\cX \backslash \sigma)$) of a 4-ball. If $z\in\partial\bB\cap\sigma_j$, then the space $\pp^{-1}(U(z))$ is homeomorphic to ($\pi_1(\cX\backslash\sigma)/\langle\varpi(\gamma_j)\rangle$ copies of 2-ball)$\times$2-ball, glued together as shown in  \figurename~\ref{f:N3}.}

\FloatBarrier


\subsubsection{Commutativity of $\pi_1 (\NR^2 \setminus \sigma)$ under some assumptions}

Throughout this work, we will use many times the fact that the bypasses about the singularities crossing  transversally commute. Namely, we use the following lemma:

\begin{lemma}
\label{le:commutativity}
Let the irreducible singularity components $\sigma_j$ be such that 
\begin{itemize}
\item 
they possesses the real property, 
\item
they have only regular points in $\NR^2$ (i.e.\ $\nabla g_j \ne {\bf 0}$),
\item
each two singularity components cross, 
\item
there are no triple (or more) crossings,
\item
each crossing is transversal.
\end{itemize}

Then the fundamental group $\pi_1 (\NR^2 \setminus \sigma)$
is generated by simple loops about $\sigma_j$, and these bypasses commute,
i.e.\ the group is Abelian.
\end{lemma}

We don't prove this lemma in details here, but the sketch of the proof is as follows. 
The possibility to represent any $\gamma \in \pi_1 (\NR^2 \setminus \sigma)$ by  a product of simple loops is discussed in \cite[V.1]{Pham2011}. 
Then, using Lemma~\ref{le:locality_x}, one  can prove that all elementary loops about the same singularity component $\sigma_j$ are homotopical. 
Finally, we establish the commutativity by using Lemma~\ref{le:locality_x} again. 

\begin{remark}
 This lemma follows from Fulton -- Delign theorem, see \cite{Fulton1980}, 
 \cite{Deligne1981}.
\end{remark}

\subsubsection{An example of $\tilde U$}
\label{ex:2Dsing}

\begin{figure}[h]
\centering
  \includegraphics[width=0.8\textwidth]{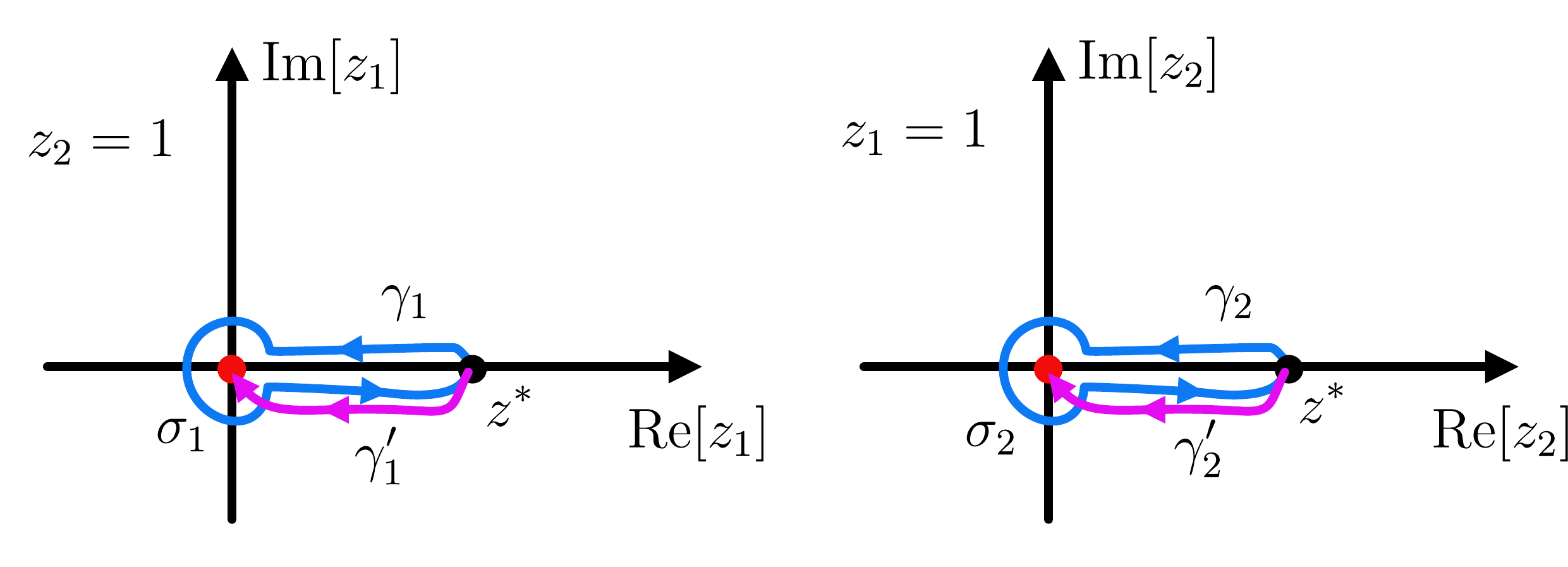}
  \caption{Contours for Example~\ref{ex:2Dsing}}
  \label{f:01003i3}
\end{figure}

Consider the case $z = (z_1, z_2) \in \mathbb{C}^2$. 
Define $\sigma$ be
\[
\sigma = \sigma_1 \cup \sigma_2, 
\qquad 
\sigma_1 : \, z_1 = 0 ,
\qquad 
\sigma_2 : \, z_2 = 0 .
\]
Let us start by building the stratum $\tilde U_2$.
Take, say,  $z^* = (1,1)$.
Introduce the simple loops $\gamma_1 , \gamma_2 \in \Pi(z^*, z^*)$
bypassing $\sigma_1$ and $\sigma_2$. These loops are shown in
\figurename~\ref{f:01003i3}. 

According to Lemma~\ref{le:locality_x}, the loops $\gamma_1$ and $\gamma_2$ commute.
Thus, 
\[
\pi_1 (\mathbb{C}^2 \setminus \sigma) = \mathbb{Z}^2, 
\]
and each element of $\pi_1$ is $\gamma_1^{k_1} \gamma_2^{k_2}$ with $k_1 , k_2 \in \mathbb{Z}$.

Let us now build the stratum $\tilde U_1$. Consider the points \purple{$z' = (0,1)$ and $z'' = (1,0)$}
belonging to $\sigma_1$ and $\sigma_2$, respectively.  
Introduce the paths $\gamma_1'$ and $\gamma_2'$ as it is shown in 
\figurename~\ref{f:01003i3}. 
According to Lemma~\ref{le:locality},
paths from $\Pi(z^*,z')$ can be written as $\gamma_2^k \gamma_1'$, $k \in \mathbb{Z}$,
and paths from $\Pi(z^*,z'')$ can be written as $\gamma_1^k \gamma_2'$.

According to Lemma~\ref{le:locality_x}, the local group for $z = (0,0)$
is the whole $\mathbb{Z}^2$, \purple{the local group for the stratum $\tilde U_1$ is $\mathbb{Z}$} and
the stratum $\tilde U_0$ consists of a single point.


\subsection{Riemann domain of a function}

Consider some possibly multivalued 
analytical function $F(z)$, $z \in \cX$ having singularities only at~$\sigma$.
Here {\em analytical\/} means that there exists analytical continuation of $F$ along any path in $\cX \setminus \sigma$.

As it is well-known, the Riemann domain of $F(z)$ over $\cX$ can be defined as a set of pairs $(z,\nu)$, where $z \in \cX \setminus \sigma$, and $\nu$ is the germ of $F$ at~$z$
(we assume that the germs are indexed somehow).
Denote the Riemann \mesNik{domain }  of $F$ by~$\hat U$. Our aim is to connect $\hat U$ with~$\tilde U$.  

\begin{proposition}
\label{pr:single-valued}
The function $F(z)$ can be defined as a single-valued function on $\tilde U_2$. 
\end{proposition}

The proof is obvious. 
Take some reference point $z^* \in \cX \setminus \sigma$ and select arbitrarily one of the germs   of $F(z)$
at $z^*$; refer to it as the reference germ.
 To each point $(z, \gamma) \in \tilde U_2$, assign the value of $F$ obtained by analytical continuation of the reference germ along~$\gamma$. This procedure cannot give incorrect results, since by definition of the universal Riemann domain, two different paths cannot end at the same point of~$\tilde U$. 

Use the reference point and the reference germ defined for the proof of the proposition and take 
$ \pi_1(\cX \setminus \sigma) = \Pi (z^* , z^*)$.
Introduce the subgroup $A \subset \pi_1(\cX \setminus \sigma)$
as the set of all elements going from the reference germ to the same reference germ. 

\begin{lemma}
\label{le:continuation}
The set of germs $\nu$ in the definition of the Riemann surface of $F$ is the set of right cosets of $A$ in $\pi_1 (\cX \setminus \sigma)$: 
\[
\nu = A \gamma, 
\qquad
\gamma \in \pi_1 (\cX \setminus \sigma).
\]
\end{lemma}

This lemma expresses the {\em principle of analytical continuation:} 
if $\gamma_1,\gamma_2 \in A$, $\gamma_3 \in \pi_1 (\cX \setminus \sigma)$, then 
$\gamma_1 \gamma_3$ and $\gamma_2 \gamma_3$ lead to the same germ (not necessarily to \purple{the reference germ}). Moreover, if $\gamma_1, \gamma_2 \in \pi_1(\cX \setminus \sigma)$ lead to the same germ, 
then $\gamma_2 \gamma_1^{-1} \in A$. 

Lemma~\ref{le:continuation} reproduces the statement from \cite[VII]{Pham2011}. Note that there are {\em left\/} cosets instead of {\em right\/}, since, in \cite{Pham2011}, convention of writing the product in $\pi_1(\cX \setminus \sigma)$ is different from ours. 

Thus, Lemma~\ref{le:continuation} establishes a map 
\begin{equation}
\Psi :\, \tilde U_2 \to \hat U, 
\qquad 
(z , \gamma) \to (z, \nu),
\label{e:map_Psi}
\end{equation}
such that the affix $z$ remains the same and the element $\gamma$ is mapped to the right coset to which it belongs.

\begin{remark}
The last statement just repeats the covering classification theorem, according to which coverings are classified by cosets of subgroups of the fundamental group of the base. 
\end{remark}


\section{Representation of surfaces through relative homologies}
\label{sec:3}
\subsection{Algebraic notations}

\subsubsection{An overview. Homology groups under consideration}
\label{sec:Overview_Homology}
Here and below we assume $\cX$ to be $\NR^2$. 
Let the singularity $\sigma$ have the real property.  
Denote 
\begin{equation}
\PI = \pi_1 (\NR^2 \setminus \sigma). 
\label{e:def_PI}
\end{equation}

We will study the following homology groups: 

\begin{itemize}

\item 
$H_2 (\tilde U_2 , \tpB)$. This is the group of admissible boundaryless integration surfaces for the integrals of class (\ref{e:i0001}). 
The elements of the homologies avoid the singularities and can extend to infinity, i.e.\ to $ \tpB$. The branched structure of $\tilde U_2$ enables one to handle arbitrary branching of any function~$F$. 

\item 
$H_2 (\tilde U ,\tpB)$. This is the group of homologies that are 
similar to $H_2(\tilde U_2 ,\tpB)$ but
allowed to pass through singularities. 
The main result of this section is the isomorphism $H_2 (\tilde U, \tpB) \simeq H_2 (\tilde U_2, \tpB)$.

\item 
$H_2(\tilde U, \tilde U' \cup \tpB)$. 
This is the relative homology group generated by polygons that may have boundaries on the strata 
of smaller dimensions.
This group has a simple structure, and its ramification (see below) admits an explicit description. 
The group 
$H_2 (\tilde U , \tpB)$
can be identified with a subgroup of boundaryless elements of $H_2(\tilde U, \tilde U' \cup \tpB)$ 
(the boundary is understood in the sense of the homomorphism from the exact sequence of homology for a triple \eqref{triple}). 

\item 
$H_1 (\tilde U' , \tpB)$.
This group contains ``boundaries'' of the elements of 
$H_2 (\tilde U , \tilde U' \cup \tpB)$ in the sense of the homomorphism
\[
H_2 (\tilde U , \tilde U' \cup \tpB)\rightarrow H_1 (\tilde U' , \tpB)
\]
of the exact sequence of the triple of spaces
\[
(\tilde U , \tilde U' \cup \tpB , \tpB).
\]
{
This homomorphism is intuitively clear: it is just taking boundaries of the polygons 
belonging to $H_2 (\tilde U , \tilde U' \cup \tpB)$. The boundaries are taken modulo~$\tpB$.}

\end{itemize}

All homology groups have coefficients in $\mathbb{Z}$. 
For all groups, only homologies sitting on a finite number of sheets of $\tilde U$ are allowed. 

Some of the groups under consideration can be organized into the following commutative diagram: 
\begin{equation}
		\begin{CD}\label{cd1}
			H_{2}(\tilde{U_2}) @>>> H_{2}(\tilde{U},\tilde{U}')\\
			@VVV @VVV \\
			H_{2}(\tilde{U_2} ,\tilde{\partial \bB}) @>>>  H_{2}(\tilde{U},\tilde{U}'\cup\tilde{\partial \bB}).\\
		\end{CD}
	\end{equation}
    
The upper horizontal arrow is defined by the embedding of absolute chains into relative ones, the lower one is similar, only for chains modulo $\tpB$, the vertical arrows are the reduction modulo $\tpB$.

Our aim is to describe the ramification of some element of $H_2(\tilde U_2, \tpB)$ under some loop $\lambda$ passed by the parameter $t$ about the Landau set. 
For that, we use the following scheme. Using the established isomorphism between $H_2 (\tilde U ,\tpB)$ and $H_2 (\tilde U_2 ,\tpB)$, 
we study the ramification of a corresponding element of $H_2(\tilde U ,\tpB)$ under the same loop~$\lambda$.
The image of the element of $H_2(\tilde U ,\tpB)$ under injection $H_2(\tilde U, \tpB) \rightarrow 
H_{2}(\tilde{U},\tilde U' \cup \tpB)$, see Proposition~\ref{pr:inclusion}  is written as a linear combination of elements of $H_2(\tilde U, \tilde U' \cup \tpB)$. Using the linearity {of homologies}, we study the ramification of the elements of 
$H_2(\tilde U, \tilde U' \cup \tpB)$, which can be represented by a matrix multiplication, and return to 
$H_2 (\tilde U_2 ,\tpB)$ through $H_2 (\tilde U , \tpB)$.

The outlined scheme is the diagram (\ref{e:015006m}), and it is explained in details later. 



\begin{figure}[h]
\centering
  \includegraphics[width=0.9\textwidth]{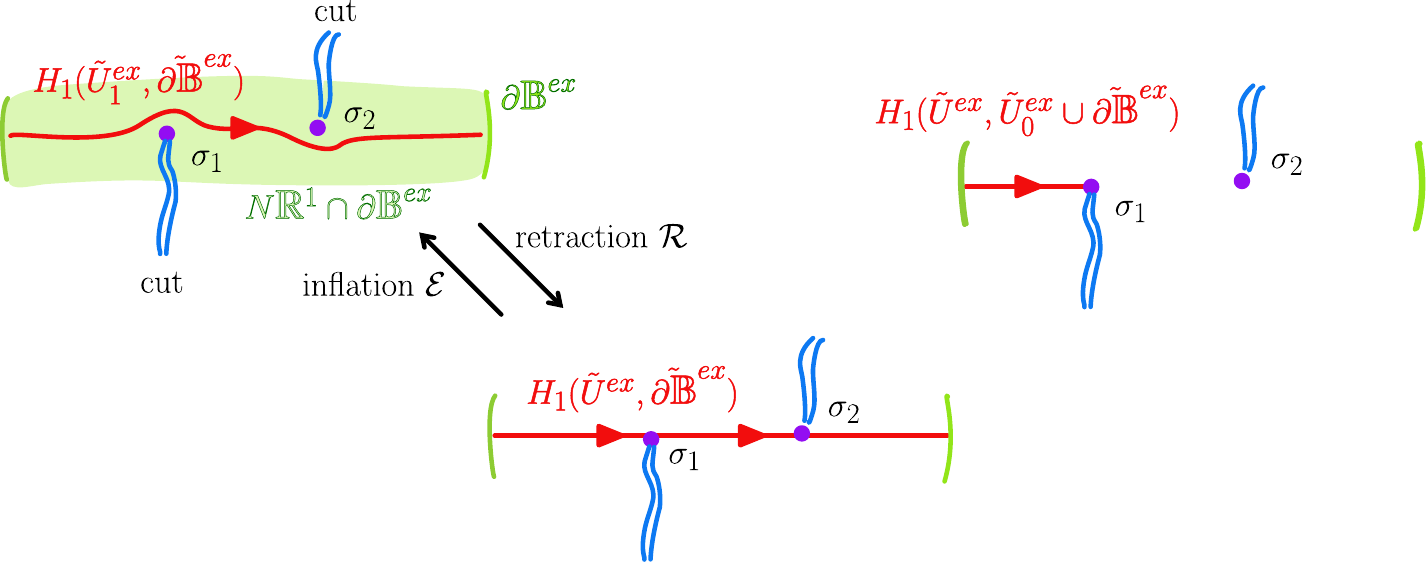}
  \caption{Examples of representatives of different homology groups in the 1D case}
  \label{f:1D_example}
\end{figure}

To give some visual understanding of the difference between homology groups introduced here, let us consider groups playing a similar role in the 1D case. In this case, we consider $\NR^1 \subset \mathbb{C}$.
Introduce also a ball $\bB^{ex}$, such that its boundary $\tpB^{ex}$ acts as the ``infinity''.
The singular set $\sigma$ is a union of several discrete points~$\sigma_j$. The universal Riemann domain 
$\tilde U^{ex}$ has two strata $\tilde U_1^{ex}$ and~$\tilde U_0^{ex}$ (the index $ex$ stands for the 1D example). 
The analogues of the groups $H_2(\tilde U_2, \tpB)$, $H_2(\tilde U , \tpB)$, $H_2(\tilde U_2, \tilde U' \cup \tpB)$ are
the groups $H_1(\tilde U_1^{ex} , \tpB^{ex})$, $H_1(\tilde U^{ex} , \tpB^{ex})$, 
$H_1(\tilde U_1^{ex}, \tilde U_0^{ex} \cup \tpB^{ex})$, respectively. Examples of representatives of these groups are shown in \figurename~\ref{f:1D_example}.
We indicate in this figure the retraction and the inflation procedures that will be introduced below.

\subsubsection{Description of $H_2(\tilde U, \tilde U' \cup\tpB)$ and the group ring $\Omega$}
\label{sec:Algebra_of_H2_overview}

Let us introduce the following notations:
\begin{equation}
\tilde U^* = \{ (z, \gamma) \in \tilde U \, \, : \, \,  z\in \mathbb{R}^2\} ,
\label{e:U_retract}
\end{equation}
\begin{equation}
\tilde U'^* = \{ (z, \gamma) \in \tilde U \, \, : \, \,  z\in \sigma\cap\mathbb{R}^2\}.
\label{e:U_retrac'}
\end{equation}

Let us make an important preliminary statement: 

\begin{proposition}
\label{pr:retract}
Let all crossings of the singularity components $\sigma_j$ be transversal and assume that there are no triple crossings.
Then we can choose a sufficiently  small tubular neighbourhood $N\mathbb{R}^2$, such that  the space $\tilde U^*$
is a strict deformation retract of~$\tilde U$; moreover, the triple 
\[
(\tilde U^*, \, \tilde U'^*\cup(\tpB \cap\tilde U^*), \, \tpB \cap\tilde U^*)
\]
is a strict deformation retract of the triple 
\[
(\tilde U, \, \tilde U'\cup \tpB , \,\tpB ). 
\]
The retraction 
$\tilde U \to \tilde U^*$ will be denoted by $\cR$. 
\end{proposition}

\textbf{Proof.} 

\noindent 
{\bf 1.}
 Let us show that the triple of spaces 
 \[
 (\mathbb{R}^2\cap \bB, \, \mathbb{R}^2 \cap (\sigma\cup \partial \bB), \, \mathbb{R}^2\cap\partial \bB)
 \] 
 is a deformation retract of the triple 
 \[
 (\NR^2 \cap \bB, \, \NR^2 \cap(\sigma\cup\partial \bB), \, \NR^2 \cap \partial \bB).
 \]
 We define a continuous family of transformations in some local coordinates $w=(w_1, w_2) \in \mathbb{C}^2$:
 \[\varphi_{\alpha}: \quad  (\text{Re}[w_1],\text{Im}[w_1],\text{Re}[w_1],\text{Im}[w_2])\rightarrow
 (\text{Re}[w_1],\alpha\text{Im}[w_1],\text{Re}[w_2],\alpha\text{Im}[w_2]),
 \]
 \[
  \qquad \qquad \qquad \qquad \qquad  \qquad \qquad \qquad \alpha\in[0,1],
 \]
  connecting the identity mapping with retraction $\varphi_0$.
 
  Let us build a retraction locally. Consider the neighborhoods of points $z \in \bB \cap\mathbb{R}^2$.
    There are five possible cases. It is easy to see that in all cases the required retraction is given by~$\varphi_0$.

\begin{enumerate}

\item    
$z \in\mathbb{R}^2\cap(\bB\setminus(\partial \bB\cup\sigma))$. We can take global coordinates $z$ as~$w$.

\item
$z \in\mathbb{R}^2\cap(\partial \bB \backslash\sigma).$ Using a local diffeomorphism $\mathbb{R}^4\simeq\mathbb{C}^2$ that preserves $\mathbb{R}^2$, we pass to the coordinates $w$, centered at the point $z$, in which $\partial \bB$ is given by $\text{Re}[w_1]=0$.

\item
$z\in \bB \cap \mathbb{R}^2\cap(\sigma^{(1)}\setminus\partial \bB),$ i.e\ $z$ is a point of the maximal strata of $\sigma$. 
By the implicit function theorem, locally one can make a biholomorphic coordinate change 
$(z_1 , z_2) \rightarrow (x,y)$, 
such that $\sigma$ becomes a graph of a holomorphic function $y=f(x)$,  and
$f(\mathbb{R})\subset\mathbb{R}$. 
Take $w_1=x$, $w_2=y-f(x)$. Then $\sigma$ will be given by $w_2=0$. 

\item
$z\in \bB\cap \mathbb{R}^2\cap\sigma^{(0)}$, i.e.\ $z$ is a point of transversal intersection of ~$\sigma$. 
Similarly, we can choose local coordinates $w$ in which $\sigma$ is defined by $w_1 w_2 = 0$.

\item
\purple{$z\in\mathbb{R}^2\cap\partial \bB\cap \sigma^{(1)}$.} In this case, there are coordinates in which $\partial \bB$ is given by $\text{Re}[w_1]=0$ and $\sigma$ is given by $w_2=0.$

\end{enumerate}

Let us define retractions arbitrarily in the neighborhoods of points of the set $\sigma^{(0)}$. It is easy to see that there are no obstructions to continuing them to strata of higher dimensions (since $\sigma' \cap\bB$ is compact, we can select a finite subcover from our cover by neighborhoods, and take the value of $\delta$ that is smallest over all neighborhoods).  

\vskip 6pt
\noindent
{\bf 2.}
We can lift the constructed retraction to a retraction  of $\tilde U$ onto $\tilde U^*$
that will also be denoted~$\cR$ (see \figurename~\ref{f:N1}--\ref{f:N3}, green lines). $\Box$

\vskip 6pt

A corollary of this proposition  is as follows: 

\begin{proposition}
\label{pr:retraction}
One can study $H_2(\tilde U^* , \tpB)$ and $H_2 (\tilde U^*, \tilde U' \cup \tpB)$ 
instead of $H_2(\tilde U , \tpB)$ and $H_2 (\tilde U, \tilde U' \cup \tpB)$, respectively: 
\begin{equation}
H_2(\tilde U , \tpB) \simeq H_2(\tilde U^* , \tpB), 
\qquad 
H_2(\tilde U, \tilde U' \cup \tpB) \simeq H_2(\tilde U^* , \tilde U' \cup \tpB). 
\label{e:retracts}
\end{equation}
\end{proposition}

(This follows from the fact that the deformation retraction is a homotopy equivalence and in view of the homotopy invariance of homology.)

Another obvious proposition: 

\begin{proposition}
As a basis of the group $H_2 (\tilde U^*, \tilde U' \cup\tpB)$, one can take the set of classes of possibly curvilinear polygons, into which the ball $\bB \cap \mathbb{R}^2$ 
is split by~$\sigma'$, each taken on all different sheets of~$\tilde U$. 
\end{proposition}

\noindent 
\purple{It} can be transferred into a similar statement about $H_2(\tilde U, \tilde U'\cup\tpB)$
as a corollary from Proposition~\ref{pr:retract}.

Let us formalize this statement.
Take a reference point $z^*$ as it is done in the definition of~$\tilde U$.
Let the lines $\sigma'_j$ split the ball $\bB \cap \mathbb{R}^2$ into $k$ polygons
$A,B,C, \dots$
(if the geometry is not fixed we can introduce aliases for these polygons by  
$Q^1  = A$, $Q^2 = B$, \dots).
The polygons are assumed to be oriented in a standard way. 
These polygons form a basis of $H_2(\mathbb{R}^2, \sigma \cup \partial \bB)$.

Take {\em reference points\/} on each such polygon (they can be arbitrary points inside the polygons). 
Denote these points by  $z_A,z_B, z_C,\dots$ (or by $z_{Q^1}, \dots , z_{Q^k}$). 
Introduce some arbitrary {\em base paths}
\[
\gamma_A \in \Pi(z^*, z_A),
\qquad 
\gamma_B \in \Pi(z^*, z_B),\dots.
\]
Each point of $\tilde U$ over, say $z_A$, is described as $(z_A, \gamma \gamma_A)$, $\gamma \in \PI$ (see Proposition~\ref{pr:U2_str}).
If point $(z_A, \gamma \gamma_A)$ belongs to some polygon~$A$ on a certain sheet of $\tilde U$, this polygon can be denoted 
$A_\gamma$ (polygon $A$ on the sheet $\gamma$). Indeed, 
$A_\gamma \in H_2(\tilde U , \tilde U' \cup \tpB)$.

Some visual examples of these notations are given in the next subsection.
It is easy to see that the index $\gamma$ does not depend on the choice of the reference point $z_A$ inside 
the polygon~$A$.

An arbitrary element $w \in H_2 (\tilde U , \tilde U' \cup \tpB)$ is a linear combination of all such elements: 
\begin{equation}
w = \sum_{l = 1}^k \sum_{\gamma \in \PI}\alpha_{l,\gamma} Q^l_\gamma , 
\qquad \alpha_{l,\gamma} \in \mathbb{Z},
\label{e:element_relative}
\end{equation}
provided that only a finite number of the coefficients $\alpha_{l,\gamma}$ are non-zero. 

Let us change the notation (\ref{e:element_relative})
to make it useful for further analysis. 
Introduce the {\em free group ring} $\Omega$  for $\PI$
over~$\mathbb{Z}$. The elements of $\Omega$ are 
formal linear combinations of $\gamma$ with integer coefficients:
\begin{equation}
\omega = \sum_{\gamma\in \PI} \beta_\gamma \gamma, 
\qquad 
\beta_\gamma \in \mathbb{Z}. 
\label{e:01318}
\end{equation}
The summation is held over all elements of $\PI$, but we assume that only a finite number of coefficients are non-zero. 
If some $\gamma$ does not participate in the sum, 
this is equivalent to $\beta_\gamma = 0$. The zero element of the ring 
is the empty sum (all coefficients are zero).  

{
The group $\PI$ is naturally embedded in $\Omega$ by the following rule 
\[
1 \gamma \in \Omega \quad \mbox{if} \quad \gamma \in \PI.
\]
}

One can define a sum of elements {of $\Omega$: if}, in addition to (\ref{e:01318}),
\[
\omega' = \sum_{\gamma \in \PI}\beta'_\gamma \gamma  
\]
then 
\[
\omega + \omega' = \sum_\gamma
(\beta_\gamma + \beta'_\gamma) \gamma. 
\]

A product of two elements {of $\Omega$ } is defined in a natural way. Say,  
\[
(\beta_1 \gamma_1 + \beta_2 \gamma_2 + \beta_3 \gamma_3)
(\beta_4 \gamma_4 + \beta_5 \gamma_5)
= 
\]
\[
\beta_1 \beta_4 \, \gamma_1 \gamma_4 + 
\beta_2 \beta_4 \, \gamma_2 \gamma_4 + 
\beta_3 \beta_4 \, \gamma_3 \gamma_4 + 
\beta_1 \beta_5 \, \gamma_1 \gamma_5 + 
\beta_2 \beta_5 \, \gamma_2 \gamma_5 + 
\beta_3 \beta_5 \, \gamma_3 \gamma_5
\]
($\gamma_j \gamma_k$ is a product of group elements).
The unit element of $\Omega$ is $1e = e$. Note that the multiplication is generally 
not commutative, so the order of factors is important.  

{
Define a multiplication 
\begin{equation}
\Omega \times H_2(\tilde U, \tilde U' \cup \tpB)  \to H_2(\tilde U, \tilde U' \cup \tpB). 
\label{e:mult_Om_H_H}
\end{equation}
For each $Q^l_\gamma$ and each $\gamma_1 \in \PI$ define
\[
\gamma_1 Q^l_{\gamma} \equiv Q^l_{\gamma_1 \gamma},
\]
in particular, one can write 
\[
Q^l_\gamma = \gamma Q^l_e. 
\]

This definition can be extended to the the whole 
$\Omega \times H_2(\tilde U, \tilde U' \cup \tpB)$
by linearity:
\[
\left( \sum_j \beta_j \gamma_j \right)
\left( \sum_l \sum_n \alpha_{l,n} Q^l_{\gamma_n} \right)
= 
\sum_j \sum_n \sum_l  \beta_j \alpha_{l,n} Q^l_{\gamma_j \gamma_n}
=
\sum_l  
\sum_j \sum_n \beta_j \alpha_{l,n} \gamma_j \gamma_n
Q^l_e
\]
for $\alpha_{l,n}, \beta_j \in \mathbb{Z}$, $\gamma_j , \gamma_n \in \PI$.

}

Return to the sum (\ref{e:element_relative}).
For  { some $l$ take arbitrary 
\begin{equation}
\omega_{l} = \sum_\gamma \alpha_{l,\gamma} \gamma, \qquad \omega_l \in \Omega.
\label{e:GR_2}
\end{equation}
Denote formally  
\begin{equation}
\omega_{l} Q^l_e \equiv \sum_\gamma \alpha_{l,\gamma} Q^l_\gamma = 
\sum_\gamma \alpha_{l,\gamma} \gamma Q^l_e.
\label{e:GR_1}
\end{equation}
}

Finally, we rewrite the general form of  $w \in H_2(\tilde U, \tilde U' \cup\tpB)$ as 
{
\begin{equation}
w = \sum_{l = 1}^k \omega_{l} Q^l_e  .
\label{e:GR_3}
\end{equation}
}
We note that the form (\ref{e:GR_3}) is nothing more than a convenient notation for a formal linear combination of 
polygons taken on different sheets of $\tilde U$ labelled by elements of~$\PI$.

We can simplify the notations even more if we note that (\ref{e:GR_3}) can be represented as a vector 
with coordinates $\omega_{1}, \dots, \omega_{k}$:
\begin{equation}
w \leftrightarrow {\bf w} = (\omega_{1} , \dots , \omega_{k}).
\label{e:GR_4}
\end{equation}
This makes sense, since we will use matrix multiplication for such vectors. 

\begin{definition}
\label{def:module}
Denote the linear space of \purple{vector-rows} of length $k$ with elements from $\Omega$ by~$\cM_k$. 
The elements of this space will be subject to summation and right-multiplication by elements of~$\Omega$.
\end{definition}


Finally, we come to the following statement: 

\begin{proposition}
Under the conditions of Proposition~\ref{pr:retract}, the elements of $H_2(\tilde U, \tilde U' \cup \tpB)$
have one-to-one correspondence with elements of $\cM_k$, where $k$ is the number of polygons, 
into which $\bB \cap \mathbb{R}^2$ is split by~$\sigma_j$.
\end{proposition}

\begin{remark}
{
Expressions (\ref{e:GR_3}) form a {\em left module\/} over~$\Omega$, since only the 
left multiplication by an element of $\Omega$ has been defined for such expressions.
The situation is more flexible for the elements of $\cM_k$: one can easily define both 
$\omega {\bf w}$ and ${\bf w} \omega$ for ${\bf w} \in \cM_k$, $\omega \in \Omega$.
}
\end{remark}


\subsubsection{Visual notations and examples}

Let us use some benefits of a rather simple geometrical situation and introduce visual notations for paths in $\NR^2 \setminus \sigma$. 
These notations are explained in detail in Appendix~\ref{app:B}.  We should note that the same notations are used in \cite[V.3.3.]{Pham2011}  with the only comment ``The interpretation of this picture is left to the imagination of the reader''.

\begin{example}
\label{ex:circle_relative}

Let $\sigma$ consist of a single component $\sigma_1$ with generating function 
\[
g_1(z) = z_1^2 + z_2^2 - 1.
\]
One can see that $\sigma_1'$ is a circle in the real plane (see \figurename~\ref{f:ex_circ}).  

\begin{figure}[h]
\centering
  \includegraphics[width=0.4\textwidth]{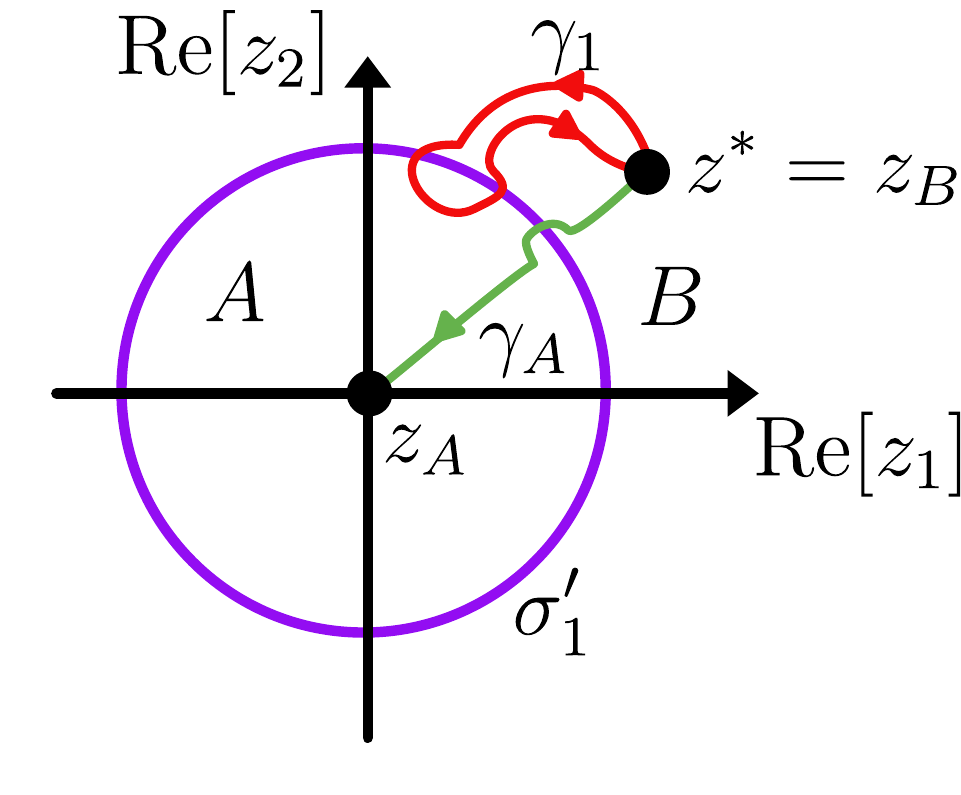}
  \caption{Notations for $H_2(\tilde U, \tilde U' \cup \tpB)$ for a circle}
  \label{f:ex_circ}
\end{figure}

The circle $\sigma'_1$ splits the real plane into two ``polygons'': inside the circle this is $A$, and outside is~$B$.
The ball $\bB$ is large, so its boundary is not visible in the figure (as well as in most of the figures below).

The visual notations can be introduced after choosing: 

\begin{itemize}
\item
a reference point $z^*$ for $\PI$, and reference points $z_A$ and $z_B$ in the polygons;

\item
some base paths $\gamma_A$, $\gamma_B$;

\item
some generator(s) of $\PI$.

\end{itemize}

 Take arbitrary reference points $z_A$ and $z_B$ in the polygons $A$ and $B$, respectively. Let the reference point $z^*$ for $\PI$ coincide with~$z_B$.  

Draw a base path $\gamma_A$ from $z^*$ to $A$ as it is shown in \figurename~\ref{f:ex_circ} (note that we are starting to use the bypass notations introduced in Appendix~\ref{app:B}). Let the base path $\gamma_B$ be trivial. 

The group $\PI$ has a single generator $\gamma_1$ that is a simple loop  
about~$\sigma_1$. Any element $\gamma \in \PI$ can be written as 
\[
\gamma = \gamma_1^n, \qquad n \in \mathbb{Z}.
\]
Obviously, $\PI$ is Abelian.

The group ring $\Omega$ for this example is the set of finite sums
\[
\omega = \sum_n \beta_n \gamma_1^n,
\qquad
\beta_n \in \mathbb{Z},
\]
i.e.\ any $\omega$ can be treated as a power sum of $\gamma_1$. The summation and the multiplication in $\Omega$ are performed just as for power sums. 

Any element $w\in H_2(\tilde U, \tilde U' \cup \tpB)$ is a linear combination of some 
samples of polygons $A$ and $B$, taken on different sheets of $\tilde U$. 
Any polygon over $A$ taken on a certain sheet is denoted~$A\gamma$. Such a notation means that the point $(z_A, \gamma \gamma_A)$ belongs to this polygon. 
Similarly, for polygons over $B$ introduce the notation~$B_\gamma$.

According to the introduced notation, 
each element $w \in H_2(\tilde U , \tilde U')$ can be written as 
\[
w = \omega_A A_e + \omega_B B_e. 
\]
Besides, we introduce the vector 
\[
{\bf w} = (\omega_A, \omega_B) \in \cM_2.
\]
Let us explain the meaning of this vector with an example. Assume, for example, that 
\[
{\bf w}  = (2\gamma_1^0 - \gamma_1^5, 3 \gamma_1). 
\]
Then 
\[
w = 2 A_{e} - A_{\gamma_1^5} + 3 B_{\gamma_1}. 
\]

\end{example}

\begin{example}
\label{ex:triangle_relative}
Consider a singularity $\sigma$ composed of three complex straight lines 
$\sigma_1$, $\sigma_2$, $\sigma_3$ in general position. 
For example, they may be the lines 
\[
\sigma_1: \quad z_2 = 0, 
\qquad 
\sigma_2: \quad z_1+ z_2 - 1 = 0,
\qquad
\sigma_3: \quad z_1 - z_2 = 0. 
\]

The three real traces $\sigma_1'$, $\sigma_2'$, $\sigma_3'$ split $\mathbb{R}^2$
into 7 polygons: $A, \dots, G$. 
Take a reference point $z^* = z_A$ in the finite triangle $A$ 
(see \figurename~\ref{f:01043}, left).

The generators of $\PI$
are simple loops 
$\gamma_1, \gamma_2, \gamma_3$ bypassing the singularities $\sigma_1, \sigma_2, \sigma_3$,
respectively, in the positive direction (see \figurename~\ref{f:01043}, left). 
According to Lemma~\ref{le:commutativity}, they commute:
\[
\gamma_1 \gamma_2 = \gamma_2 \gamma_1, 
\qquad 
\gamma_1 \gamma_3 = \gamma_3 \gamma_1,
\qquad 
\gamma_2 \gamma_3 = \gamma_3 \gamma_2, 
\]
and $\PI = \mathbb{Z}^3$.

Each element of $\PI$ is thus 
\begin{equation}
\gamma = \gamma_1^{m_1} \gamma_2^{m_2} \gamma_3^{m_3},
\qquad m_1, m_2, m_3 \in \mathbb{Z}.
\label{e:01065b}
\end{equation}

\begin{figure}[h]
  \centering{\includegraphics[width=0.9\textwidth]{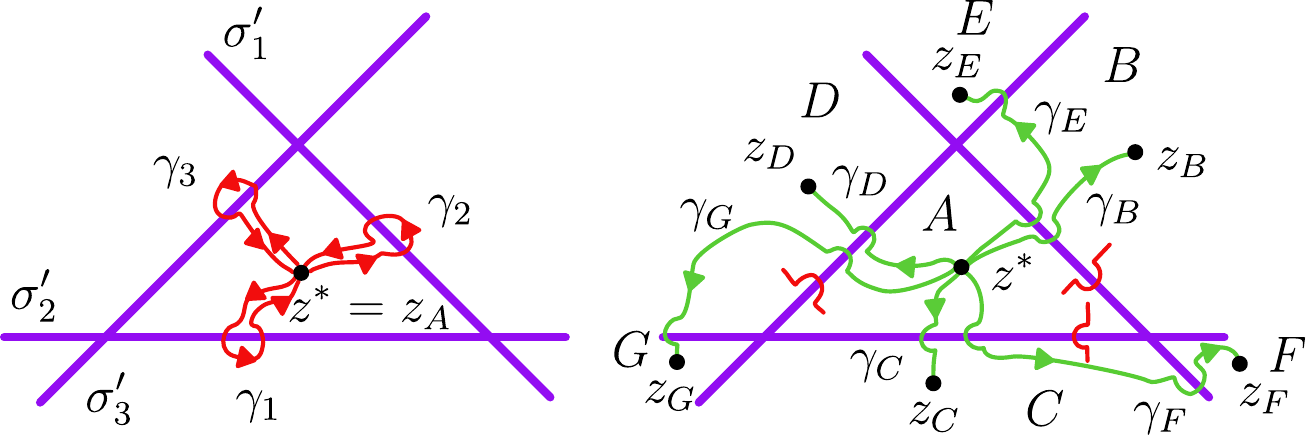}}
  \caption{Reference points and paths connecting them}
  \label{f:01043}
\end{figure}

Take a reference point in each polygon
(call them $z_B, z_C , \dots, z_G$).
Draw some base paths $\gamma_B, \gamma_C , \dots , \gamma_G$ going from $z^*$ to the reference points
($\gamma_A$ is trivial). The base paths are shown in 
\figurename~\ref{f:01043}, right.

The choice of the base paths is arbitrary, but there exists a convenient 
way to choose them. Namely, choose some surface $\Sigma$ that is a real plane slightly
indented from the singularities. According to Subsection~\ref{s:vis_surf} of Appendix~\ref{app:B},  this surface can be shown by the bridge symbols 
(red in \figurename~\ref{f:01043}, right). Assume that all base paths belong to~$\Sigma$.
One can see that the green base paths match in the figure with the red bridge notations.
The role of the surface $\Sigma$ will be explained in the next subsection.

The group ring $\Omega$ is a set of finite  formal sums 
\[
\omega = \sum_{m_1, m_2, m_3\in\mathbb{Z}} \beta_{m_1, m_2, m_3} \gamma_1^{m_1}\gamma_2^{m_2} \gamma_3^{m_3} ,
\qquad
\beta_{m_1, m_2, m_3}\in\mathbb{Z}.
\]
They can be treated as power sums of  $\gamma_1$, $\gamma_2$, $\gamma_3$:
the summation and multiplication are performed as for power sums of several variables. 
Any element $w \in H_2 (\tilde U , \tilde U' \cup\tpB)$ corresponds to a vector 
\[
{\bf w} = (\omega_A , \dots , \omega_G)  \in \cM_7.
\]
\end{example}



\subsubsection{The change of the reference point and the base paths}

Notations for the elements of $H_2 (\tilde U , \tilde U' \cup \tpB)$ depend on the choice of the reference point $z^*$ and the base 
paths $\gamma_{Q^1} , \dots , \gamma_{Q^k}$. 
As we noted, the change of position of the reference points $z_{Q^n}$ within the polygon $Q^{n}$ does not change the notations. As for the point $z^*$,
it is only important to know which polygon it belongs to. 
Let us discuss how the notations change when one changes the reference points and the base paths more strongly.

\begin{lemma}
\label{le:trans_contours}

Let $z^*$ be a real reference point, and $\gamma_{Q^1}, \dots , \gamma_{Q^k}$
be some base paths. Keeping $\sigma$ unchanged,
introduce a new reference point $\tilde z^*$
and new base paths: 
$\tilde \gamma_{Q^1}, \dots , \tilde \gamma_{Q^k}$.
Let the path connecting $z^*$ with $\tilde z^*$ in $\NR^2 \setminus \sigma$ be~$\gamma_*$.

Let some element of $H_2^*(\tilde U , \tilde U_1)$ 
be a polygon over $Q^n$ 
having notation $Q^n_\gamma$ 
for the ``old'' reference point and base paths.  
Then, for the ``new'' reference point and base paths, the notation of the same element is   
$Q^n_{\gamma_*^{-1} \gamma \gamma_{Q^n} \tilde \gamma_{Q^n}^{-1}}$.
\end{lemma}

\begin{figure}[h]
  \centering{ \includegraphics[width=0.4\textwidth]{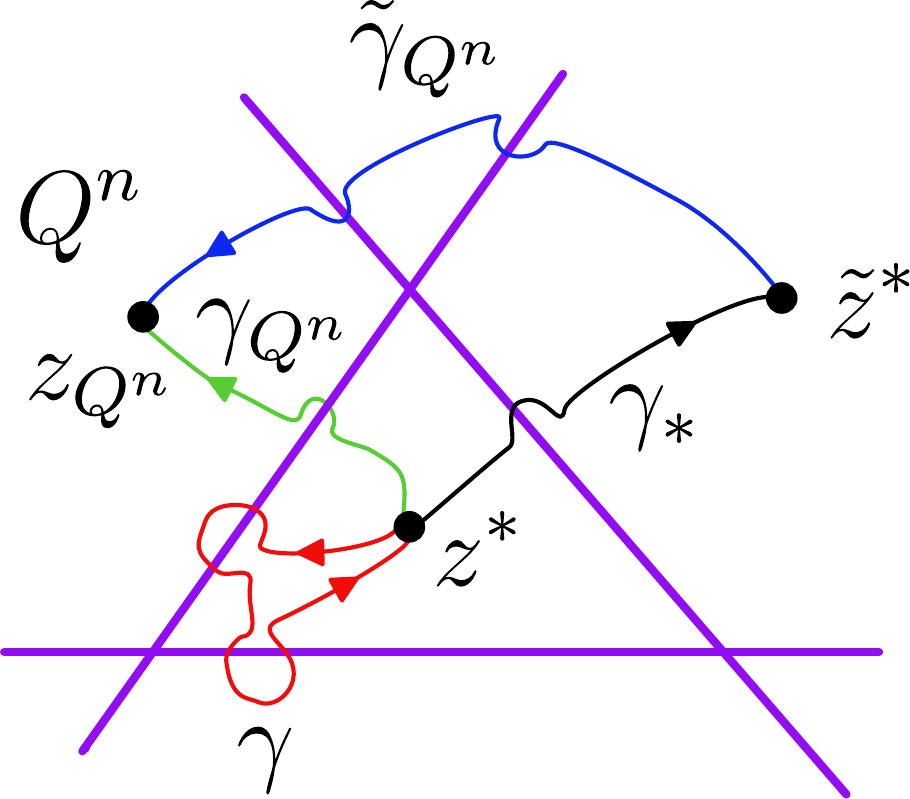} }
  \caption{To the change of contours}
  \label{f:01061x}
\end{figure}

The proof is trivial (see \figurename~\ref{f:01061x}).

Lemma~\ref{le:trans_contours} has three important corollaries, given by Proposition~\ref{pr:surface_Sigma},~\ref{pr:surface_Sigma_trivial}, and \ref{pr:path_change}. 

\begin{proposition}
\label{pr:surface_Sigma}
Let $\Sigma \subset \NR^2 \setminus \sigma$ be a slightly deformed real plane.
Let the paths $\gamma_{Q^n}$, $\tilde \gamma_{Q^n}$, and $\gamma_*$ lie in~$\Sigma$. Then 

a) $\gamma_{Q^n} \tilde \gamma_{Q^n}^{-1} = \gamma_*$;

b) if the ``old'' notation for some polygon is $Q^n_\gamma$, 
then its ``new'' notation is $Q^n_{\gamma_*^{-1} \gamma \gamma_*}$. 
\end{proposition}

The proposition 
follows from the fact that $\Sigma$ is \purple{simply}-connected, and thus the loop 
$\gamma_{Q^n} \tilde \gamma_{Q^n}^{-1} \gamma_*^{-1}$ is contractible. 

\begin{remark}
The paths $\gamma_{Q^n}$, $\tilde \gamma_{Q^n}$, and $\gamma_*$ in \figurename~\ref{f:01061x} cannot be put into any surface $\Sigma$
since $\gamma_*$ and $\gamma_{Q^n} \tilde \gamma_{Q^n}^{-1}$ bypass singularity components differently. 
\end{remark}

Let the conditions of Proposition~\ref{pr:surface_Sigma} be valid, and assume that $\PI$ is Abelian.
There are two realizations of $\PI$: with $z^*$ and $\tilde z^*$ taken as the reference point. 
For both realizations we can introduce common notations: we take simple  loops $\gamma_j$ about the singularity components $\sigma_j$ as the generators, and denote the elements of $\PI$ as $\gamma_1^{m_1} \gamma_2^{m_2}\dots$. Since $\PI$ is Abelian, the particular choice of the simple loops does not matter. Thus, any element of $\PI$ is mapped to a multiplet $(m_1, m_2, \dots)$. Note that the map described in Proposition~\ref{pr:surface_Sigma}, namely 
\[
\gamma \to \gamma_*^{-1} \gamma \gamma_*,
\]
does not change this multiplet, since the number of bypasses about each $\sigma_j$ by an element of $\PI$ is not changed by this map. 
Thus, we can formulate the following proposition:

\begin{proposition}
\label{pr:surface_Sigma_trivial}
Let the conditions of Proposition~\ref{pr:surface_Sigma} be valid, and let $\PI$ be Abelian. Then the transformation from the ``old''
to the ``new'' notations is trivial: 
\[
\tilde {\bf w} = {\bf w}.
\]
\end{proposition}

Because of Proposition~\ref{pr:surface_Sigma_trivial},
one can see that for an Abelian $\PI$ one can  uniquely set the notations for $H_2(\tilde U, \tilde U')$
simply by setting the surface $\Sigma$, i.e.\ by indicating how $\Sigma$ bypasses the singularity components, say, by bridges (see Appendix~\ref{app:B}).
The position of $z^*$ and the detailed shapes of $\gamma_{Q^n}$ make no difference. 

\begin{proposition}
\label{pr:path_change}
Assume that the reference points 
$z^*$ and $\tilde z^*$  
coincide, that the reference points for the polygons coincide, but that base paths 
$\gamma_{Q^n}$ and 
$\tilde \gamma_{Q^n}$ differ.
Then the change of the reference paths leads to the following change of notations: 
\begin{equation}
Q^n_\gamma 
\longrightarrow
Q^n_{\gamma \gamma_{Q^n} \tilde \gamma_{Q^n}^{-1}}.
\label{e:change_notations}
\end{equation}
\end{proposition}

We will use this \purple{proposition} a lot for practical computations.


\subsection{The homomorphism $H_2(\tilde U,\tpB) \to H_2(\tilde U, \tilde U' \cup \tpB)$}

\begin{proposition}
\label{pr:inclusion}
Under the conditions of Proposition~\ref{pr:retract},
there is an inclusion
\[
H_2(\tilde U,\tpB) \hookrightarrow H_2(\tilde U, \tilde U'\cup\tpB),
\]
i.e.\ $H_2(\tilde U ,\tpB)$ is a subgroup of $H_2(\tilde U, \tilde U' \cup \tpB)$.
\end{proposition}

\textbf{Proof.} 
Let us consider a fragment of the long exact sequence of homology for the triple 
\[
(\tilde U, \, \tilde U'\cup \tpB ,  \, \tpB) .
\]

\begin{equation}\label{triple}
    \dots\rightarrow H_2(\tilde U' ,\tpB)\rightarrow H_2(\tilde U,\tpB)\rightarrow 
    H_2(\tilde U,\tilde U'\cup\tpB)\rightarrow\dots
\end{equation}

Since by Proposition~\ref{pr:retract} a pair $(\sigma\cap\mathbb{R}^2,\sigma\cap\partial \bB\cap\mathbb{R}^2)$ is a deformation retract of a pair $(\sigma,\sigma\cap\partial \bB)$, and this retraction lifts to a retraction of $(\tilde U',\tilde U'\cap\tpB)$ onto $(\tilde U'^*,\tilde U'^*\cap\tpB)$, then $H_2(\tilde U',\tpB) \simeq H_2(\tilde U'^*,\tpB)=0$, which yields the statement of the proposition. 
$\square$

\vskip 6pt

Proposition~\ref{pr:inclusion} means that any element of $H_2(\tilde U ,\tpB)$ can be written as a vector ${\bf w} \in \cM_k$ introduced for $H_2(\tilde U, \tilde U' \cup\tpB)$. In practice, such a notation can be obtained by retracting a realization of  an element of $H_2 (\tilde U , \tpB)$
to the real plane and studying the resulting set of polygons. 

\begin{figure}[h]
  \centering{ \includegraphics[width=0.5\textwidth]{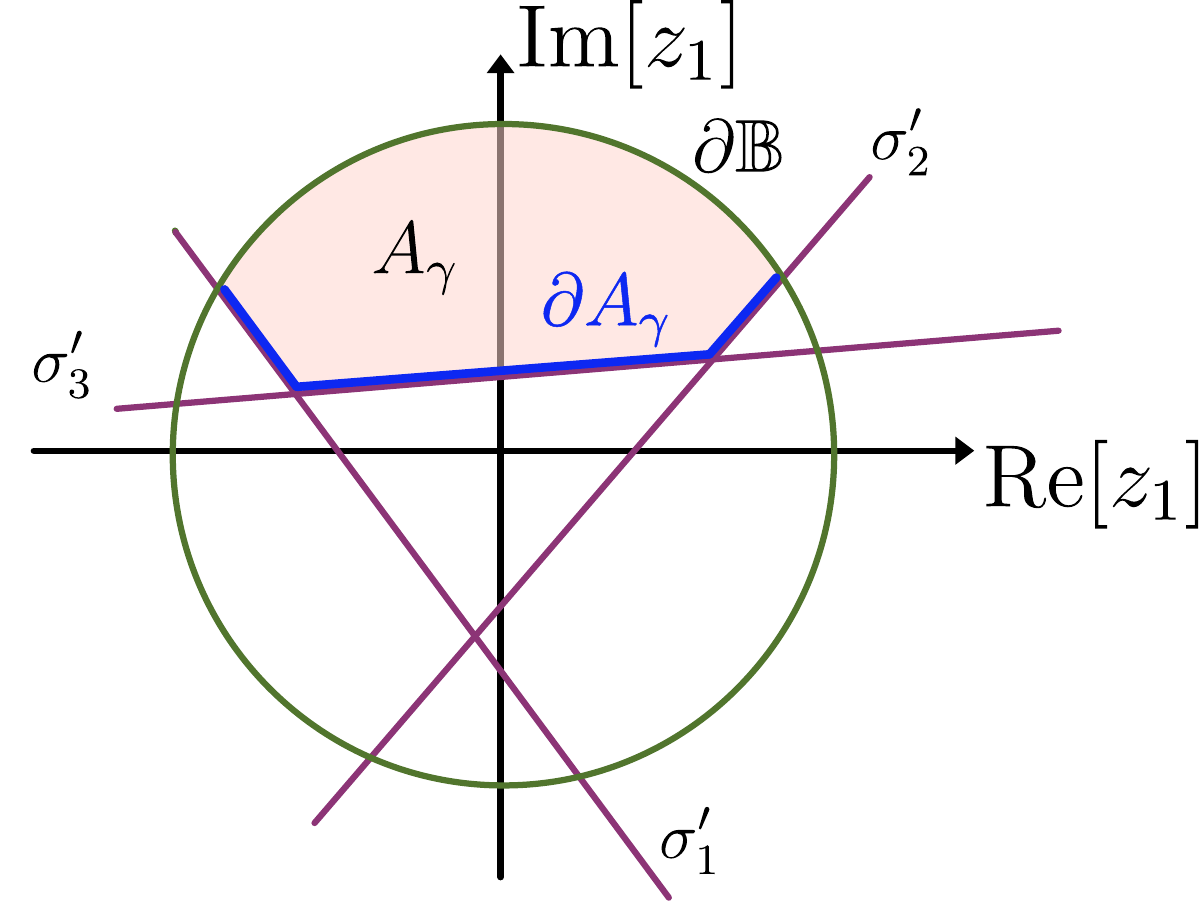} }
  \caption{Illustration of the boundary homomorphism}
  \label{f:boundary_homomorphism}
\end{figure}

Let us describe the elements of $H_2(\tilde U , \tilde U' \cup \tpB)$ that are images of the elements of 
$H_2(\tilde U ,\tpB)$. One can see that the condition is as follows: 
\begin{equation}
H_2(\tilde U ,\tpB) = \{ w \in H_2(\tilde U, \tilde U \cup \tpB)
\, : \, \ptl w = 0 \},
\label{e:cond_H2U}
\end{equation}
where 
\[
\ptl: \quad H_2(\tilde U,\tilde U'\cup\tpB)\rightarrow H_1(\tilde U',\tpB)
\]
is a homomorphism from the exact sequence \eqref{triple}. 
Intuitively, the operator $\partial$ maps a 
relative cycle (polygon) to the part of its boundary that lies in~$\tilde U'$. 
We refer to  $\partial$ as a {\em ``boundary'' homomorphism}.
An illustration of the ``boundary'' homomorphism is shown in \figurename~\ref{f:boundary_homomorphism}. 
Here $A_\gamma$ (shaded in pink) is a representative of an element of $H_2(\tilde U , \tpB)$ (we take it also belonging to 
$H_2(\tilde U^* , \tpB)$, i.e.\ retracted). 
The boundary of this representative consists of two parts: a) the part with $z \in \partial \bB$; b) the part with 
$z \in \sigma \setminus \ptl \bB$. The latter part is a representative of~$\ptl A_\gamma$ (a blue line).
This ``boundary'' belongs to $H_1 (\tilde U' , \tpB)$ and consists of three elements (segments) of 
$H_1 (\tilde U' , \tilde U^0 \cup \tpB)$. The path-indices of the segments are obtained from $\gamma$
by continuity (see Example~\ref{ex:triangle_H2U}). 

The necessity to formulate the condition (\ref{e:cond_H2U}) was the reason to introduce the universal Riemann domain that includes the points over the singularities. 

\begin{example}
\label{ex:triangle_H2U}
Consider the triangle $A$ shown in \figurename~\ref{f:01043}. Denote the sides of the triangle belonging to $\sigma_1, \sigma_2, \sigma_3$ by $a,b,c$, respectively. 
The orientation of the sides matches the orientation of~$A$.
These sides are elements of $H_1(\sigma, \sigma^{(0)} \cup \partial \bB)$. Elevate $a,b,c$ to $\tilde U_1$ by considering the pairs like $(a, \gamma)$, where 
$\gamma \in \Pi(z^*, z)$, $z \in a$. Indeed, for this we should introduce the base path $\gamma_a$, $\gamma_b$, $\gamma_c$ (see \figurename~\ref{f:sides}). 
Each path $\gamma \in \Pi(z^* , z)$ can then be written as $\gamma = \gamma' \gamma_a$, 
$\gamma' \in \PI$. We remind that due to Lemma~\ref{le:locality} the paths $\gamma'$ may be considered up to a right multiplication by a corresponding local subgroup. 

\begin{figure}[h]
  \centering{ \includegraphics[width=0.5\textwidth]{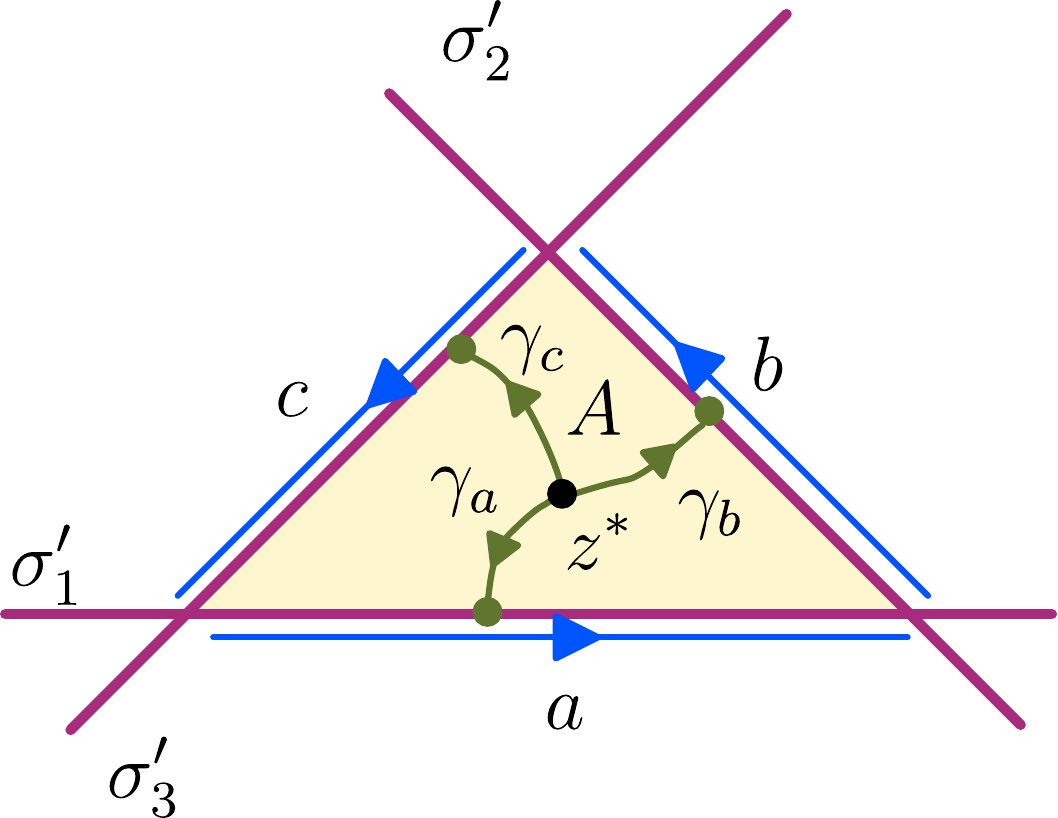} }
  \caption{Notations for sides and their base paths}
  \label{f:sides}
\end{figure}

Introduce the notation $a_{\gamma'}$, $b_{\gamma'}$, $c_{\gamma'}$
for the elements of $H_1 (\tilde U', \tilde U_0 \cup \tpB)$ ($\gamma'$ are corresponding classes of equivalence). 

First, consider the element $w' = A_e \in H_2(\tilde U, \tilde U' \cup \tpB)$. 
Indeed, 
\[
\ptl w' = a_e + b_e + c_e \ne 0,
\]
thus $w'$ does not belong to $H_2(\tilde U , \tpB)$. 

Second, consider the element 
\begin{equation}
w =
A_e - A_{\gamma_1} - A_{\gamma_2} - A_{\gamma_3} + 
A_{\gamma_1 \gamma_2} + A_{\gamma_2 \gamma_3} + A_{\gamma_1 \gamma_3} - 
A_{\gamma_1 \gamma_2 \gamma_3},
\label{e:triangle_def}
\end{equation}
\purple{where $\gamma_1$, $\gamma_2$, $\gamma_3$ are defined in \figurename~\ref{f:01043}.}
It is easy to establish that $\ptl w = 0$, and thus $w \in H_2(\tilde U , \tpB)$. To do this, study the boundary of each term. For example, 
\[
\ptl A_{\gamma_1 \gamma_2} = a_{\gamma_2} + b_{\gamma_1} + c_{\gamma_1 \gamma_2}
\]
(the sides ``inherit'' the path index of the polygon, and then Lemma~\ref{le:locality} is applied). The local subgroups for sides $a$, $b$, and $c$ have generators $\gamma_1$, $\gamma_2$, and $\gamma_3$, respectively.  
Similarly, 
\[
\ptl A_e = a_e + b_e + c_e, 
\qquad 
\ptl (-A_{\gamma_1}) = - a_e - b_{\gamma_1} - c_{\gamma_1},
\]
\[
\ptl (-A_{\gamma_2}) = - a_{\gamma_2} - b_{e} - c_{\gamma_2},
\qquad 
\ptl (-A_{\gamma_3}) = - a_{\gamma_3} - b_{\gamma_3} - c_{e},
\]
\[
\ptl A_{\gamma_2 \gamma_3} =  a_{\gamma_2 \gamma_3} + b_{\gamma_3} + c_{\gamma_2},
\qquad 
\ptl A_{\gamma_1 \gamma_3} =  a_{\gamma_3} + b_{\gamma_1 \gamma_3} + c_{\gamma_1},
\]
\[
\ptl A_{\gamma_1 \gamma_2 \gamma_3} = -a_{\gamma_2 \gamma_3} - b_{\gamma_1 \gamma_2} - c_{\gamma_1 \gamma_2}.
\]
We see that the boundary terms compensate each other pairwise. 

Topologically, $w$ is a sphere. 
Connecting the triangles by the sides compensating each other, we build 
an octahedron (see \figurename~\ref{f:01056}).  

\begin{figure}[h]
  \centering{\includegraphics[width=0.8\textwidth]{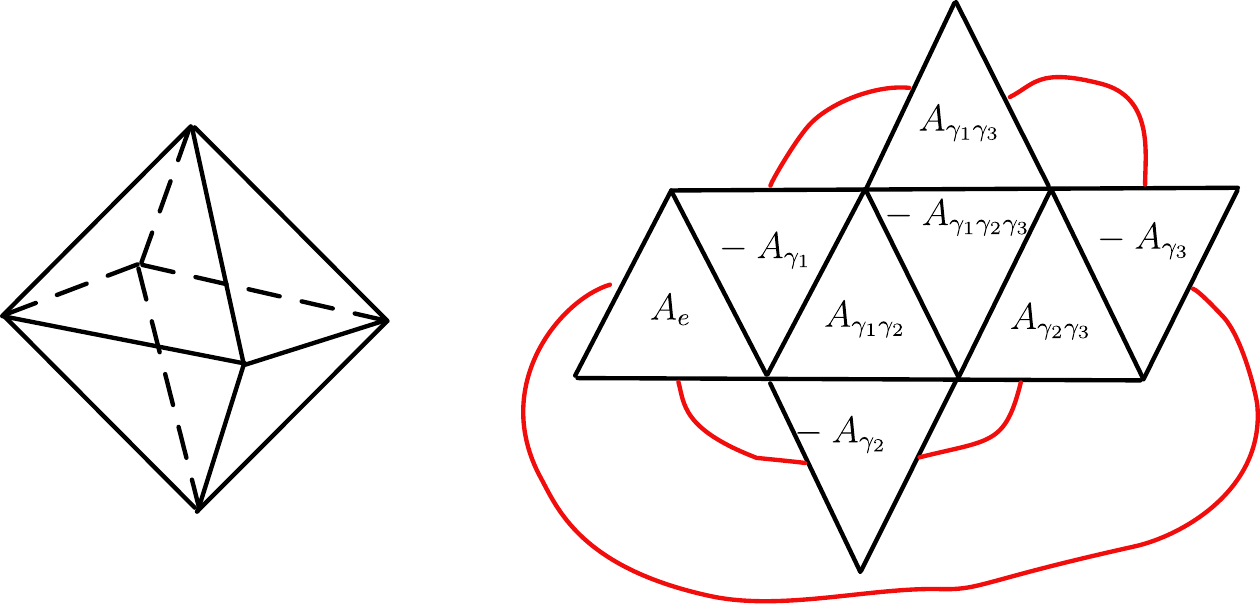}}
  \caption{The cycle $w$ as an octahedron}
  \label{f:01056}
\end{figure}
\end{example}

\begin{remark}
The elements of $H_2 (\tilde U , \tpB)$ form a submodule in $\cM_k$. The structure of this submodule is potentially important for the performed study, but it is beyond the scope of the current paper. 
\end{remark}


\subsection{The ``inflation'' theorem}

\subsubsection{Formulation and discussion}

\begin{theorem}[Inflation]
\label{th:inflattability}
Under the conditions of Proposition~\ref{pr:retract} (only transversal crossings, no triple crossings of $\sigma_j$)
the natural homomorphism
\begin{equation}
H_2(\tilde U_2 , \tpB) \to H_2(\tilde U,\tpB)
\label{e:inflation_homomorphism}
\end{equation}
is an isomorphism. 
\end{theorem}

A proof of this theorem is given in Appendix~\ref{app:C}. Here we put a discussion of this statement. 

The theorem comprises the injectivity and the surjectivity of the homomorphism (\ref{e:inflation_homomorphism}). 
Let us discuss the surjectivity. It means that any homology from $H_2(\tilde U , \tpB)$ can be slightly ``inflated'' into a homology 
from $H_2 (\tilde U_2 , \tpB)$, i.e.\ to a cycle
not passing through the singularities. In fact, the operation of inflation is somewhat inverse to the retraction~$\cR$.
We denote the inflation operation by $\cE$. Due to (\ref{e:retracts}), we can say that $\cE$ acts as 
\begin{equation}
\cE: \quad H_2(\tilde U^*, \tpB) \to H_2 (\tilde U_2 , \tpB), 
\label{e:inflation_def}
\end{equation}
i.e.\ inflates the retracted (flattened) homologies. 
For the 1D example studied above (see Section~\ref{sec:Overview_Homology}), the inflation is shown in \figurename~\ref{f:1D_example}. 

We will need the following statement: 

\begin{proposition}
Let the singularity $\sigma$ be a slight perturbation 
of a singularity having the real property. Let the other conditions of Proposition~\ref{pr:retract} be valid. Then Theorem~\ref{th:inflattability} still holds. 
\end{proposition}

This proposition is quite natural: a slight perturbation of singularities cannot make a topological statement invalid while no topological collapse occurs. Formally, this follows from Thom's first isotopy lemma  \cite{Berghoff2022} (see also a discussion in \S\ref{s:overview}).
The statement is needed when the singularity  $\sigma$ depends on~$t$,
and $t$ leaves the real space. 

The inflation theorem is close to the statement of \cite[VI.3]{Vassiliev2002}, the section named ``Regularization of non-compact cycles''. 
It studies forms having a special structure: 
\[
g_1^{\alpha_1}(z)   \dots  g_m^{\alpha_m}(z) \, dz,
\]
i.e.\ the branching is simple: the form is multiplied by a constant as a result of a bypass about each of the singularities. 
The concept of a {\em local system\/} is introduced, which can be interpreted as a homology group having coefficients in a complicated algebraic structure. Instead of the homomorphism (\ref{e:inflation_homomorphism}), the author of \cite{Vassiliev2002} studies the homomorphism 
(VI.19) of \cite{Vassiliev2002}, proving that it is an isomorphism. The proof is constructive (by ``patchworking''), resemble Appendix~\ref{app:C} here. 

Note that instead of studying the universal Riemann domain, \cite{Vassiliev2002} studies the proper Riemann domain of a certain form. The work contains a condition under which the homomorphism (VI.19) of \cite{Vassiliev2002} is an isomorphism. 

We should remark that our consideration can be partly interpreted in terms of local systems from \cite{Vassiliev2002} since our representation of $H_2(\tilde U, \tilde U' \cup \tpB)$ can be seen as a relative homology 
group $H_2(\NR^2 , \sigma \cup \partial \bB)$
with coefficients from~$\Omega$. 

Finally, we can conclude that in spite of a slightly different approach and a different subject, the inflation theorem here has much in common with  \cite{Vassiliev2002}. 

Besides, the inflation procedure is used for a less general case in 
\purple{\cite[VII.2.2]{Pham2011}}.


\subsubsection{Examples of inflation}

Here we consider several examples of application of Theorem~\ref{th:inflattability}. These examples are important, since they represent typical integration surfaces.

\begin{example}
\label{ex:W_circle}

Consider the singularity studied in Example~\ref{ex:circle_relative}, i.e.\ a circle, and define
\[
w = A_e - A_{\gamma_1} \in H_2 (\tilde U, \tilde U' \cup \tpB).
\]
The local subgroup for any point of $\sigma$ is the whole $\PI$, thus for any 
$z \in \sigma$ there exists a single point in $\tilde U_1$ over it. This yields
\[
\ptl w = 0,
\]
the condition (\ref{e:cond_H2U}) is fulfilled, and $w \in H_2(\tilde U , \tpB)$.
According to Theorem~\ref{th:inflattability} it can be deformed into~$\mathcal{E}(w) \in H_2 (\tilde U_2 , \tpB)$.
Let us describe the result of this inflation.

There is a single edge~$\sigma'$ and no vertices. 
According to the proof of Theorem~\ref{th:inflattability},
a representative of $\mathcal{E} (w)$ consists of three parts: two face parts $A_e$,$-A_\gamma$, and the edge part. The 
face parts are circles of radius $1-\epsilon$, where $\epsilon$ is the width of the 
narrow strip drawn near the singularity. 
The edge part is the product of $\sigma'$ and a small cut circle. The 
radii of the torus are $\epsilon$ and $1$.
The scheme 
of $\mathcal{E}(w)$ is shown in \figurename~\ref{f:01049}. 
The edges labeled by the same numbers should be attached to each other. 

\begin{figure}[h]
  \centering{\includegraphics[width=0.3\textwidth]{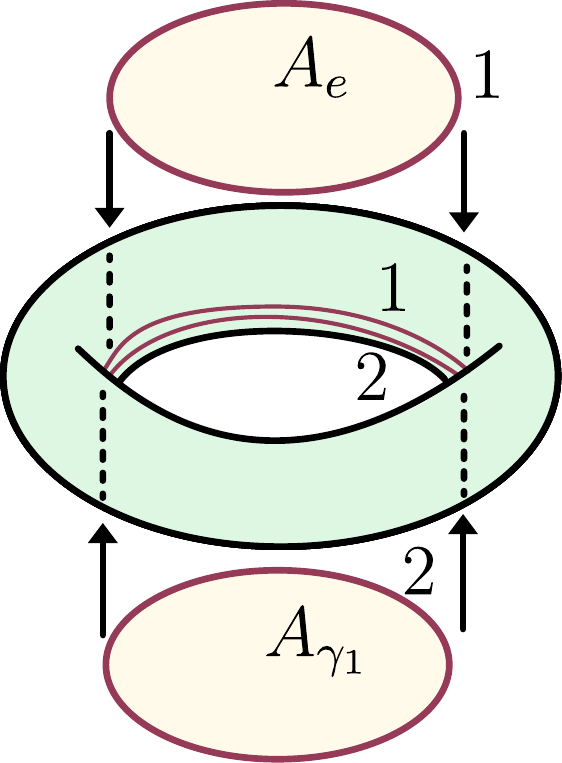}}
  \caption{Inflation of $w = A_e - A_{\gamma_1}$}
  \label{f:01049}
\end{figure}

\end{example}

\begin{example}
\label{sec:W2_cross}

Let the singular set be $\sigma = \sigma_1 \cup \sigma_2$,
\begin{equation}
\sigma_1: \, \, z_1 = 0, \qquad \sigma_2: \, \, z_2 = 0.
\label{e:01068a}
\end{equation}
The real traces $\sigma_1'$, $\sigma'_2$ split the real plane into four quadrants: $A,B,C,D$
(see \figurename~\ref{f:01050}). Take the reference point $z^* = z_A$ in~$A$,
and the reference points  $z_B, z_C, z_D$ in~$B,C,D$.
Define also the base path $\gamma_B$, $\gamma_C$, $\gamma_D$. 
The base path $\gamma_A$ is trivial. The base paths are located in some surface $\Sigma$ bypassing the singularities in accordance with the red bridges (see \figurename~\ref{f:01050}).

\begin{figure}[h]
  \centering{\includegraphics[width=0.4\textwidth]{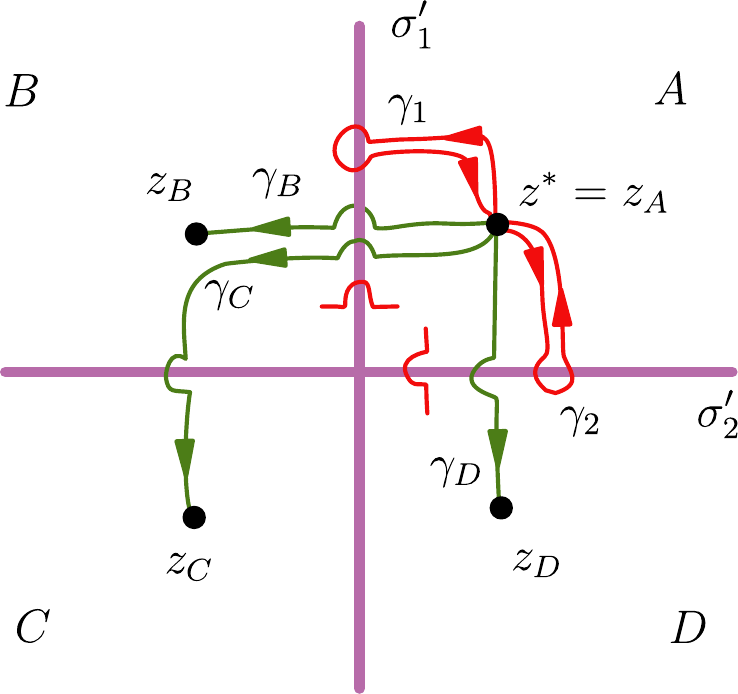}}
  \caption{Geometry for Example~\ref{sec:W2_cross}}
  \label{f:01050}
\end{figure}

As it follows from Lemma~\ref{le:commutativity},
group $\PI$ is $\mathbb{Z}^2$ with generators $\gamma_1$, $\gamma_2$ shown in \figurename~\ref{f:01050}. 

Consider the element 
\begin{equation}
w = A_e + B_e + C_e + D_e \in H_2 (\tilde U , \tilde U' \cup \tpB).
\label{e:01068}
\end{equation} 
By a direct computation similar to that of Example~\ref{ex:triangle_relative},
one can deduce that $\ptl w = 0$.
Thus, $w \in H_2 (\tilde U , \tpB)$, and one can build $\cE (w) \in H_2(\tilde U_2 , \tpB)$. 

The surface \purple{$\mathcal{E}(w)$} consists of four face parts, four edge parts, and one vertex part 
(see \figurename~\ref{f:01051}, left). The edge parts are half-cylinders of a small radius,
and the vertex part is a quarter of a torus (a product of two arcs of angle~$\pi$).    
Indeed, building  $\mathcal{E}(w)$ is rather simple  in this case: it  is a product of  
contours $\tilde \gamma_1$ and $\tilde \gamma_2$ in the planes $z_1$ and $z_2$
shown in \figurename~\ref{f:01051}, right.  

\begin{figure}[h]
  \centering{\includegraphics[width=0.8\textwidth]{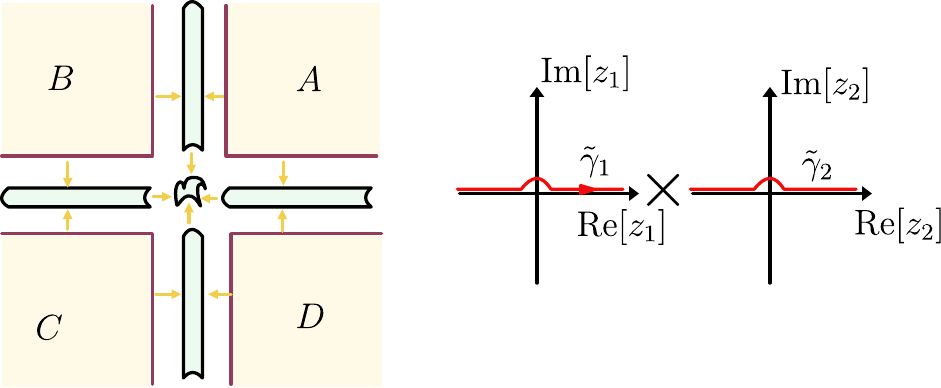}}
  \caption{$\mathcal{E}(A_e + B_e + C_e + D_e)$}
  \label{f:01051}
\end{figure}

\end{example}

\begin{example}

\label{ex:crossing_1}

Let  the singularity $\sigma$ be  defined by (\ref{e:01068a}) again, and the notations for the quadrants, base paths and generators for $\PI$ as above. 
Consider now the element
\begin{equation}
w = A_e - A_{\gamma_1} - A_{\gamma_2} + A_{\gamma_1 \gamma_2} \in H_2(\tilde U, \tilde U' \cup \tpB).  
\label{e:01069}
\end{equation}
One can see that $\ptl w = 0$.

%
%

The inflated homology  $\mathcal{E}(w)$ consists of  
four face parts (sectors), four edge parts (cut half-infinite cylinders), and a single vertex part that is a torus cut along two circles (see \figurename~\ref{f:01053}, left). 
\begin{figure}[h]  \centering{\includegraphics[width=0.8\textwidth]{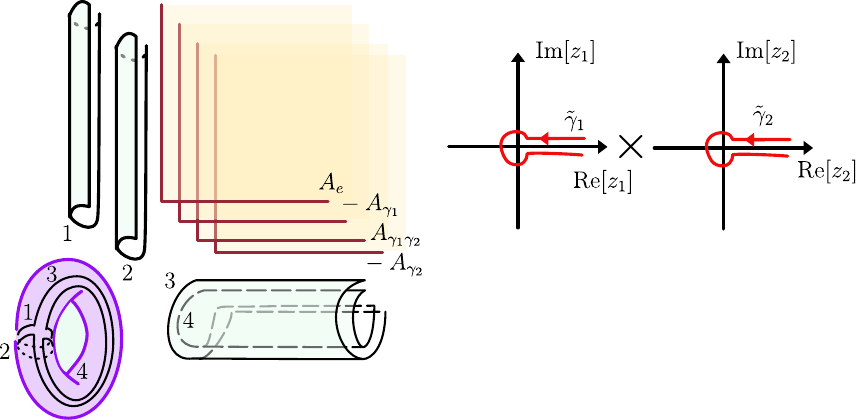}}
  \caption{$\mathcal{E}(A_e - A_{\gamma_1} - A_{\gamma_2} + A_{\gamma_1 \gamma_2})$}
  \label{f:01053}
\end{figure}
Indeed, the surface is a product $\tilde \gamma_1 \times \tilde \gamma_2$ shown in \figurename~\ref{f:01053}, right.
\end{example} 

\begin{remark}
Take a plane (complex line) $p: \, z_1 + z_2 = 1$. The intersection $p \cap \cE(w)$ is a Pochhammer's contour bypassing the points $(1,0)$
and $(0,1)$.
\end{remark}

\begin{example}
\label{ex:biangle}
Consider the singular set 
\[
\sigma = \sigma_1 \cup \sigma_2, 
\qquad 
\sigma_1: \, z_1 = 0, 
\qquad
\sigma_2: \, z_1 = z_2^2 - 1. 
\]
The lines $\sigma_1'$ and $\sigma_2'$ split the plane $\mathbb{R}^2$ into six  curvilinear polygons. 
Let $A$ be the finite ``biangle'' (see \figurename~\ref{f:01054}, left). 
Introduce the reference point $z^* \in A$ and the simple loops $\gamma_1$, $\gamma_2$ about the
singularities. These simple loops are generators of $\PI = \mathbb{Z}^2$.
Consider the element 
\[
w = A_e - A_{\gamma_1} - A_{\gamma_2} + A_{\gamma_1 \gamma_2} \in H_2(\tilde U,\tilde U' \cup \tpB). 
\] 
Using the same logic as above, one can prove that $\ptl w =0$. Moreover, 
locally near the vertices, the structure of $w$ is similar to that of 
Example~\ref{ex:crossing_1}.  
The structure of $\mathcal{E}(w)$ can be easily built as before; it is shown in 
\figurename~\ref{f:01054}, right. 

\begin{figure}[h]
  \centering{\includegraphics[width=0.6\textwidth]{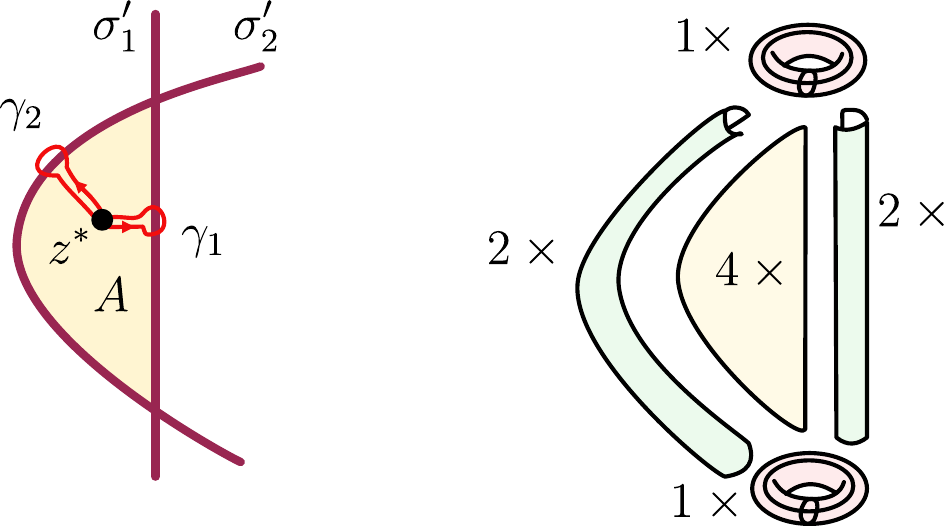}}
  \caption{$\mathcal{E}(A_e - A_{\gamma_1} - A_{\gamma_2} + A_{\gamma_1 \gamma_2})$ for a biangle}
  \label{f:01054}
\end{figure}

Note that the surface $\mathcal{E}(w)$ cannot be presented as a product of two contours. Topologically,  
$w$ and $\mathcal{E}(w)$ are spheres.  

\end{example}

%
%

\begin{example}
\label{ex:W_triangle}
Consider the singularities of Example~\ref{ex:triangle_H2U}, i.e.\ a triangle. 
Let us build $\cE (w)$ for $w$ defined by (\ref{e:triangle_def}).
One can see that $\cE(w)$ is described by the scheme
shown in \figurename~\ref{f:01055}. It consists of 8 triangles, 12 cut cylinders, 
and 6 cut tori.

\begin{figure}[h]
  \centering{\includegraphics[width=0.5\textwidth]{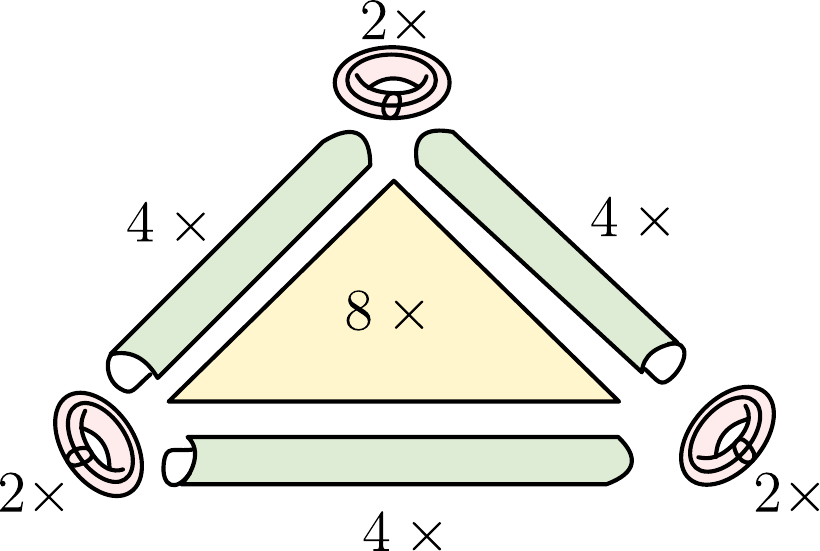}}
  \caption{$\mathcal{E}(w)$ for $w=  A_e - A_{\gamma_1} - A_{\gamma_2} - A_{\gamma_3}+ 
  A_{\gamma_1 \gamma_2}+ A_{\gamma_2 \gamma_3} + A_{\gamma_1 \gamma_3} - 
  A_{\gamma_1 \gamma_2 \gamma_3}$}
  \label{f:01055}
\end{figure}

\end{example}

\begin{remark}
The singularities under consideration (a circle, a biangle, and a triangle) are the main types of vanishing cycles 
from \cite{Pham2011} (page 91, middle column there).
\end{remark}


\subsection{Homologies for Riemann surfaces of certain functions}

\label{s:reduction}

In all examples above, the homologies $\mathcal{E}(w)$ are built in the universal covering~$\tilde U_2$.
The structure of this covering assumes that all components of the singularity 
are logarithmic branch lines. Typically, however, we use these cycles to integrate a function
$F(z)$
having the same singularity set, but, possibly, a simpler structure.

One may wonder whether it is possible to use the Riemann domain $\hat U$ of $F$ instead of the 
universal Riemann domain $\tilde U$ from the very beginning. However, it is not possible to  
prove an analog of Theorem~\ref{th:inflattability} in this case (at least unless some additional 
restrictions are imposed).
The proof fails at the vertex parts: the structure of sheets of the Riemann domain should be $\mathbb{Z}^2$
near the crossings of singularities. The following example illustrates this difficulty. 

\begin{example}
Consider the Riemann domain of the function 
\begin{equation}
F(z) = \sqrt{z_2 (z_1 + z_2 - 1)(z_1 - z_2)},
\label{e:triangle_sqrt}
\end{equation}
so that the components of the singularity $\sigma'$ form a triangle, the same as in \figurename~\ref{f:01043}.
One can introduce the Riemann domain $\hat U$ of this function as it is done above. $\hat U$ has two sheets 
over the regular points, and a single sample of the singularity set~$\hat U'$. 
Denote a bypass about any of the three singularities by~$\gamma_1$.
Introduce the relative homologies $H_2 (\hat U, \hat U' \cup \hat{\partial \bB})$, 
where $\hat{\partial \bB}$ is the lifting of $\partial \bB$ to~$\hat U$.
Let $A_e$ and $A_{\gamma_1}$ be triangles on the two different sheets of~$\hat U$; both 
triangles belong to $H_2 (\hat U, \hat U' \cup \hat{\partial \bB})$. One can see that 
\[
\ptl (A_e - A_{\gamma_1}) = 0.
\] 
One can prove, however, that there exists no element $w \in H_2 (\hat U_2, \hat{\partial \bB})$ such that $\cR (w) = A_e - A_{\gamma_1}$
(this may be not elementary, however).
\end{example}

\begin{remark}
As we mentioned earlier, the author of \cite{Vassiliev2002} works with Riemann surfaces of functions having 
a special ``non-resonance'' property. In our terms, the Riemann surface should have the structure of $\mathbb{Z}$
 near the singularities and of $\mathbb{Z}^2$ near the crossings of singularities.
\end{remark}

Let the function $F(z)$ have some singularity set $\sigma$. Build 
the universal Riemann domain $\tilde U$ and the Riemann domain of $F$ denoted~$\hat U$.
There exists a map $\Psi: \tilde U \to \hat U$ (see (\ref{e:map_Psi})), and this map can be continued to  
homomorphisms of homology groups (we denote them by the same letter $\Psi$):
\[
\Psi: \quad 
H_2 (\tilde U, \tpB) \to H_2(\hat U, \hat{\partial \bB}),
\quad 
H_2 (\tilde U_2, \tpB) \to H_2(\hat U_2, \hat{\partial \bB}),
\]

Remembering the definition (\ref{e:U_retract}) of $\tilde  U^*$, $w$ belong to $H_2 (\tilde  U^*, \tpB)$, and let $\cE (w)$ be built. There should exist 
$\Psi(\cE (w)) \in H_2 (\hat U_2, \hat{\partial \bB})$, and, 
moreover, since $\hat U$ can have ``a smaller amount of sheets'' comparatively 
to $\tilde U$, the surface $\Psi(\cE (w))$ can be a reduced version of~$\cE (w)$. 
This reduction is made by cancelling all parts of 
$\cE (w)$ that bear the same function $F$ but have the opposite signs
as oriented surfaces. 
Some examples of this reduction are given below.


\begin{example}\label{ex:red_triangle_1}
Consider the singularities of Example~\ref{sec:W2_cross}. Let a function $F(z)$ have simple poles at $\sigma_1$
and~$\sigma_2$ (and no branching).
Let $w$  be defined by (\ref{e:01069}).
The map $\Psi$ simplifies $\cE (w)$ significantly. 
First, note that 
the face parts $A_e$ and $-A_{\gamma_1}$ cancel each other, and so do 
$-A_{\gamma_2}$ and~$A_{\gamma_1 \gamma_2}$. 
Looking at \figurename~\ref{f:01053}, edge parts for  $\sigma_1$ (cut cylinders)
cancel each other, and  
so do the edge parts for~$\sigma_2$. Finally, $\Psi(\cE(w))$
is just a small torus near the crossing. The torus is no longer cut, and it produces 
the {\em composed residue\/} for the crossing of the singularities (see \cite{Shabat2}). 
\end{example}

\begin{example}
\label{ex:red_triangle_2}
Consider the singularities and the element $w$ of Example~\ref{ex:biangle}. 
Let $\sigma_2$ be a logarithmic branch line for some function $F(z)$, 
and $\sigma_1$ be a simple polar set. 
We will now attempt to reduce the homology $\mathcal{E}(w)$ shown in \figurename~\ref{f:01054}. 

The face parts $A_{e}$ and $-A_{\gamma_1}$ cancel each other; so do 
$-A_{\gamma_2}$ and $A_{\gamma_1 \gamma_2}$. 
The edge parts related to $\sigma_2$ cancel each other as well. 
The remaining parts are two cylinders near $\sigma_1'$ (they are not cut anymore), 
and two tori cut along a single circle. 
These components are shown in \figurename~\ref{f:01057}, left.
 
\begin{figure}[h]
  \centering{\includegraphics[width=0.8\textwidth]{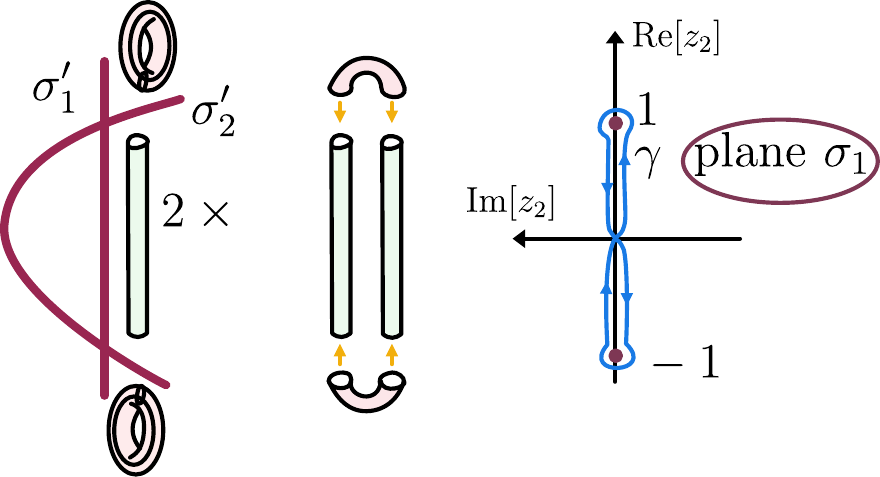}}
  \caption{The reduced homology for the biangle}
  \label{f:01057}
\end{figure}
 
The remaining edge and vertex parts are attached to each other as it is shown in the 
middle part of the figure. The resulting topology is a torus. Indeed, this torus is 
the Leray coboundary 
(see \cite{Pham2011}) of 
some contour $\gamma \subset \sigma_1$.
The set $\sigma_1$ can be parametrized by the variable $z_2$; 
the intersection $\sigma_1 \cap \sigma_2$ corresponds to $z_2 = \pm 1$. The contour $\gamma$
is shown in the right part of the figure. 
The integral of $F$ over $\Psi(\cE (w))$ can be computed using the Leray residue. 
\end{example}

\begin{example}
Consider the singularities and $w$ from Example~\ref{ex:W_triangle}. Assume that the function 
$F(z)$ have branching at $\sigma_3$ and simple poles at $\sigma_1, \sigma_2$.

The reduced homology $\Psi(\cE(w))$ consists of two tori 
located near $\sigma_1 \cap \sigma_2$ (see \figurename~\ref{f:01058}, left).  

\begin{figure}[h]
  \centering{\includegraphics[width= 0.8\textwidth]{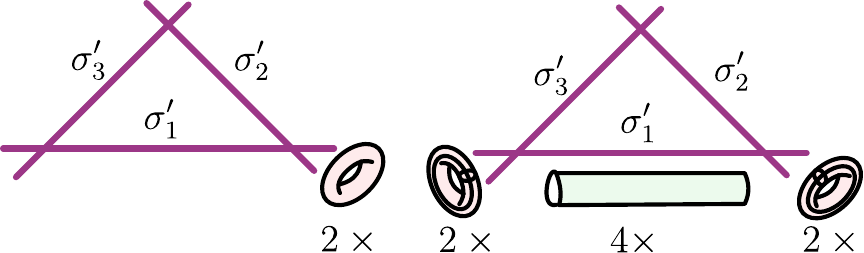}}
  \caption{Reduced homologies for Examples~\ref{ex:red_triangle_1} and~\ref{ex:red_triangle_2}}
  \label{f:01058}
\end{figure}

\end{example}

\begin{example}
Consider again the singularities and $w$ from Example~\ref{ex:W_triangle}.
Let $\sigma_1$ be a polar set, and $\sigma_2$, $\sigma_3$ be branch lines.
Then the reduced homology consists of 4 cut tori, and 4 edge cylinders 
going along $\sigma_1'$
(see \figurename~\ref{f:01058}, right).
It is a Leray coboundary of 
a Pochhammer's contour lying in the plane $\sigma_1$.
\end{example}

\begin{example}
Consider once more the singularities and $w$ from Example~\ref{ex:W_triangle}. Let $F$ have branching of order~2 at each of the three singularity components, i.e.\ this corresponds to (\ref{e:triangle_sqrt}).
Surprisingly, the homology $\Psi (\cE(w))$ cannot be reduced: its structure is still (\ref{f:01055}). We claim that there exists no ``simpler'' homology, having, say, two or four face components.  
\end{example}



\section{Ramification of relative homologies}
\label{sec:4}
\label{s:ramification_main}

\subsection{Landau set in the space of parameters}

\subsubsection{Definition of Landau set}


All sections above were preparatory, and now we return to the matter of Picard--Lefschetz theory, namely to branching of homologies from $H_2 (\tilde U_2 , \tpB)$. 
We recall that the position of at least some components of singularity $\sigma_j$ depends on the parameter $t = (t_1, \dots , t_M)$ (the values $t_j$ are complex; $M = 1, 2$ or 3 in our examples). 
Thus, the spaces $\tilde U$, $\tilde U_2$ and $\tilde U'$ depend on~$t$, and so do $\PI$ and $\Omega$:
\[
\tilde U = \tilde U(t), 
\quad 
\tilde U_2 = \tilde U_2(t), 
\quad 
\tilde U' = \tilde U'(t),
\quad 
\PI = \PI(t),
\quad 
\Omega = \Omega(t).
\]


The whole domain $\NR^M$ of $t$ is split into two sets: the regular points 
and the singular points. The set of singular points 
is usually referred to as the {\em Landau set} denoted~$\bL$. 
The formal definition of the Landau set is quite complicated (see \cite{Pham2011,Hwa1966,Berghoff2022}), 
and we do not use it here.
Instead, we use a rather vague ``definition'' working for $z$ in $\NR^2$: 

\begin{definition}
\label{def:Landau}
The singular set (the Landau set) $\bL$ is the set of $t\in\NR^M$ for which $\sigma(t)$ degenerates, i.e.\ some of the following takes place: 
\begin{enumerate}
\item  Some singularity component has the form $w_1^2 + w_2^2 = t'$ in some local coordinates 
$(w_1, w_2)$, $t'$ is some function of  $t_1, \dots , t_M$, and $t'$ takes the value~0. 

\item  Tangency of two intersecting singularity components. 
\item  Intersection of three or more singularity components.
\end{enumerate}
\end{definition}

\begin{remark}
The first type of degeneration corresponds to a degeneration of a circle in appropriate coordinates. 
Alternatively, {\em locally,} one can think of this degeneration as a Morse rearrangement of components of the singularity \cite{Milnor1963}.
\end{remark}

In all cases, for any $t'\in\cL$ there exists a relative cycle $\Delta(t)$ realizing some
element of $H_2(\NR^2 , \sigma(t))$ that vanishes as $t\rightarrow t'$. The simplest examples of such local degenerations are the Pham's singularities of types $P_1,P_2,P_3$ (see \cite{Pham2011}), which are studied also in this paper by our methods.

We remark that $\bL$ is an object totally different from the 
singular set in the $z$-space (that is $\sigma(t)$). 

We assume everywhere (and check on several examples) that $\bL$ is an analytic set 
whose main stratum has complex codimension~1, i.e.\ it has real dimension $2M - 2$.
Irreducible components of the main stratum of $\cL$ will be denoted by $\sigma^t_j$:
\[
\cL = \cup_j \sigma^t_j.
\]
We denote also the set of regular points by 
\begin{equation}
\cB \equiv \NR^M \setminus \cL. 
\label{e:def_B}
\end{equation}

{
Note that the number of irreducible components of the singularity $\sigma_j$ in the $z$-space
is not necessarily the same as the number of irreducible components $\sigma^t_j$ in the $t$-space. 
}

\begin{remark}
\label{rem:Thom_restriction}
There are some restrictions imposed on the space of parameters that are supposed to be valid, but we prefer not to 
write down everywhere for the sake of clarity. Namely, to be able to apply the first Thom's isotopy lemma (see Section~\ref{s:overview}), we need 
the singularity $\sigma(t)$ to be transversal to the boundary $\ptl (\NR^2 \cap \bB)$ for all $t$ considered. 
For this to be valid, the parameter $t$ should be restricted to the domain 
\[
\{ t\in \mathbb{C}^M \, \, : \, \, |t| < R', \,  ({\rm Im }[t_1])^2 +\dots + ({\rm Im }[t_M])^2  < \delta' 
\},
\]
where $R'$ and $\delta'$ are slightly smaller the $R$ and $\delta$ used for the $z$-space. Rigorously speaking, 
this domain should be taken instead of $\NR^M$. 

An example for which we cannot apply our methods is a singularity locally defined by 
\[
\sigma(t) : \quad z_1^2 - z^2_2 = t.
\]
Fix some value $\delta$ in \eqref{e:r2_def}. 
Let us try to make a bypass $\lambda$ in the complex plane of a single variable~$t$ 
about $\cL=\{0\}$. If $\lambda$ is a small circle of radius $r \ll \delta$ then $\sigma(t)$ 
is transversal to $\partial (N\mathbb{R}^2)$ for all $t \in \lambda$, but 
the singularity set $\sigma(t)$ cannot be retracted to its real part: $\sigma(t)$ is homeomorphic to a cylinder, while the real part is a pair of curves. Particularly, $\sigma(t)$ cannot be retracted to the real part for the starting point 
of~$\lambda$.

On the other hand, let us consider $\lambda$ with a real starting point far enough from~0. Now, by Proposition~\ref{pr:retract}, $\sigma(t)$ can be retracted onto the real part for the starting point of $\lambda$, but there exists some (small) point $t'$, at which the transversality condition is violated, and Thom's first isotopy lemma does not work.
\end{remark}

\subsubsection{Examples of $\bL$}

\label{s:examples_L}

We consider three main general cases of the degeneration:    
a vanishing triangle, biangle, and circle. 
In the examples below we assume that the variables are chosen in 
a special way providing the simplest formulae for the singular sets. Indeed, this 
choice of variables does not reduce the generality of the consideration. 

\begin{example}
\label{ex:tr_br}
Let be $t = (t_1 , t_2 , t_3)$
\begin{equation}
\sigma_1(t): \, z_1 = t_1 ,
\qquad 
\sigma_2(t): \, z_2 = t_2 ,
\qquad 
\sigma_3(t): \, z_1 + z_2  = t_3.
\label{e:015001}
\end{equation}
Let us find the set $\bL \subset \NR^3$.

\begin{figure}[h]
  \centering{\includegraphics[width= 0.9\textwidth]{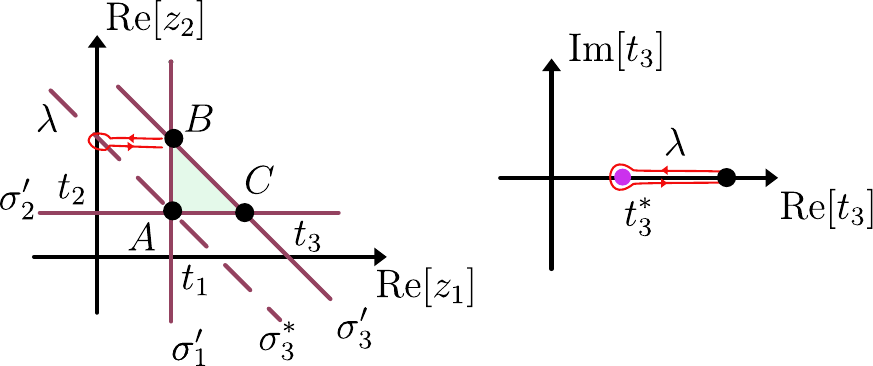}}
  \caption{The vanishing relative homology for a triangle}
  \label{f:01065}
\end{figure}

The singularities $\sigma_j$ have three crossing points: 
${\rm A} = (t_1, t_2)$, ${\rm B} = (t_1 , t_3-t_1)$, ${\rm C} = (t_2, t_3- t_1)$ 
(see \figurename~\ref{f:01065}, left).
The triangle ${\rm ABC} \in H_2 (\NR^2 , \sigma(t))$ vanishes (degenerates into a single point)
when
\begin{equation}
t_3 = t_1 + t_2. 
\label{e:015002}
\end{equation}
The relation (\ref{e:015002}) defines a set $\sigma^t_1 = \bL$ of complex codimension~1. 

The space 
$\cB$ (see \ref{e:def_B})
is not simply connected. Its 
fundamental group is $\mathbb{Z}$. Its generator is a simple loop about~$\bL$.

One can imagine the bypass about $\bL$ as follows. Fix the variables 
$t_1$ and $t_2$, i.e.\ consider the cross-section of the whole domain of $t$. 
There is a selected point $t_3^* = t_1 + t_2$ in this cross-section. 
This point corresponds to the special position of the singularity $\sigma_3$
with respect to $\sigma_1$ and $\sigma_2$. This special position is shown in 
the left part of \figurename~\ref{f:01065}, left.
Perform a bypass $\lambda$ shown in the right part of the figure. 
One can see that $\sigma_3$ bypasses about the position of $\sigma_3^*$
as it is shown on the left. 

One can also fix the position of $\sigma_1$ and $\sigma_3$, and carry
$\sigma_2$ about its critical position (see \figurename~\ref{f:01066}, left). Alternatively, 
one can fix the position of  $\sigma_2$ and $\sigma_3$, and carry
$\sigma_1$ (see \figurename~\ref{f:01066}, right). What  important is that, 
the result will always be the same due to the topology of $\cB$: the value 
$t_1 + t_2 - t_3$ makes a single bypass about zero in the positive direction in all three cases, so 
all such variants of realization of $\lambda$ are homotopical in~$\cB$.

\begin{figure}[h]
  \centering{\includegraphics[width= 0.7\textwidth]{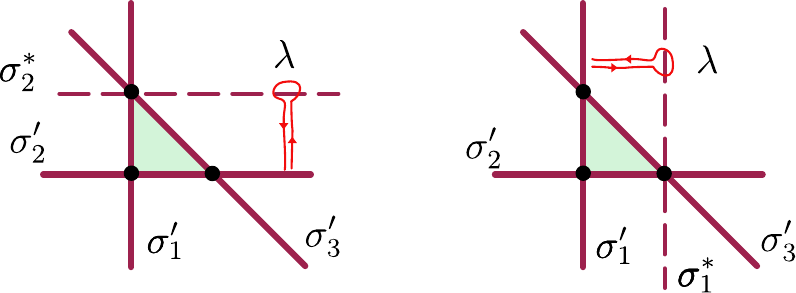}}
  \caption{Variants of realization of bypass $\lambda$}
  \label{f:01066}
\end{figure}
\end{example}
  
\begin{example}
\label{ex:bi_br}
Consider the singularities 
\begin{equation}
\sigma_1 (t): \, z_1 = t_1,
\qquad 
\sigma_2 (t): \, z_1 - z_2^2 = t_2.
\label{e:015003}
\end{equation}
The space of parameters is~$\NR^2$
{with a restriction ${\rm Re}[t_1 - t_2] > - \delta'$ for 
a sufficiently small positive $\delta'$.  The latter restriction follows 
from Remark~\ref{rem:Thom_restriction}}.
These singularities form a biangle ${\rm AB}$ (see \figurename~\ref{f:01067}, left). 
This biangle vanishes if 
\begin{equation}
t_1 - t_2 =0.
 \label{e:015004}
\end{equation}
This equation defines $\bL$, which is a  set  
of complex codimension~1.

\begin{figure}[h]
  \centering{\includegraphics[width= 0.8\textwidth]{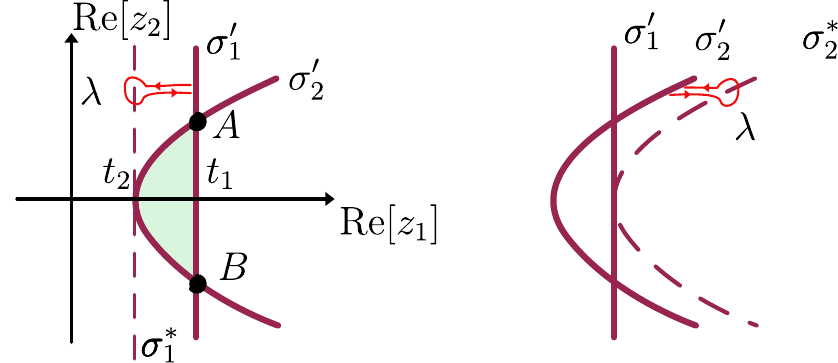}}
  \caption{Vanishing of a biangle}
  \label{f:01067}
\end{figure}
 
The elementary bypass about $\cL$ can be made in two ways
shown in \figurename~\ref{f:01067}, right and left. The result of both bypasses is the same.  
\end{example}

\begin{example}
Finally, consider a singularity defined by the circle 
\begin{equation}
z_1^2 + z_2^2 = t.
\label{e:015005}
\end{equation}
The space of parameters is $\NR^1$ 
with a restriction ${\rm Re}[t] > -\delta'$ following from Remark~\ref{rem:Thom_restriction}.
The circle vanishes at a single point $\cL = \{0\}$.
\end{example}


\subsection{Overview of the procedure}

\label{s:overview}

Let $\lambda$ be some path in $\cB$ starting and ending at some $t_0$, and $w$ be an element of $H_2(\tilde U_2(t_0) , \tpB(t_0))$. 
Informally, we introduce $\psi_\lambda (w)$ as a result of continuous deformation of \mesNik{ the cycle }$w(t)$ as $t$ moves along 
$\lambda$ and \mesNik{obtain a certain class in the } initial group $H_2(\tilde U_2(t_0) , \tpB(t_0))$: 
\begin{equation}
w \stackrel{\lambda}{\longrightarrow} \psi_\lambda(w).
\label{e:variation}
\end{equation}
\mesNik{The operator $\psi_\lambda$ is called {\em monodromy operator.} }Of course, it is possible that $\psi_\lambda (w) \ne w$.
We also introduce the {\em variation\/} of $w$:
\begin{equation}
{\rm var}_\lambda (w) \equiv \psi_\lambda (w) - w.
\label{e:def_var}
\end{equation}
If some ramification of $H_2(\tilde U_2, \tpB)$ occurs, then there exist elements with a non-zero variation. 

Formally, this can be expressed as follows. 
According to Thom's first isotopy lemma \cite{Pham1967IntroductionAL}, all types of homology groups from \eqref{cd1} are identified with each other along some paths in $\cB$, but this identification depends on the homotopy class of the path. More precisely, the natural projection $\mathbb{E}\rightarrow\cB$, where $\mathbb{E} \equiv \cB\times N\mathbb{R}^2$, 
defines a fiber bundle of pairs 
$(\mathbb{E}, \mathbb{K})\rightarrow\cB$, where the set $\mathbb{K}\subset \mathbb{E}$ consists of pairs of the form 
($t\in\cB$, $z\in\sigma(t)$). 
Using this projection and Definition~\ref{def:URD}, define a fiber bundle of triples $(\tilde U ,\tilde U'\cup\tpB,\tpB)$ over $\cB$. Then the corresponding homology groups are identified with each other via the \textit{Gauss-Manin connection} in the associated homology fiber bundle, see \cite{Vassiliev2002}. In particular, the groups
\[
H_2(\tilde U_2(t),\tpB(t)), \qquad t \in \cB
\] 
form a fiber bundle over $\cB$. \mesNik{The representatives 
of the classes can be chosen varying continuously with~$t$, i.e.\ the monodromy over any path in $\cB$ can be realized as a composition of diffeomorphisms of the pair $(\tilde U_2,\tpB)$ that are fixed in the vicinity of $\tpB$}.

The action of the path $\lambda$ (i.e.\ (\ref{e:variation})) defines an automorphism of the group
$H_2(\tilde U_2(t_0),\tpB(t_0))$, and
$\psi_\lambda(w)$ denotes the image of $w$ under this automorphism. Naturally, \mesNik{the monodromy operator $\psi_\lambda$ } \purple{commutes} with \eqref{e:inflation_homomorphism} and arrows from \eqref{cd1},\eqref{triple}.

Note that, due to a continuity argument, if the loop $\lambda$ can be contracted in $\cB$
into a point, then there should be no variation of $H_2(\tilde U_2(t), \tpB(t))$ along~$\lambda$.
Therefore, one can study the ramification of $H_2(\tilde U_2(t), \tpB(t))$ as $t$ bypasses the components of~$\bL$.
This ramification is the main subject of the Picard--Lefschetz theory. 
We assume that the reader is familiar with 
the concept of ramification of homologies (see \cite[V]{Pham2011}).

A key statement of the Picard--Lefschetz theory is that 
a surface 
\[
\Gamma \in H_2(\tilde U_2(t),\tpB(t))
\]
changes as a result of some bypass $\lambda$
if and only if the corresponding vanishing relative homology 
{\em pinches\/} it. 
``Pinches'' means that $\Gamma$ has a non-zero intersection index with the corresponding relative homology. 
The intersection index is a topological invariant, therefore one cannot deform $\Gamma$
in such a way that it passes far enough from the vanishing homologies. So, some sort of catastrophe 
happens with $\Gamma$ when the relative homology vanishes.

To describe the variation (\ref{e:variation}) we use the approach shown in the following diagram
\begin{equation}
\begin{tikzcd}
H_2(\tilde U_2, \tpB) \arrow{r}{\psi_\lambda}  \arrow[swap]{d}{\cR}  & H_2(\tilde U_2, \tpB)  \\
H_2(\tilde U^*, \tpB) \arrow[swap]{d}{a}& H_2(\tilde U^*, \tpB) \arrow[swap]{u}{\cE} \\
\cM_k \arrow{r}{\times T} & \cM_k \arrow[swap]{u}{b}
\end{tikzcd}
\label{e:015006m}
\end{equation}

The arrows labelled by $\cR$ and $\cE$ denote isomorphisms of homology groups, induced by retraction and inflation. The arrow labelled by ``$a$'' is 
a composition of the inclusion 
$H_2 (\tilde U^* ,\tpB) \hookrightarrow H_2(\tilde U^* , \tilde U' \cup \tpB)$
and the isomorphism $H_2(\tilde U^* , \tilde U' \cup \tpB) \simeq \cM_k$.

The arrow labelled by ``$\times T$'' is the description of ramification of relative homologies from $H_2 (\tilde U, \tilde U' \cup \tpB)$. This description will be given in the current section, and as we will see, the ramification is described by multiplication by a $k \times k$ matrix of elements from ~$\Omega$.

The arrow labelled by ``$b$'' is the inversion of the inclusion 
$H_2 (\tilde U^* , \tpB) \hookrightarrow H_2(\tilde U^* , \tilde U' \cup \tpB)$. The possibility to 
perform this inversion follows from the fact that if 
$w(t) \in H_2(\tilde U (t) , \tilde U'(t)  \cup \tpB(t))$, and $\ptl w(t_0) = 0$ (see \eqref{e:cond_H2U}) at the start of $\lambda$, then $\ptl w(t) = 0$ at each~$t$ during the transformation~$\psi_\lambda$. This follows from continuity. 

If one studies a transformation of a homology from $H_2 (\hat U_2 , \hat{\partial \bB})$ for some Riemann domain of a given function, 
one should add one more level atop the diagram 
(\ref{e:015006m}):
\begin{equation}
\begin{tikzcd}
H_2 (\hat U_2 , \hat{\partial \bB} ) \arrow{r}{\psi_\lambda} \arrow[swap]{d}{\Psi^{-1}}  
& H_2(\hat U_2 , \hat{\partial \bB}) \\
H_2(\tilde U_2 , \tpB) \arrow{r}{\psi_\lambda}  \arrow{d} & H_2(\tilde U_2 , \tpB)  \arrow[swap]{u}{\Psi} \\
\dots & \dots \arrow{u}
\end{tikzcd}
\label{e:015006mm}
\end{equation}
As one can see, by our method we can only describe  the homologies from $H_2 (\hat U_2 , \hat{\ptl \bB})$ that are represented as 
$\Psi (w)$ for some $w \in H_2 (\tilde U_2 , \tpB)$. 


\subsection{Ramification of the fundamental group}

Before we start with the ramification of the relative homologies
$H_2(\tilde U , \tilde U' \cup \tpB)$
 as the singularity set 
$\sigma(t)$ is deformed, let us study the ramification of the fundamental group $\PI(t) \equiv \pi_1 (\NR^2 \setminus \sigma(t))$. 

The change of $\PI(t_0)$ induced by a transformation $\lambda$ can be described by  an automorphism 
\begin{equation}
\psi_{\lambda}: 
\, 
\PI(t_0) \stackrel{\lambda}{\longrightarrow} \PI(t_0).
\label{e:psi_lambda_def}
\end{equation}
Indeed, this automorphism generates an automorphism 
\begin{equation}
\psi_{\lambda}: 
\, 
\Omega(t_0) \stackrel{\lambda}{\longrightarrow} \Omega(t_0)
\label{e:psi_lambda_def_1}
\end{equation}
denoted by the same symbol.

There is a seemingly surprising property of ramification of $\PI(t)$
following from an argument based on dimensions:

\begin{lemma}
\label{pr:fundamental_ramification}
If a vanishing homology in $H_2 (\NR^2 , \sigma(t))$ belongs to the cases considered in Subsection~\ref{s:examples_L} (i.e.\ if the vanishing homology is a triangle, biangle, or a circle), the group $\PI(t)$ is not ramifying: 
\begin{equation}
\psi_{\lambda} (\gamma) = \gamma
\label{e:PI_ramification}
\end{equation}
for any $\lambda \in\pi_1(\cB)$ and $\gamma\in\pi_1(\NR^2\backslash\sigma)$.
\end{lemma}  

\begin{figure}[h]
  \centering{\includegraphics[width= 0.4\textwidth]{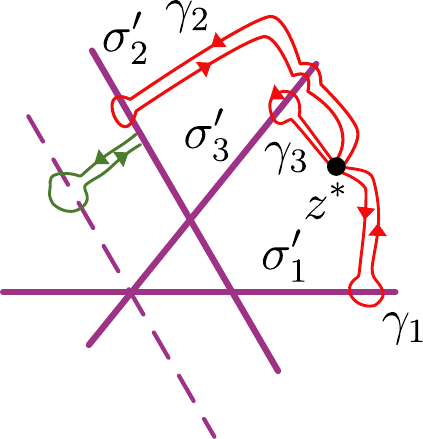}}
  \caption{Generators of $\PI$ that are not affected by $\lambda$}
  \label{f:non-ramification}
\end{figure}

The sketch of the proof is as follows: generally, one can homotopically deform a path 
$\gamma$ 
to make it passing far from the vanishing homology. After such a deformation, 
$\gamma$ ``doesn't feel'' the bypass~$\lambda$.
For example, consider the vanishing of a triangle. Select the representatives of the 
classes $\gamma_1$, $\gamma_2$, $\gamma_3$ (the generators of $\PI$) as it is shown in
\figurename~\ref{f:non-ramification}. Indeed, a bypass $\lambda$ shown in the figure 
does not change $\gamma_1$, $\gamma_2$, $\gamma_3$, and thus it does not change any element of~$\PI$.

A more formal reasoning is that 
one cannot correctly define an {\em intersection index\/} for
a 1D object (contour $\gamma$)
and a 2D object (a vanishing relative homology from $H_2(\NR^2 , \sigma(t))$) in a four-dimensional space.



\begin{remark}

Let us give an example of violation of the condition of Lemma~\ref{pr:fundamental_ramification}. 
Let be 
\[
\sigma_1: \, z_1 = 0, 
\qquad 
\sigma_2(t): \, z_1 = t, 
\]
i.e.\ the components of the singularity are parallel lines. 
Consider parameter $t$ that varies in $\NR^1$ (or in $\mathbb{C}$), and the Landau set is 
a point $\bL = \{ 0\}$. Indeed, this type of vanishing is not described in Subsection~\ref{s:examples_L}.
The group 
$\PI$ is a free group with generators $\gamma_1$, $\gamma_2$ shown in \figurename~\ref{f:01067a}, left.  

\begin{figure}[h]
  \centering{\includegraphics[width= 0.7\textwidth]{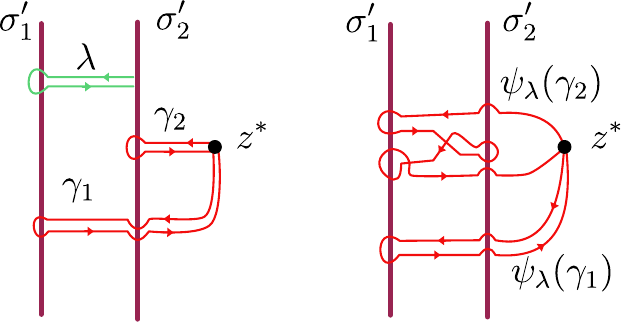}}
  \caption{Graphical finding of $\psi_\lambda (\gamma)$}
  \label{f:01067a}
\end{figure}

Let $\lambda$ be a simple loop in the $t$-space bypassing $t = 0$ one time in a positive direction.
The change of the generators $\gamma_1$ and $\gamma_2$ under $\lambda$ is shown in \figurename~\ref{f:01067a}, right:
\[
\psi_\lambda (\gamma_1) = \gamma_2 \gamma_1 \gamma_2^{-1} , 
\qquad 
\psi_\lambda (\gamma_2) = \gamma_2 \gamma_1 \gamma_2 \gamma_1^{-1} \gamma_2^{-1}. 
\]
One can see that the statement (\ref{e:PI_ramification}) is not valid. 

Mathematically, the reason of the breaking of the proof of Lemma~\ref{pr:fundamental_ramification} is as follows. 
Instead of vanishing of a 2D homology, a 3D homology vanishes in this case. 
Namely, the vanishing homology belongs to $H_3 (\NR^2 , \sigma(t) \cup \tpB)$, and it is
 the product of the segment $[0,t]$ in the $z_1$-plane and the whole $z_2$-plane. 
This homology is not local; moreover, one can define its intersection index with some $\gamma \in \PI(t)$.

Below we consider {\em only\/} the cases for which Lemma~\ref{pr:fundamental_ramification} is applicable, and the fundamental group does not ramify. 
However, we should admit that the condition of Lemma~\ref{pr:fundamental_ramification} is restrictive, since in practical application 
the integrals with parallel singularity components do exist. We plan to address this issue in a subsequent paper. 
\end{remark}


\subsection{Linear representation of ramification of elements of $H_2 (\tilde U , \tilde U' \cup \tpB)$}
\label{LR}

As explained in Section~\ref{sec:Algebra_of_H2_overview}, the elements $Q^1_e \dots Q^k_e$ form a basis of $H_2(\tilde U, \tilde U' \cup \tpB)$ over $\Omega$, i.e.\ any element 
$w \in H_2(\tilde U, \tilde U' \cup \tpB)$ can be written as 
\begin{equation}
w = \sum_l \omega_l Q^l_e \equiv {\bf w} \, {\bf Q},
\qquad 
{\bf w} = (\omega_1 , \dots , \omega_k), 
\qquad 
{\bf Q} = \left( \begin{array}{c}
Q^1_e \\
\vdots \\
Q^k_e
\end{array} \right).
\label{e:vec_form}
\end{equation}

Let $\lambda$ be some loop in~$\cB$, and  $\psi_\lambda$ be the corresponding map (\ref{e:psi_lambda_def}). 
Let $\lambda$ transform the basis as
\begin{equation}
{\bf Q} \stackrel{\lambda}{\longrightarrow} {\bf T} \, {\bf Q}, 
\label{e:T_trans_1}
\end{equation}
where 
\begin{equation}
{\bf T} = 
\left( \begin{array}{ccc}
\omega_{1,1} &  \cdots & \omega_{1, k} \\
\vdots       & \ddots  & \vdots   \\
\omega_{k,1} &  \cdots & \omega_{k, k}
\end{array} \right) 
\label{e:T_def}
\end{equation}
is a $k \times k$ matrix of elements of $\Omega$.
Our aim is to describe how $\lambda$ acts on~${\bf w}$ assuming that ${\bf T}$ is known.
This description is given by the following lemma:

\begin{lemma}
\label{le:linear_representation}
Let a transformation of the form (\ref{e:T_trans_1}) be known for some $\lambda$, and let the condition of Lemma~\ref{pr:fundamental_ramification} be valid.
Then for any $w$ described by (\ref{e:vec_form}) the vector ${\bf w} \in \cM_k$ transforms as 
\begin{equation}
{\bf w} \stackrel{\lambda}{\longrightarrow} {\bf w}\, {\bf T}. 
\label{e:T_trans}
\end{equation}
\end{lemma}

\begin{proof}
Let $\lambda$ be some loop in~$\cB$, and  $\psi_\lambda$ be the corresponding map (\ref{e:psi_lambda_def}). 
{
Use the notation (\ref{e:variation}) for for elements $w$ of $H_2 (\tilde U , \tilde U' \cup \tpB)$.
Extend also the notation $\psi_\lambda$ from $\PI$ to $\Omega$ by linearity. 
Note that by construction of multiplication (\ref{e:mult_Om_H_H}),
\[
\psi_\lambda (\omega \, w)  = \psi_\lambda (\omega ) \psi_\lambda ( w)  , 
\qquad 
\omega \in \Omega, 
\quad 
w \in H_2 (\tilde U , \tilde U' \cup \tpB).
\]

Rewrite (\ref{e:T_trans_1}) as 
\begin{equation}
 \psi_{\lambda} (Q^n_e) = 
\sum_{l = 1}^k \omega_{n,l} Q^l_e .  
\label{e:01309a7a}
\end{equation}

Let $\gamma$ be an arbitrary element of $\PI$. Then
\begin{equation}
\psi_\lambda ( Q^l_\gamma ) 
= 
\psi_\lambda ( \gamma Q^l_e )
=
\psi_\lambda ( \gamma ) \psi_\lambda ( Q^l_e ) 
= 
\sum_{l = 1}^k \psi_\lambda (\gamma) \omega_{n,l} Q^l_e .
\label{e:01309a7b}
\end{equation}

}


Since the condition of Lemma~\ref{pr:fundamental_ramification} is valid, one can use (\ref{e:PI_ramification}). 
The formula (\ref{e:01309a7b}) becomes simplified: 
\begin{equation}
Q^n_\gamma 
 \stackrel{\lambda}{\longrightarrow}
{ \sum_{l = 1}^k \gamma \, \omega_{n,l} Q^l_e. }
\label{e:simp_1}
\end{equation}
By linearity, one can generalize this relation: for any $\omega_n \in \Omega$
\begin{equation}
\omega_n Q^n_e  
 \stackrel{\lambda}{\longrightarrow}
\sum_{l = 1}^k \omega_n\, \omega_{n,l} Q^l_e . 
\label{e:simp_2}
\end{equation}

Again, by linearity this gives 
\begin{equation}
{\bf  w} \, {\bf Q} 
\stackrel{\lambda}{\longrightarrow} 
{\bf w} \, ({\bf T} \, {\bf Q} ) .
\label{e:vec_form_1}
\end{equation}
Using the associativity of the matrix multiplication, we obtain
\[
{\bf  w} \, {\bf Q} 
\stackrel{\lambda}{\longrightarrow} 
({\bf w} \, {\bf T} )\, {\bf Q}  ,
\]
i.e.\ (\ref{e:T_trans}).
\end{proof}

The formula (\ref{e:T_trans}) corresponds to the arrow labeled by ``$\times T$'' in the diagram (\ref{e:015006m}).


\subsection{Elementary transformations}\label{Elt}

\begin{figure}[h]
  \centering{\includegraphics[width= 0.8\textwidth]{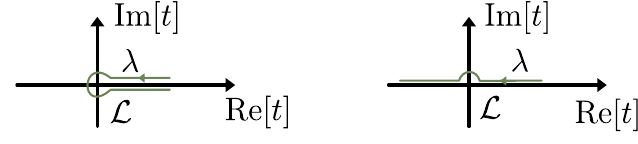}}
  \caption{A loop (left) and a jump (right) for $\lambda$}
  \label{f:loop_jump}
\end{figure}

Here we describe elementary transformations of relative homologies as $t$ travels along the loop $\lambda$ in~$\cB$.
There are two sorts of transformations that  should be studied: a loop about a component of $\cL$ and a jump over such a component
(example shown in \figurename~\ref{f:loop_jump} corresponds to 1D $t$-space). For a loop, a relative homology of $H_2(\tilde U, \tilde U')$  vanishes. This homology may be a triangle, a biangle, or a circle. A jump generally corresponds to a vanishing of a triangle.
Thus, we should consider 4 cases in total.

The transformation matrices obtained here are elementary building blocks for any more complicated matrices. Thus, formulas in this subsection provide a ``computational core'' of the method.

\subsubsection{The triangle}

\begin{figure}[h]
  \centering{ \includegraphics[width=0.8\textwidth]{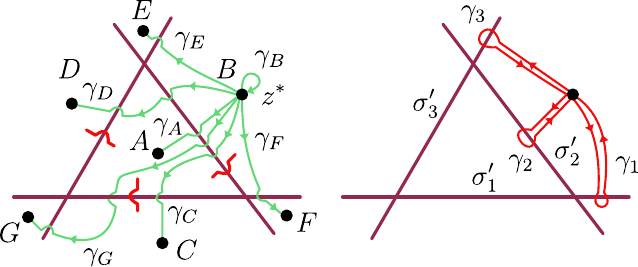} }
  \caption{Reference point and base paths for a triangle}
  \label{f:01061a1}
\end{figure}

Consider three singularity components $\sigma = \sigma_1\cup \sigma_2 \cup \sigma_3$ that are lines having the real property. 
Introduce the notations enabling us to study the situation algebraically. 
Namely, denote the polygons of $H_2 (\mathbb{R}^2 , \sigma \cup \partial \bB)$
by the letters $A, B, C, D, E, F, G$ (ordered in this way) as it is shown in \figurename~\ref{f:01061a1}, left.
Choose the orientation of the polygons in the natural way. 
Assume that the singularity $\sigma_2$ is movable, $\sigma_2 = \sigma_2 (t)$, that the parameter $t$ is  complex, and that the motion is a parallel displacement: 
\[
\sigma_2 (t) : \qquad a z_1 + b z_2 = t.
\]

The triangle $A$ is a vanishing homology. Take a reference point $z^*$ 
in any of  non-vanishing polygons, say in~$B$.

Introduce the base paths $\gamma_{A}$, \dots ,
$\gamma_{G}$ (the path $\gamma_B$ is trivial)
 as it is
shown in \figurename~\ref{f:01061a1}, left. The paths are shown in green. 
One can see that these paths belong to the surface $\Sigma$ shown by the 
red bridge symbols. This choice of base paths is in agreement with the condition of Lemma~\ref{le:trans_contours}
and its corollary, thus the green base paths can be omitted in favour 
of the bridge notations. 
Introduce also some simple loops $\gamma_1, \gamma_2 , \gamma_3$ that are generators of
$\PI (t)$ 
(see \figurename~\ref{f:01061a1}, right). 

\begin{remark}
We transported the reference point 
from $A$ (see \figurename~\ref{f:01043}) to $B$ along the corresponding base path.
As it follows from Proposition~\ref{pr:surface_Sigma_trivial}, such a change does not affect the path-indices of the polygons. 
However, taking a reference point far away from the vanishing cycle is preferable, since the reference point there 
experiences no transformation as $\sigma_2(t)$ evolves with~$t$. 
\end{remark}


\begin{figure}[h]
  \centering{ \includegraphics[width=0.35\textwidth]{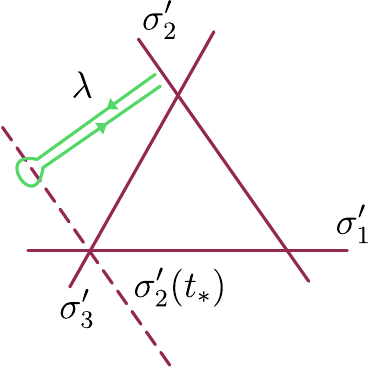} }
  \caption{A basic bypass for triangle}
  \label{f:01061a}
\end{figure}

The Landau set $\cL$ consists of a single value $t = t_*$, for which the line $\sigma_2(t_*)$ passes through 
the crossing point of $\sigma_1$ and $\sigma_3$ (see \figurename~\ref{f:01061a}). Let $\lambda$ be a simple loop in the $t$-space
passing about~$t_*$ once in the positive direction.

%


\begin{theorem}
\label{the:PLP_triangle}
The basis shown in \figurename~\ref{f:01061a1} is transformed by the loop $\lambda$
shown in \figurename~\ref{f:01061a} as
\begin{equation}
{\bf Q} \stackrel{\lambda}{\longrightarrow} {\bf T}_3 {\bf Q},
\qquad 
{\bf Q} = \left( \begin{array}{c}
A_e \\
\vdots \\
G_e
\end{array} \right), 
\label{e:01331a}
\end{equation}
\begin{equation}
{\bf T}_3 = \left( \begin{array}{ccccccc}
\gamma_1 \gamma_2 \gamma_3  & 0 & 0 & 0 & 0 & 0 & 0 \\ 
\gamma_2 - \gamma_1 \gamma_2 \gamma_3 & e & 0 & 0 & 0 & 0 & 0 \\
\gamma_1 - \gamma_1 \gamma_2 \gamma_3 & 0 & e & 0 & 0 & 0 & 0 \\
\gamma_3 - \gamma_1 \gamma_2 \gamma_3 & 0 & 0 & e & 0 & 0 & 0 \\
-\gamma_2 \gamma_3 + \gamma_1 \gamma_2 \gamma_3 & 0 & 0 & 0 & e & 0 & 0 \\
-\gamma_1 \gamma_2 + \gamma_1 \gamma_2 \gamma_3 & 0 & 0 & 0 & 0 & e & 0 \\
-\gamma_1 \gamma_3 + \gamma_1 \gamma_2 \gamma_3 & 0 & 0 & 0 & 0 & 0 & e 
\end{array} \right) .
\label{e:01331}
\end{equation}
\end{theorem}

The proof can be found in Appendix~\ref{app:D}.

Note that according to Lemma~\ref{le:linear_representation}, the matrix ${\bf T}_3$ can be used for the transformation of ${\bf w}$ describing certain $w \in H_2(\tilde U, \tilde U' \cup \tpB)$.

\begin{remark}
In Appendix~\ref{app:D} we propose an {\em ad hoc\/} reasoning valid only for dimension~2, too much linked to our notations, and not going in line with the mainstream of the Picard--Lefschetz theory. So we note here that the statement of Theorem~\ref{the:PLP_triangle} (as well as Theorems~\ref{the:PLP_biangle} and~\ref{the:PLP_circle}) follows from the following Picard--Lefschetz formula for relative homologies. 

\begin{proposition}[Pham, \cite{Pham1965}]
\label{pr:PL_relative}
 Let $\lambda$ be a small simple loop in $\cB$ bypassing (once in the positive direction) the component of $\cL$ corresponding to the vanishing of a polygon  $Q$ (which is a triangle formed by singularities $\sigma_1, \sigma_2, \sigma_3$, 
a biangle formed by singularities $\sigma_1, \sigma_2$, or a circle formed by $\sigma_1$). Then, for any $w \in H_2(\tilde U , \tilde U' \cup \tpB)$
\begin{equation}
{\rm var}_{\lambda} (w) = \sum_{\gamma \in \PI}
\langle \cE {(\gamma \omega Q_e  )} \, | \, w \rangle Q_\gamma
\label{e:proper_PL}
\end{equation}
where $\langle \cE (Q \gamma \omega) \, | \, w \rangle$
is the intersection index between corresponding homologies, and, 
\begin{equation}
\omega = \left\{ \begin{array}{lll}
(e - \gamma_1^{-1})(e - \gamma_2^{-1}) (e - \gamma_3^{-1}) & \mbox{for a triangle} \\
(e - \gamma_1^{-1})(e-\gamma_2^{-1}) & \mbox{for a biangle} \\
e - \gamma_1^{-1} & \mbox{for a circle} 
\end{array} \right.
\end{equation}
and $\gamma_1, \gamma_2, \gamma_3$ are simple loops about $\sigma_1, \sigma_2, \sigma_3$. 
\end{proposition}

The intersection index $ \langle w^1 \, | w^2  \rangle $ is defined as follows. Let
some representatives of the classes $w_1$ and $w_2$ be chosen in such a way that 
they intersect in a discrete set of points $p_j$ and the crossings are transversal. At each point $p_j$ take a pair of vectors $v^{m}_{n} = (v^m_{n,1} , v^m_{n,2})$ tangent to the representatives of homologies. The pair is ordered with respect to the orientation of~$w$; index $m = 1,2$ shows the link with $w^m$, and $n= 1,2$ is the number of the element in the pair. 
Then (see \cite{Pham2011}), 
\begin{equation}
\langle w^1 \, | w^2 \rangle = \sum_{p_j} {\rm sign}\, {\rm det}
\left( \begin{array}{cccc}
{\rm Re}[v^1_{1,1}] & {\rm Re}[v^1_{2,1}] & {\rm Re}[v^2_{1,1}] & {\rm Re}[v^2_{2,1}] \\
{\rm Im}[v^1_{1,1}] & {\rm Im}[v^1_{2,1}] & {\rm Im}[v^2_{1,1}] & {\rm Im}[v^2_{2,1}] \\
{\rm Re}[v^1_{1,2}] & {\rm Re}[v^1_{2,2}] & {\rm Re}[v^2_{1,2}] & {\rm Re}[v^2_{2,2}] \\
{\rm Im}[v^1_{1,2}] & {\rm Im}[v^1_{2,2}] & {\rm Im}[v^2_{1,2}] & {\rm Im}[v^2_{2,2}]
\end{array}\right)
\label{e:intersection_index}
\end{equation}

In Appendix~\ref{app:F} we present a scheme of 
calculation of intersection indices for a triangle (Pham's degeneration~$P_3$).

Moreover, in Appendix~\ref{app:G} we demonstrate a connection of the approach based on the relative homologies developed in the 
current paper with the approach based on the local systems from \cite{Vassiliev2002}. Besides, we again compute the variation 
of homologies in the case of the triangle. So, in total, there are three different proofs of (\ref{e:01331a})-(\ref{e:01331})
in the paper. 
\end{remark}

\subsubsection{The biangle}

\begin{figure}[h]
  \centering{ \includegraphics[width=0.6\textwidth]{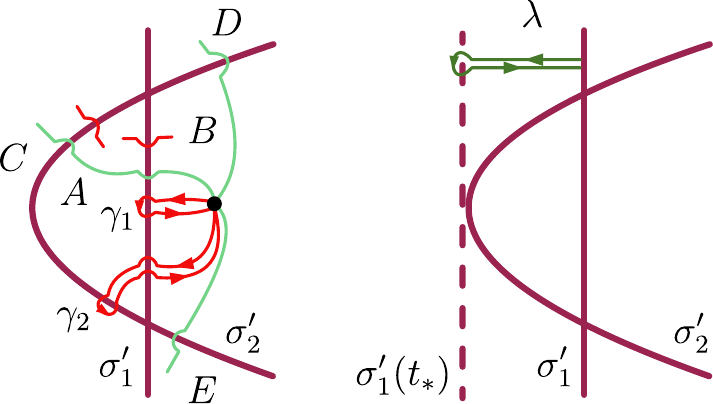} 
  }
  \caption{Notations for the biangle}
  \label{f:01061d}
\end{figure}

Let be $\sigma = \sigma_1 \cup \sigma_2$, such that the traces 
of the singularities form a biangle in $\mathbb{R}^2$ (see \figurename~\ref{f:01061d}). 
The  basic contours are shown in the same figure.
The group $\PI$ is $\mathbb{Z}^2$ with generators $\{ \gamma_1 , \gamma_2 \}$
shown in \figurename~\ref{f:01061d}. 
Introduce 
the polygons =$A, B, C, D , E$ (ordered this way).

For example, take the singularities 
\begin{equation}
\sigma_1(t) : \,\, z_1 = t, \qquad \sigma_2 : \,\, z_1 - z_2^2 = 0.
\label{e:biangle_sing}
\end{equation}
The particular form of the singularities is not important, the formulas 
(\ref{e:biangle_sing}) are used only for clarity.

The singular value of $t$ in this case is $t_* = 0$, for which the biangle $A$ vanishes.
Note that the line $\sigma_1 (t_*)$ is tangent to $\sigma_2$. 

\begin{figure}[h]
  \centering{ \includegraphics[width=0.3\textwidth]{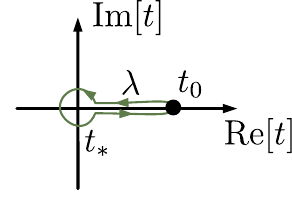}  }
  \caption{A typical loop $\lambda$ in the $t$-space}
  \label{f:01037c}
\end{figure}

Consider a simple loop 
 $\lambda$ in the $t$-space $\mathcal{B}$ bypassing $t_*$ once in the positive direction
(see \figurename~\ref{f:01037c}). 

\begin{theorem}
\label{the:PLP_biangle}
As the result of the bypass $\lambda$, the basis
transforms as follows:  
\begin{equation}
{\bf Q} \stackrel{\lambda}{\longrightarrow} {\bf T}_2 {\bf Q},
\qquad 
{\bf Q} = \left( \begin{array}{c}
A_e \\
\vdots \\
E_e
\end{array} \right), 
\qquad 
{\bf T}_2 = \left( \begin{array}{ccccc}
-\gamma_1 \gamma_2  & 0 & 0 & 0 & 0 \\ 
\gamma_1 + \gamma_1 \gamma_2  & e & 0 & 0 & 0 \\
\gamma_2 + \gamma_1 \gamma_2  & 0 & e & 0 & 0 \\
-\gamma_1 \gamma_2  & 0 & 0 & e & 0 \\
-\gamma_1 \gamma_2  & 0 & 0 & 0 & e 
\end{array} \right). 
\label{e:01333}
\end{equation}
\end{theorem}

The proof is given in Appendix~\ref{app:D}.


\subsubsection{The circle}

\begin{figure}[h]
  \centering{ \includegraphics[width=0.3\textwidth]{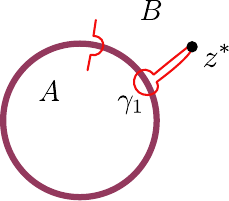}  }
  \caption{Notations for a circle}
  \label{f:01061g}
\end{figure}

Finally, consider a circle. Namely, let $\sigma = \sigma_1 (t)$ be defined by 
\begin{equation}
\sigma_1 (t): \, z_1^2 + z^2_2 = t,
\label{e:01315}
\end{equation}
and let $t$ move along some contour $\lambda$ bypassing zero once in the positive direction (see \figurename~\ref{f:01037c}).
The real trace $\sigma'$ splits $\mathbb{R}^2$ into the parts $A$ and $B$ as shown in 
\figurename~\ref{f:01061g}.

\begin{theorem}
\label{the:PLP_circle}
The basis is transformed as follows: 
\begin{equation}
{\bf Q} \stackrel{\lambda}{\longrightarrow} {\bf T}_1 {\bf Q},
\qquad 
{\bf Q} = \left( \begin{array}{c}
A_e \\
B_e
\end{array} \right), 
\qquad 
{\bf T}_1 = \left( \begin{array}{cc}
\gamma_1 & 0  \\ 
0 & e
\end{array} \right). 
\label{e:01334}
\end{equation}
\end{theorem}

The proof is given in Appendix~\ref{app:D}.


\subsubsection{A jump}

\begin{figure}[h]
  \centering{ \includegraphics[width=0.9\textwidth]{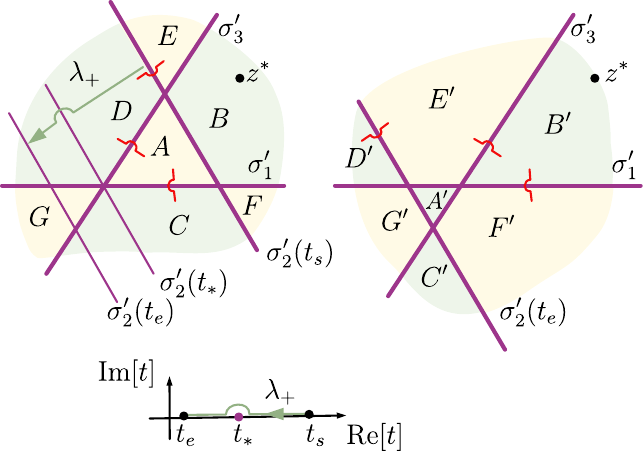}  }
  \caption{Jump $\lambda_+$}
  \label{f:01061a2a}
\end{figure}

Let the singularity $\sigma$ be a union of three lines $\sigma_1$, $\sigma_2$, $\sigma_3$, and assume that
$\sigma_2$ is movable: $\sigma_2 = \sigma_2 (t)$, where $t$ is a scalar parameter. 
Consider a variation $\lambda_+$ of $t$ shown in \figurename~\ref{f:01061a2a}, bottom. Let this variation 
correspond to the displacement of $\sigma_2$ shown in the left part of the figure. 
The starting value of $t$ is $t_s$, the end value is $t_e$, and the singular value is~$t_*$ ($\sigma_2 (t_*)$ passes through the crossing point of $\sigma_1$ and $\sigma_3$).

Introduce the polygons $A, \dots , G$ oriented in a natural way, the reference point $z^*$ in $B$, and the base paths corresponding to the surface 
$\Sigma$ indicated by bridge symbols (see the left figure). 

After the bypass, the plane becomes split into polygons in a different way (see the right part of the figure). Denote these 
polygons by $A', \dots , G'$. Let them also be oriented in a natural way. Introduce the notations for them using the surface $\Sigma$ described by red bridge symbols.

\begin{theorem}
\label{the:PLP_jump}
The basis is transformed as follows: 
\begin{equation}
{\bf Q} \stackrel{\lambda_+}{\longrightarrow} {\bf T}_{\lambda_+} {\bf Q}'.
\label{e:jump_b1}
\end{equation}
\begin{equation}
{\bf Q} = 
\left( \begin{array}{c}
A_e \\
B_e \\
C_e \\
D_e \\
E_e \\
F_e \\
G_e 
\end{array}\right) ,
\quad 
{\bf Q}' = 
\left( \begin{array}{c}
A'_e \\
B'_e \\
C'_e \\
D'_e \\
E'_e \\
F'_e \\
G'_e 
\end{array}\right),
\quad 
{\bf T}_{\lambda_+} = 
\left( \begin{array}{rcccccc}
e  & 0 & 0 & 0 & 0 & 0 & 0 \\
-e & e & 0 & 0 & 0 & 0 & 0 \\
-e & 0 & e & 0 & 0 & 0 & 0 \\
-e & 0 & 0 & e & 0 & 0 & 0 \\
e  & 0 & 0 & 0 & e & 0 & 0 \\
e  & 0 & 0 & 0 & 0 & e & 0 \\
e  & 0 & 0 & 0 & 0 & 0 & e 
\end{array}\right)
.
\label{e:jump_b}
\end{equation}

\end{theorem}

See the proof in Appendix~\ref{app:D}. 


\subsection{Building matrices ${\bf T}$ for more complicated cases }\label{BMT}
\label{sec:parabola}

\subsubsection{Using elementary cases as ``building blocks''}

\begin{figure}[h]
  \centering{ \includegraphics[width=0.5\textwidth]{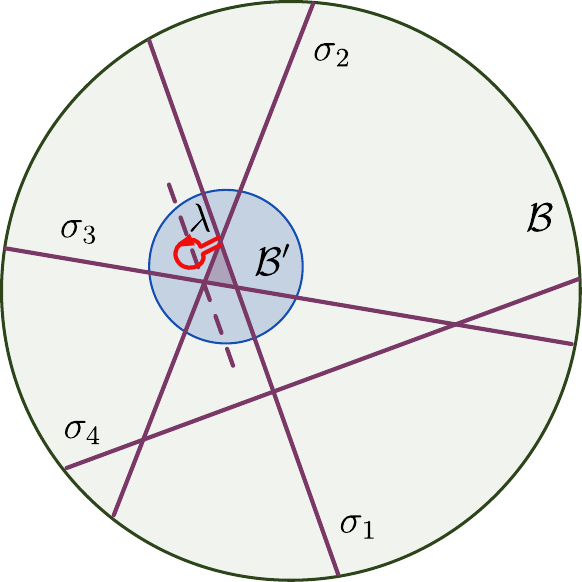}  }
  \caption{A ``large ball''  $\bB$ and a ``small ball'' $\bB'$}
  \label{f:small_ball}
\end{figure}

In the subsection above, we have described the transformation of relative homologies for three simplest basic cases (triangle, biangle, circle, jump). Here we consider more complicated cases. Our aim is to derive the matrices ${\bf T}$ for these cases, using 
the elementary cases as ``building blocks.'' For this, we use the following ideas:

\begin{itemize}

\item 
The locality concept. Let a transformation $\lambda$ (simple loop in $\cB$) corresponds to the vanishing of a certain triangle or  biangle (one of many polygons forming a basis of $H_2(\NR^2 , \sigma \cup \ptl \bB)$). Then only polygons adjacent to the vanishing one are affected by~$\psi_\lambda$, i.e.\ only a  submatrix of ${\bf T}$ is non-trivial, and this submatrix is given by (\ref{e:01331}), 
(\ref{e:01333}), (\ref{e:01334}), (\ref{e:jump_b}). ``Adjacent'' here means having common side or a common vertex.

The local configurations of singularities are studied in their ``small balls'', while the full configuration is contained in a 
``large ball''. This is shown schematically in \figurename~\ref{f:small_ball}. A ``large ball'' $\bB$ is green there, a ``small ball''
$\bB'$ is blue, and the vanishing triangle is red (it is formed by the components $\sigma_1$, $\sigma_2$, $\sigma_3$). A considered 
bypass $\lambda$ is shown by a red line. Only the polygons that have intersection with $\bB'$ are transformed by~$\lambda$.

\item 
If it is necessary to change the base paths related to the vanishing polygon, one can use Proposition~\ref{pr:path_change}. 

\item 
Several bypasses performed one by one are described by a product of matrices. Namely, if 
{
$\lambda$ is a concatenation of two bypasses, $\lambda_1$  and $\lambda_2$, i.e.\ }
$\lambda = \lambda_1 \lambda_2$, then 
\begin{equation}
{\bf T}_\lambda = {\bf T}_{\lambda_1} {\bf T}_{\lambda_2},
\label{e:matrix_mult}
\end{equation}
{
where matrices ${\bf T}_{\lambda_1}$ and ${\bf T}_{\lambda_2}$ are related to these bypasses.
}

\end{itemize}

Let us comment on the last point.
{
The formula (\ref{e:matrix_mult}) can be easily proven as follows. 
Using the notation (\ref{e:variation}) and the definition of the matrix 
${\bf T}_{\lambda_1}$, one can write}
\[
\psi_{\lambda_1}({\bf Q}) = {\bf T}_{\lambda_1} {\bf Q}, 
\qquad
\psi_{\lambda_2}({\bf Q}) = {\bf T}_{\lambda_2} {\bf Q}. 
\]
{
Note that here we extend the definition of $\psi_{\lambda}$ from the elements of 
$H_2 (\tilde U , \tilde U' \cup \tpB)$ to a vector-column ${\bf Q}$ of such elements;
the operator $\psi_\lambda$ is applied element-wise. }

Apply the transformation $\psi_{\lambda_2}$ to this identity: 
\[
\psi_{\lambda_2}(\psi_{\lambda_1}({\bf Q})) = 
\psi_{\lambda_2}({\bf T}_{\lambda_1}) \, \psi_{\lambda_2} ({\bf Q}) = 
{\bf T}_{\lambda_1} {\bf T}_{\lambda_2} {\bf Q}.
\]
The last relation follows from 
\[
\psi_2({\bf T}_{\lambda_1}) = {\bf T}_{\lambda_1}
\]
(the group ring does not ramify \purple{because of Lemma~\ref{pr:fundamental_ramification}}),
and 
\[
\psi_{\lambda_2}({\bf Q}) = {\bf T}_{\lambda_2} {\bf Q}. 
\]
The identity (\ref{e:matrix_mult}) is the main benefit of the matrix approach.



\subsubsection{Derivation of (\ref{e:01331}) from (\ref{e:jump_b})}

Let us show that the formula for a triangle (\ref{e:01331}) follows from the formula for a jump (\ref{e:jump_b}). 
Using this example, we: a) demonstrate how to implement 
Proposition~\ref{pr:path_change}, b) obtain one more proof of Theorem~\ref{the:PLP_triangle}, c) show an example of matrix-vector computations. The proof is not completely independent from that of Appendix~\ref{app:D}, but nevertheless it demonstrates some consistency of the method proposed in the paper.

\begin{figure}[h]
  \centering{ \includegraphics[width=0.8 \textwidth]{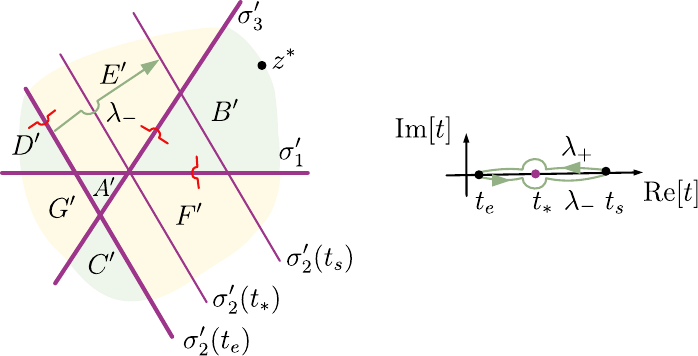}  }
  \caption{Jump $\lambda_-$}
  \label{f:jump_2}
\end{figure}

Note that the bypass $\lambda$ from \figurename~\ref{f:01061a} is a concatenation of the path $\lambda_+$ from 
\figurename~\ref{f:01061a2a} and path $\lambda_-$ from \figurename~\ref{f:jump_2}, {right}. The transformation of the basis of $H_2(\tilde U, \tilde U')$ under $\lambda_+$ is given by (\ref{e:jump_b}), so let us describe the transformation produced by $\lambda_-$, and then combine $\lambda_+$ with~$\lambda_-$.

{The configuration of the polygons $A, \dots, G$ becomes changed after 
the bypass~$\lambda_+$. The new polygons are denoted $A', \dots , G'$ and are shown in 
\figurename~\ref{f:jump_2}, left.}

 
To describe the effect of $\lambda_-$, use Theorem~\ref{the:PLP_jump} after some geometrical changes. 
One can see that the theorem cannot be applied directly, since 
the mutual position of the base paths and the bypass $\lambda_-$ 
differs from the conditions of the theorem. 
To change this, introduce new base paths,   
apply Theorem~\ref{the:PLP_jump}, and then apply Lemma~\ref{le:trans_contours}.

\begin{figure}[h]
  \centering{ \includegraphics[width=0.9\textwidth]{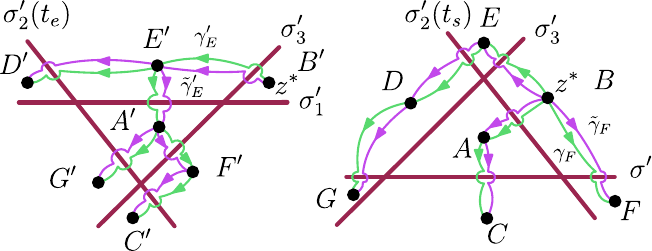} }
  \caption{``Old'' and ``new'' base contours}
  \label{f:01061a3}
\end{figure}

The new paths $\tilde \gamma'_A , \dots , \tilde \gamma'_G$ for the singularities 
$\sigma_1, \sigma_{2} (t_e), \sigma_3$ (\figurename~\ref{f:01061a3}, left)
and  $\tilde \gamma_A , \dots , \tilde \gamma_G$ 
for  $\sigma_1, \sigma_2 (t_s), \sigma_3$ (\figurename~\ref{f:01061a3}, right).
All paths should be understood as going from $z^*$ to each particular polygon along the corresponding tree (green for 
the ``old'' notations and magenta for the new notations).
The reference point $z^*$ is common for both figures (see
Proposition~\ref{pr:surface_Sigma_trivial}).

One can see by rotation of the figure that Theorem~\ref{the:PLP_jump} describes the 
bypass $\lambda_-$ if the basic contours $\tilde \gamma'_Q$,
are chosen. 
Thus, for these contours, 
\[
A'_e \stackrel{\lambda_-}{\longrightarrow} A_e,
\]
\[
B'_e \stackrel{\lambda_-}{\longrightarrow} B_e + A_e,
\]
\[
C'_e \stackrel{\lambda_-}{\longrightarrow} C_e + A_e,
\]
\[
D'_e \stackrel{\lambda_-}{\longrightarrow} D_e + A_e,
\]
\[
E'_e \stackrel{\lambda_-}{\longrightarrow} E_e - A_e,
\]
\[
F'_e \stackrel{\lambda_-}{\longrightarrow} F_e - A_e,
\]
\[
G'_e \stackrel{\lambda_-}{\longrightarrow} G_e - A_e.
\]

Let us return to the initial base contours $\gamma_Q$ and $\gamma'_Q$. For this, apply 
Proposition~\ref{pr:path_change}: 
\begin{equation}
A'_{\tilde \gamma'_A (\gamma'_A)^{-1}} 
\stackrel{\lambda_-}{\longrightarrow} 
A_{ \tilde \gamma_A  \gamma_A^{-1}},
\label{e:013061a10}
\end{equation}
\begin{equation}
B'_{\gamma'_B (\gamma'_B)^{-1}} 
\stackrel{\lambda_-}{\longrightarrow} B_{ \tilde\gamma_B \gamma_B^{-1}}+ 
A_{ \tilde \gamma_A  \gamma_A^{-1}},
\label{e:013061a11}
\end{equation}
\[
\vdots
\]
As one can see from \figurename~\ref{f:01061a3}, 
\[
\tilde \gamma'_A (\gamma'_A)^{-1} = \gamma_1^{-1} \gamma_3^{-1},
\qquad 
\tilde \gamma_A  \gamma_A^{-1} = \gamma_2 ,
\qquad 
\tilde   \gamma_B \gamma_B^{-1} = e.
\]
Thus, 
\[
A'_{\gamma_1^{-1} \gamma_3^{-1}} 
\stackrel{\lambda_-}{\longrightarrow} 
A_{ \gamma_2},
\]
\[
\vdots
\]
Taking into account (\ref{e:simp_1}), we can 
left-multiply the identities by corresponding elements of $\PI$
and rewrite (\ref{e:013061a10}) and (\ref{e:013061a11}) as 
\begin{equation}
A'_{e} 
\stackrel{\lambda_-}{\longrightarrow} 
A_{\gamma_1 \gamma_2  \gamma_3},
\label{e:01309a12}
\end{equation} 
\begin{equation}
B'_{e} 
\stackrel{\lambda_-}{\longrightarrow} B_{e}+ 
A_{ \gamma_2},
\label{e:013061a13}
\end{equation}
Similarly, obtain the relations 
\begin{equation}
C'_{e} 
\stackrel{\lambda_-}{\longrightarrow} C_{e}+ 
A_{ \gamma_1},
\label{e:013061a14}
\end{equation}
\begin{equation}
D'_{e} 
\stackrel{\lambda_-}{\longrightarrow} D_{e}+ 
A_{ \gamma_3},
\label{e:013061a15}
\end{equation}
\begin{equation}
E'_{e} 
\stackrel{\lambda_-}{\longrightarrow} E_{e} - 
A_{  \gamma_2 \gamma_3},
\label{e:013061a16}
\end{equation}
\begin{equation}
F'_{e} 
\stackrel{\lambda_-}{\longrightarrow} F_{e} - 
A_{ \gamma_1 \gamma_2},
\label{e:013061a17}
\end{equation}
\begin{equation}
G'_{e} 
\stackrel{\lambda_-}{\longrightarrow} G_{e} - 
A_{ \gamma_1 \gamma_3}.
\label{e:013061a18}
\end{equation}
The whole transformation can be written as 
\begin{equation}
\left( \begin{array}{c}
A'_e \\
B'_e \\
C'_e \\
D'_e \\
E'_e \\
F'_e \\
G'_e 
\end{array}\right) 
\stackrel{\lambda_-}{\longrightarrow}
{\bf T}_{\lambda_-}
\left( \begin{array}{c}
A_e \\
B_e \\
C_e \\
D_e \\
E_e \\
F_e \\
G_e 
\end{array}\right),
\qquad 
{\bf T}_{\lambda_-}
=
\left( \begin{array}{rcccccc}
\gamma_1 \gamma_2 \gamma_3  & 0 & 0 & 0 & 0 & 0 & 0 \\
\gamma_2 & e & 0 & 0 & 0 & 0 & 0 \\
\gamma_1 & 0 & e & 0 & 0 & 0 & 0 \\
\gamma_3 & 0 & 0 & e & 0 & 0 & 0 \\
-\gamma_2 \gamma_3  & 0 & 0 & 0 & e & 0 & 0 \\
-\gamma_1 \gamma_2  & 0 & 0 & 0 & 0 & e & 0 \\
-\gamma_1 \gamma_3  & 0 & 0 & 0 & 0 & 0 & e 
\end{array}\right).
\label{e:jump_d}
\end{equation}
Rewrite this equation as 
\begin{equation}
{\bf Q'}  \stackrel{\lambda_-}{\longrightarrow} {\bf T}_{\lambda_-} {\bf Q}. 
\label{e:jump_e}
\end{equation}

Now apply $\lambda_-$ to (\ref{e:jump_b1}). The result is 
\begin{equation}
{\bf Q} \stackrel{\lambda_+ \lambda_-}{\longrightarrow}
{\bf T}_{\lambda_+} {\bf T}_{\lambda_-} {\bf Q}. 
\label{e:jump_f}
\end{equation}

By performing multiplication, we obtain that 
\begin{equation}
{\bf T}_{\lambda_+} {\bf T}_{\lambda_-} = {\bf T}_3,
\label{e:jump_g}
\end{equation}
where ${\bf T}_3$ is defined by (\ref{e:01331}).


\subsubsection{Two lines and parabola}

Let the singularities have the form 
\[
\sigma(t) = \sigma_1(t) \cup \sigma_2(t) \cup \sigma_3,
\]
\begin{equation}
\sigma_1(t): \, z_1 = t_1, 
\quad 
\sigma_2(t): \, z_2 = t_2, 
\quad
\sigma_3: z_1 - z_2^2 = 0,
\label{e:01352}
\end{equation}
where $t = (t_1, t_2) \in \NR^2$. 
One can see that $\PI(t) = \mathbb{Z}^3$.

\begin{figure}[h]
  \centering{ \includegraphics[width=0.9\textwidth]{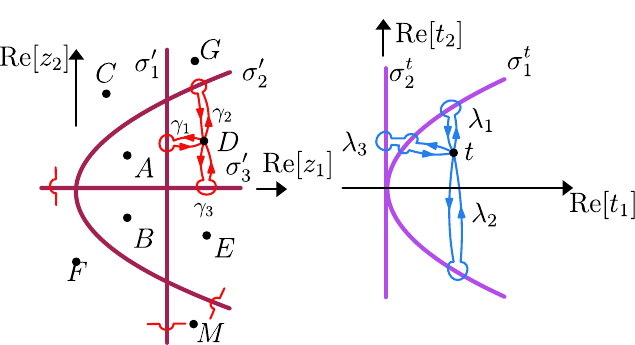}  }
  \caption{Singularities in the $(z_1, z_2)$-space (left) and in the $(t_1 , t_2)$-space (right)}
  \label{f:01061k}
\end{figure}

The singularities in the $(z_1, z_2)$-space are shown in \figurename~\ref{f:01061k}, left. 
As above, we introduce the polygons $A, \dots , M$, a reference point in $D$, and some generators of 
$\PI$ that are $\gamma_1 , \gamma_2, \gamma_3$. The base paths belong to the surface $\Sigma$
indicated by the red bridge symbols. 
According to the procedure developed above, the group $H_2 (\tilde U , \tilde U' \cup \tpB )$ is represented by 
vectors of the form
\begin{equation}
{\mathbf{w}} = (\omega_1 , \dots , \omega_8 ),
\qquad \omega_j \in \Omega, 
\label{e:01356}
\end{equation}
$\Omega$ is the group ring of $\PI$ over~$\mathbb{Z}$. 
{
The indices $1,\dots , 8$ correspond to the polygons ordered as $A, B,C,D,E,F,G,M$.
}

The structure of $H_2 (\NR^2 ,  \sigma \cup \ptl \bB)$
degenerates when $t_1 - t_2^2 = 0$ or $t_1 = 0$. 
Thus, the Landau set is 
\begin{equation} 
\bL = \sigma_1^t \cup \sigma_2^t,
\qquad
\sigma_1^t : \,\, t_1 - t_2^2 = 0, 
\qquad 
\sigma_2^t : \,\, t_1  = 0.
\label{e:01353}
\end{equation}
The singularities in the $t$-space are shown in
\figurename~\ref{f:01061k}, right.  

The singularities $\sigma_1^t$ and $\sigma_2^t$ are not crossing transversally: they have a quadratic touch at $t = (0,0)$. The fundamental group of $\cB = \NR^2 \setminus \cL$ is thus not necessarily 
Abelian. Introduce some loops $\lambda_1, \lambda_2, \lambda_3$ as shown in \figurename~\ref{f:01061k}. 
Note that $\lambda_3$ is not a simple loop. 

Our aim will be to build some matrices ${\bf T}_{\lambda_1}$,
${\bf T}_{\lambda_2}$, ${\bf T}_{\lambda_3}$ describing the ramification of $H_2(\tilde U, \tilde U' \cup \tpB)$
under the bypasses $\lambda_1$, $\lambda_2$, $\lambda_3$, respectively: 
\begin{equation}
{\bf w} \stackrel{\lambda_j}{\longrightarrow}
{\bf w} {\bf T}_{\lambda_j}.
\label{e:01357}
\end{equation}
Equivalently, the matrices ${\bf T}_{\lambda_1}$,
${\bf T}_{\lambda_2}$, ${\bf T}_{\lambda_3}$ describe the ramification of the basis of $H_2 (\tilde U, \tilde U' \cup \tpB)$:
\begin{equation}
{\bf Q}  \stackrel{\lambda_j}{\longrightarrow}
{\bf T}_{\lambda_j} {\bf Q},
\qquad 
{\bf Q}
= 
\left( \begin{array}{c}
A_e \\
\vdots \\
M_e
\end{array} \right) 
\end{equation}

The simplest computation leads to the matrix ${\bf T}_{\lambda_1}$. The bypass $\lambda_1$
corresponds to the vanishing of the triangle~$A$. All polygons except $M$ are adjacent to~$A$, thus, $M_e$ is not affected by~$\lambda_1$. 

\begin{figure}[h]
  \centering{ \includegraphics[width=0.7\textwidth]{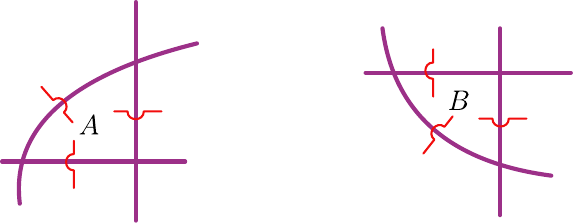}  }
  \caption{Orientation of $\Sigma$ near the triangles $A$ and $B$}
  \label{f:orientation}
\end{figure}

Study the surface $\Sigma$ near the triangle $A$ (\figurename~\ref{f:orientation}, left). Note that the bridge symbols are oriented in the same way as in \figurename~\ref{f:01061a1}. 
Thus, the submatrix of ${\bf T}_{\lambda_1}$ corresponding to the polygons $A, \dots, G$ is the same as (\ref{e:01331}) (up to rearranging of the elements):
\begin{equation}
{\bf T}_{\lambda_1} = \left( \begin{array}{cccccccc}
\gamma_1 \gamma_2 \gamma_3    & 0 & 0 & 0 & 0 & 0 & 0 & 0 \\
\gamma_3- \gamma_1 \gamma_2 \gamma_3 & e & 0 & 0 & 0 & 0 & 0 & 0 \\
\gamma_2- \gamma_1 \gamma_2 \gamma_3  & 0 & e & 0 & 0 & 0 & 0 & 0 \\
\gamma_1- \gamma_1 \gamma_2 \gamma_3 & 0 & 0 & e & 0 & 0 & 0 & 0 \\
-\gamma_1 \gamma_3 + \gamma_1 \gamma_2 \gamma_3 & 0 & 0 & 0 & e & 0 & 0 & 0 \\
-\gamma_2 \gamma_3 + \gamma_1 \gamma_2 \gamma_3 & 0 & 0 & 0 & 0 & e & 0 & 0 \\
-\gamma_1 \gamma_2 + \gamma_1 \gamma_2 \gamma_3 & 0 & 0 & 0 & 0 & 0 & e & 0 \\
0        & 0 & 0 & 0 & 0 & 0 & 0 & e 
\end{array} \right).
\label{e:01358}
\end{equation}

\begin{figure}[h]
  \centering{ \includegraphics[width=0.5\textwidth]{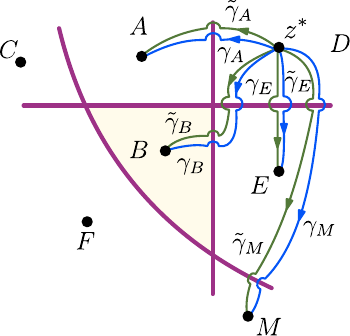}  }
  \caption{``New'' and ``old'' base paths near the polygon $B$}
  \label{f:new_old}
\end{figure}

Consider the bypass $\lambda_2$. It corresponds to the vanishing of~$B$. All polygons except 
$G$ are adjacent to $B$, so due to the locality idea our aim is to build a submatrix of ${\bf T}_{\lambda_2}$
corresponding to the lines $A,B,C,D,E,F,H$, and this matrix is closely linked with (\ref{e:01331}). However, 
this submatrix is not just a rearranged (\ref{e:01331}), since the orientation of the surface $\Sigma$ near $B$
(see \figurename~\ref{f:orientation}, right) differs from that of \figurename~\ref{f:01061a1}.
Thus, the matrix should be changed using Proposition~\ref{pr:path_change}.
This is done in a way similar to that of the previous subsubsection. Draw the ``new'' base path $\tilde \gamma_A , \dots , \tilde \gamma_M$ corresponding to the choice of $\Sigma$ 
in \figurename~\ref{f:01061k} and the ``old'' base path $\gamma_A , \dots , \gamma_M$
corresponding to the reference situation in \figurename~\ref{f:01061a1}.
Some of these paths are shown in \figurename~\ref{f:new_old} (the other ones are constructed similarly).

Use the ``old'' paths to describe the transformation of the polygons. According to formula
(\ref{e:01331}),
\[
A_e \stackrel{\lambda_2}{\longrightarrow} A_e + B_{\gamma_3} - B_{\gamma_1 \gamma_2 \gamma_3},
\]
\[
B_e \stackrel{\lambda_2}{\longrightarrow} B_{\gamma_1 \gamma_2 \gamma_3},
\]
\[
C_e \stackrel{\lambda_2}{\longrightarrow} E_e - B_{\gamma_1 \gamma_2} + B_{\gamma_1 \gamma_2 \gamma_3},
\]
\[
\dots
\]
Apply Proposition~\ref{pr:path_change} and rewrite these relations using the ``new'' base paths.
Note that 
\[
\gamma_A \tilde \gamma_A^{-1} = e,  
\qquad
\gamma_B \tilde \gamma_B^{-1} = \gamma_3^{-1}, 
\qquad \dots. 
\]
The result is 
\[
A_e \stackrel{\lambda_2}{\longrightarrow} A_e + B_{e} - B_{\gamma_1 \gamma_2},
\]
\begin{equation}
B_{\gamma_3^{-1}} \stackrel{\lambda_2}{\longrightarrow} B_{\gamma_1 \gamma_2},
\label{e:sample}
\end{equation}
\[
C_{e} \stackrel{\lambda_2}{\longrightarrow} C_{e} - B_{\gamma_2} + B_{\gamma_1 \gamma_2},
\]
\[
\dots
\]
If necessary, modify the obtained relation using (\ref{e:simp_1}) to make them describe the transformation of $Q^m_e$.
For example, (\ref{e:sample}) can be written as 
\[
B_{e} \stackrel{\lambda_2}{\longrightarrow} B_{\gamma_1 \gamma_2 \gamma_3}.
\]
Finally, the result is
\begin{equation}
{\bf T}_{\lambda_2} = \left( \begin{array}{cccccccc}
 e & e- \gamma_1 \gamma_2                    & 0 & 0 & 0 & 0 & 0 & 0 \\
 0 & \gamma_1 \gamma_2 \gamma_3              & 0 & 0 & 0 & 0 & 0 & 0 \\
 0 & -\gamma_2  + \gamma_1 \gamma_2          & e & 0 & 0 & 0 & 0 & 0 \\
 0 & -\gamma_1 + \gamma_1 \gamma_2           & 0 & e & 0 & 0 & 0 & 0 \\
 0 & \gamma_1 - \gamma_1 \gamma_2 \gamma_3   & 0 & 0 & e & 0 & 0 & 0 \\
 0 & \gamma_2 - \gamma_1 \gamma_2 \gamma_3   & 0 & 0 & 0 & e & 0 & 0 \\
 0 &       0 & 0 & 0 & 0 & 0 & e & 0 \\
0  & - \gamma_1 \gamma_2 + \gamma_1 \gamma_2 \gamma_3 & 0 & 0 & 0 & 0 & 0 & e 
\end{array} \right)
\label{e:01359}
\end{equation}

\begin{figure}[h]
  \centering{ \includegraphics[width=0.8\textwidth]{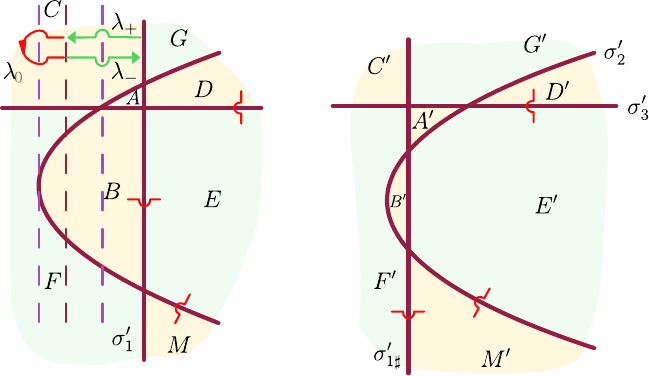}  }
  \caption{Intermediate step of building ${\bf T}_{\lambda_3}$}
  \label{f:01061s}
\end{figure}

The computation of ${\bf T}_{\lambda_3}$ is more complicated. 
One can see that $\lambda_3$ consists of a jump $\lambda_+$ over $\sigma_1^t$, a loop $\lambda_0$ about $\sigma_2^t$, and then, again, a jump $\lambda_-$ over~$\sigma_1^t$, 
see \figurename~\ref{f:01061s}, i.e.
\[
\lambda_3 = \lambda_+ \lambda_0 \lambda_-.
\]
We will use Theorems~\ref{the:PLP_biangle} and~\ref{the:PLP_jump} to describe these three parts one-by-one, and then multiply the obtained matrices. 

We will move the singularity~$\sigma_1$.
In the course of the bypass $\lambda_+$, the singularity $\sigma_1$ passes
one of the crossing points of
$\sigma_2$ and $\sigma_3$. We choose $A$ as the vanishing triangle.   
As a result of the bypass $\lambda_+$, we obtain a configuration 
shown in \figurename~\ref{f:01061s}, right, 
with an intermediate position of $\sigma_1$ denoted by $\sigma_{1 \sharp}$. 
For this configuration, we introduce the base contours by setting the bridge notations
(shown in red). We assume that the base contours are chosen on the surface
 $\Sigma$ corresponding to these bridges. 

According to Theorem~\ref{the:PLP_jump}, the jump $\lambda_+$
is described by the following matrix relation: 
\begin{equation}
{\bf Q} \stackrel{\lambda_+}{\longrightarrow} {\bf T}_{\lambda_+} {\bf Q'},
\label{e:01359a}
\end{equation}
\begin{equation}
{\bf Q}  = 
\left( \begin{array}{c}
A'_e \\
B'_e \\
C'_e \\
D'_e \\
E'_e \\
F'_e \\
G'_e \\
M'_e
\end{array} \right) ,
\quad
{\bf T}_{\lambda_+} =
\left( \begin{array}{cccccccc}
 e        &   0    & 0 & 0 & 0 & 0 & 0 & 0 \\
 -e     &  e        & 0 & 0 & 0 & 0 & 0 & 0 \\
 -e      &  0       & e & 0 & 0 & 0 & 0 & 0 \\
 -e     &  0       & 0 & e & 0 & 0 & 0 & 0 \\
 e      &   0       & 0 & 0 & e & 0 & 0 & 0 \\
 e      &   0       & 0 & 0 & 0 & e & 0 & 0 \\
 e      &   0     & 0 & 0 & 0 & 0 & e & 0 \\
 0      &   0     & 0 & 0 & 0 & 0 & 0 & e 
\end{array} \right).
\label{e:01359a1}
\end{equation}

Then the singularity $\sigma_1$ travels along $\lambda_0$ about the line parallel to $\sigma_1$ and tangent 
to~$\sigma_2$. For this bypass, the biangle $B'$ vanishes. Thus, the bypass can be described in a matrix form, and the 
matrix is a rearranged (\ref{e:01333}):
\begin{equation}
{\bf Q}' \stackrel{\lambda_0}{\longrightarrow} {\bf T}_{\lambda_0} {\bf Q}',
\qquad
{\bf T}_{\lambda_0} =
\left( \begin{array}{cccccccc}
 e     & - \gamma_1 \gamma_2    & 0 & 0 & 0 & 0 & 0 & 0 \\
 0     & - \gamma_1 \gamma_2    & 0 & 0 & 0 & 0 & 0 & 0 \\
 0     &  0       & e & 0 & 0 & 0 & 0 & 0 \\
 0     &  0       & 0 & e & 0 & 0 & 0 & 0 \\
 0     & \gamma_1 + \gamma_1 \gamma_2  & 0 & 0 & e & 0 & 0 & 0 \\
 0     & \gamma_2 + \gamma_1 \gamma_2  & 0 & 0 & 0 & e & 0 & 0 \\
 0     &   0      & 0 & 0 & 0 & 0 & e & 0 \\
 0     &   - \gamma_1 \gamma_2  & 0 & 0 & 0 & 0 & 0 & e 
\end{array} \right)
\label{e:01359b}
\end{equation}

Finally, we move $\sigma_2$ to its initial position along the path $\lambda_-$. This 
operation is described by 
\begin{equation}
{\bf Q}' \stackrel{\lambda_-}{\longrightarrow} {\bf T}_{\lambda_-} {\bf Q},
\qquad 
{\bf T}_{\lambda_-} =
\left( \begin{array}{cccccccc}
 \gamma_1 \gamma_2 \gamma_3        &    0   & 0 & 0 & 0 & 0 & 0 & 0 \\
 \gamma_3        &  e        & 0 & 0 & 0 & 0 & 0 & 0 \\
 \gamma_2         &  0       & e & 0 & 0 & 0 & 0 & 0 \\
 \gamma_1     &  0       & 0 & e & 0 & 0 & 0 & 0 \\
 - \gamma_1 \gamma_3      &   0       & 0 & 0 & e & 0 & 0 & 0 \\
 - \gamma_2 \gamma_3      &   0       & 0 & 0 & 0 & e & 0 & 0 \\
 - \gamma_1 \gamma_2      &   0     & 0 & 0 & 0 & 0 & e & 0 \\
 0      &   0     & 0 & 0 & 0 & 0 & 0 & e 
\end{array} \right),
\label{e:01359d}
\end{equation}
which is a rearranged version of (\ref{e:jump_d}).

Finally, we obtain 
\begin{equation}
{\bf T}_{\lambda_3} = {\bf T}_{\lambda_+} {\bf T}_{\lambda_0} {\bf T}_{\lambda_-}  = \left( \begin{array}{cccccccc}
 0        &    - \gamma_1 \gamma_2   & 0 & 0 & 0 & 0 & 0 & 0 \\
 - \gamma_1 \gamma_2 \gamma_3     &  0        & 0 & 0 & 0 & 0 & 0 & 0 \\
 \gamma_2        &  \gamma_1 \gamma_2       & e & 0 & 0 & 0 & 0 & 0 \\
 \gamma_1        &  \gamma_1 \gamma_2       & 0 & e & 0 & 0 & 0 & 0 \\
 \gamma_1 \gamma_2 \gamma_3      &   \gamma_1       & 0 & 0 & e & 0 & 0 & 0 \\
 \gamma_1 \gamma_2 \gamma_3      &   \gamma_2       & 0 & 0 & 0 & e & 0 & 0 \\
 -\gamma_1 \gamma_2      &   - \gamma_1 \gamma_2     & 0 & 0 & 0 & 0 & e & 0 \\
-\gamma_1 \gamma_2 \gamma_3      &   - \gamma_1 \gamma_2     & 0 & 0 & 0 & 0 & 0 & e 
\end{array} \right)
\label{e:01360}
\end{equation}

The matrices ${\bf T}_{\lambda_1}$, ${\bf T}_{\lambda_2}$, ${\bf T}_{\lambda_3}$ 
defined by (\ref{e:01358}), (\ref{e:01359}), (\ref{e:01360}), respectively, describe the ramification of relative homologies for the case of two lines and a parabola shown in 
\figurename~\ref{f:01061k}.




\section{Application of matrix formalism}
\label{sec:5}
Above, 
we have developed a computational framework  for the ramification of homolologies from $H_2(\tilde U, \tilde U')$ 
and $H_2(\tilde U)$. 
This framework consists of the basic formulas (\ref{e:01331}), (\ref{e:01333}), (\ref{e:01334}), and some instructions for their application. 
In this section we demonstrate that this framework gives some benefits, namely, it enables one to compute topological values in a fast and easy way.

\subsection{Picard--Lefschetz theory. Computation of the intersection index}

\subsubsection{Key result of Picard--Lefschetz theory}

Ramification of homologies from $H_2(\tilde U_2, \tpB)$, i.e.\ of integral surfaces of functions having branch singularities,
is studied in \cite{Pham2011,Pham1965}. According to \cite[VII,2.3]{Pham2011}, the key result for the 2D case is:

\begin{theorem}[Pham] 
\label{th:Pham}
Let $\lambda$ be a simple loop
about a component of $\cL$ corresponding to a vanishing polygon $Q$, 
which may be a circle, a biangle, or a triangle formed by singularity components $\sigma_1$, 
$\{ \sigma_1, \sigma_2 \} $, or $\{ \sigma_1 , \sigma_2 , \sigma_3 \} $, respectively. 
Let $\gamma_1, \gamma_2, \gamma_3 \in \PI$ be simple loops about $\sigma_1, \sigma_2, \sigma_3$.
Consider an integration surface $\Gamma \in H_2 (\tilde U_2 , \tpB)$.
Then 
\begin{equation}
{\rm var}_\lambda (\Gamma)= 
\sum_{\gamma \in \PI} \langle Q_\gamma \, | \, 
\Gamma \rangle \mathcal{E}( {\gamma \omega Q_e}   ),
\label{e:013351a}
\end{equation}
where $ \langle Q_\gamma | \Gamma \rangle $ is the intersection index of the corresponding homologies, and
\begin{equation}
\omega =  \left\{ 
\begin{array}{ll}
e - \gamma_1 & \mbox{for a circle} \\
(e - \gamma_1)(e - \gamma_2) & \mbox{for a biangle} \\
(e - \gamma_1)(e - \gamma_2)(e - \gamma_3) & \mbox{for a triangle} 
\end{array}
\right.
\label{e:013351b}
\end{equation}
\end{theorem}

Note that Pham considers a Riemann domain that is not necessarily universal, and the summation is held over the sheets of the corresponding Riemann domain. 

As it is stressed in \cite{Pham2011}, although  the sum in (\ref{e:013351a}) formally contains an infinite number of terms, only a finite number of them may be non-zero, since $\Gamma$ may intersect $Q$ only on a finite number of sheets of $\tilde U$. 

Combinations of the form ${\gamma \omega Q_e }$ with $\omega$ defined by (\ref{e:013351b}) play an important role: 
one can see that 
\[
\ptl ({ \gamma \omega Q_e} )  =0,
\]
and thus one can apply the inflation procedure to such an element of $H_2 (\tilde U, \tilde U' \cup \tpB)$. 
For a triangle, such an element is studied in Example~\ref{ex:triangle_H2U}.

The formula (\ref{e:013351a}) is mathematically neat but is not easy to apply since one has to compute intersection indices, and this may not be a simple task. 
As we will see below, the matrix formalism developed thus far will automatically take care of the intersection indices computations. 



\subsubsection{Computing intersection indices through matrix formalism}

\begin{example}
Consider the configuration of polygons and base paths shown in \figurename~\ref{f:01061a1}. Let the singularity $\sigma_2$ move along the path $\lambda$ shown in \figurename~\ref{f:01061a} (i.e.\ the situation corresponds to the conditions of Theorem~\ref{the:PLP_triangle}). 

Consider a surface $\Gamma$ to be a naturally oriented real plane slightly deformed near the singularities according to the bridge symbols shown in 
\figurename~\ref{f:01061h}, left. 
Assume that $\Gamma$ passes through the reference point located in~$B$. 

\begin{figure}[h]
  \centering{ \includegraphics[width=0.8\textwidth]{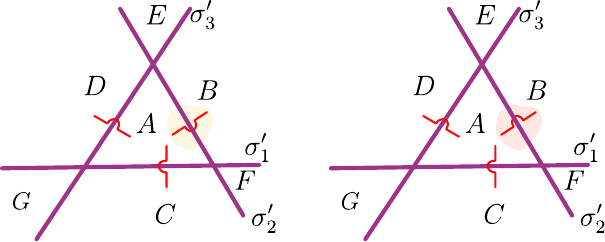}  }
  \caption{Two possible bridge notations for $\Gamma$ }
  \label{f:01061h}
\end{figure}

Let us compute ${\rm var}_{\lambda} (\Gamma)$ using the  procedure outlined in Subsection~\ref{s:overview}. 
Introduce
\[
w = \cR (\Gamma) \in H_2 (\tilde U^* , \tpB).
\]
One can see that the inclusion $H_2(\tilde U^* , \tpB) \hookrightarrow H_2(\tilde U, \tilde U' \cup \tpB)$
maps $w$ to 
\begin{equation}
w = A_e + B_e + C_e + D_e + E_e + F_e + G_e. 
\label{e:01335}
\end{equation}
Indeed, by construction
\[
\ptl w = 0.
\]
The element $w$ corresponds to the vector
\[
{\bf w}  = (e,e,e,e,e,e,e) \in \cM_7. 
\]
According to our scheme, and taking into account Lemma~\ref{le:linear_representation} we have 
\begin{equation}
{\rm var}_{\lambda} (\Gamma) = \cE ({\bf w} ({\bf T}_3 - {\bf I})),
\label{e:var_comp}
\end{equation}
where ${\bf I}$ is the $7 \times 7$ identity matrix (with $e$ on the diagonal), and 
${\bf T}_3$ is defined by (\ref{e:01331}).

Performing simple matrix computations, we obtain 
\begin{equation}
{\bf w} ({\bf T}_3 - {\bf I})  = 
( \omega , 0, 0, 0, 0, 0, 0),\quad \text{where}
\label{e:01337}
\end{equation}
\begin{align*}
\omega &= 
- e
+ \gamma_1 + \gamma_2 + \gamma_3
- \gamma_1 \gamma_2 - \gamma_2 \gamma_3 - \gamma_1 \gamma_3 
+ \gamma_1 \gamma_2 \gamma_3 \\
&= - (1- \gamma_1)(1- \gamma_2)(1- \gamma_3). 
\end{align*}
Thus, (\ref{e:var_comp}) gives the same result as (\ref{e:013351a}),
provided that
\begin{equation}
\langle A_\gamma | \Gamma \rangle = \left\{ \begin{array}{rl}
-1, & \gamma = e , \\
 0, & \gamma \ne e.
\end{array} \right.
\label{e:inters_1}
\end{equation}
One can check that (\ref{e:inters_1}) is correct by a direct computation.
\end{example}

\begin{example}
Let the singularities $\sigma_j$ and the bypass $\lambda$ correspond again to the 
conditions of Theorem~\ref{the:PLP_triangle}. 
Let $\Gamma$ be a naturally oriented real plane deformed near the singularities
according to the bridges shown in \figurename~\ref{f:01061h}, right.
Let us compute ${\rm var}_{\lambda} (\Gamma)$. 

Let us find $w = \cR (\Gamma)$. By applying Proposition~\ref{pr:path_change}, we obtain 
\begin{equation}
w = A_e + B_{\gamma_2^{-1}} + C_e + D_e + E_{\gamma_2^{-1}} + F_{\gamma_2^{-1}} + G_e.
\label{e:01339}
\end{equation}
The element (\ref{e:01339}) corresponds to the vector 
\begin{equation}
{\bf w} = (e  , \gamma_2^{-1} , e , e , \gamma_2^{-1}, \gamma_2^{-1}, e) \in \mathcal{M}_7.
\label{e:01340}
\end{equation}

A direct computation leads to 
\begin{equation}
{\bf w} ({\bf T}_3 - {\bf I})  = 
( 0 , 0, 0, 0, 0, 0, 0),
\label{e:01341}
\end{equation}
i.e.\ the surface $\Gamma$ is not ramifying under the bypass~$\lambda$.

Therefore, keeping in mind (\ref{e:var_comp}), in terms of Theorem~\ref{th:Pham}, this result corresponds to 
\[
\langle A_\gamma | \Gamma \rangle = 0 
\]
for all $\gamma\in \PI$, i.e.\ $\Gamma$ is not ``pinched'' by the singularities $\sigma_1, \sigma_2, \sigma_3$.

\end{example}

From the examples above, one can conclude that the matrix ${\bf T}_3$ (as well as ${\bf T}_2$ and ${\bf T}_1$) implicitly contains some formula enabling one to compute the intersection index between elements of $H_2(\tilde U_2 , \tpB)$ 
and $H_2 (\tilde U, \tilde U' \cup \tpB)$.


\subsection{Quadratic touch of components of $\cL$}

Consider the singularities (\ref{e:01352}) shown in \figurename~\ref{f:01061k}, left.
Such a configuration leads to the Landau set 
consisting of the parabola $\sigma^t_1$
and the line $\sigma^t_2$ (see (\ref{e:01353}) and \figurename~\ref{f:01061k}, right). These two components have a quadratic touch at $t = (0,0)$. The fundamental group of $\cB = \NR^2 \setminus \cL$
has a rather complicated structure (it is not Abelian). The matrices ${\bf T}_{\lambda_1}$
${\bf T}_{\lambda_2}$, ${\bf T}_{\lambda_3}$
given by (\ref{e:01358}), (\ref{e:01359}), (\ref{e:01360}) should provide  a representation of $\pi_1 (\cB)$, or at least of some of its subgroups. 

In Appendix~\ref{app:E} we show that the following lemma is valid
(see also \cite[V,3.2]{Pham2011}): 

\begin{lemma}
The fundamental group $\pi_1 (\NR^2 \setminus \cL)$, for $\cL$ defined by (\ref{e:01353}), is 
defined by the generators $\lambda_1$, $\lambda_2$, $\lambda_3$ (shown in \figurename~\ref{f:01061k}, right) and additional relations 
\begin{equation}
\lambda_2 \lambda_3 \lambda_1^{-1} \lambda_3^{-1} = e,
\qquad 
\lambda_1 \lambda_3 \lambda_2^{-1} \lambda_3^{-1} = e.
\label{e:relations_quad}
\end{equation}
\end{lemma}

Note that the same is valid for the whole $\mathbb{C}^2 \setminus \cL$.

By simple matrix computations in $\Omega$, one can easily check that 
the identities 
\begin{equation}
{\bf T}_{\lambda_2} {\bf T}_{\lambda_3} {\bf T}_{\lambda_1}^{-1} {\bf T}_{\lambda_3}^{-1} = {\bf I},
\qquad 
{\bf T}_{\lambda_1} {\bf T}_{\lambda_3} {\bf T}_{\lambda_2}^{-1} {\bf T}_{\lambda_3}^{-1} = {\bf I}
\label{e:relations_quad_1}
\end{equation}
are valid,
where ${\bf I}$ is the $8 \times 8$ identity matrix. Indeed, these identities guarantee (\ref{e:relations_quad}). 

Besides, one can check that the resulting group is non-commutative. For example, 
\begin{equation}
{\bf T}_3  {\bf T}_1 \ne {\bf T}_1 {\bf T}_3.
\label{e:01362b}
\end{equation}

Thus, the matrices 
(\ref{e:01358})--(\ref{e:01360})
contain some important information about the topology of~$\cB$.


\subsection{The concept of additive crossing}

\subsubsection{Configuration of singularities}

Consider the singularities 
\[
\sigma = \sigma_1 (t) \cup \sigma_2 (t) \cup \sigma_3,
\qquad 
t = (t_1, t_2) \in \NR^2,
\]
\begin{equation}
\sigma_1 : \quad z_1 = t_1,
\qquad 
\sigma_2 : \quad z_2 = t_2,
\qquad 
\sigma_3 : \quad z_1^2 + z_2^2 -1 = 0.
\label{sing_c2l}
\end{equation}

\begin{figure}[h]
  \centering{ \includegraphics[width=0.6\textwidth]{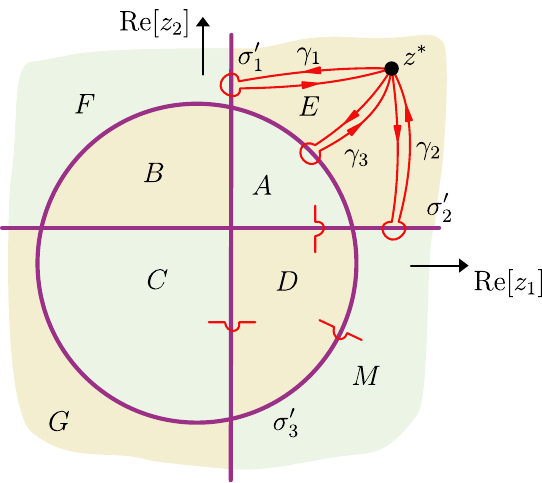}  }
  \caption{Polygons and $\Sigma$ for two lines and a circle}
  \label{f:01063}
\end{figure}

According to Remark~\ref{rem:Thom_restriction}, we require also that 
$|{\rm Re}[t_j] |< 1 + \delta'$ for $\delta'$ small enough. 
The real traces of the singularities are shown in \figurename~\ref{f:01063}.

Introduce the polygons $A,\dots , M$,
the reference point $z^*$ in $E$ far away from the singularities,
the generators $\gamma_1, \gamma_2 , \gamma_3$ of $\PI$, and the
base paths corresponding to the surface $\Sigma$ illustrated by the red bridges. 
As we have demonstrated, this set of graphical objects is enough 
to use algebraic notations in a non-ambiguous way. 

In general (i.e. for most values of $t$), the singularities cross each other tangentially, so $\PI = \mathbb{Z}^3$
with the generators~$\gamma_j$.


\subsubsection{Structure of $\cL$}

As $t$ varies in $\NR^2$, the configuration of the polygons may 
become different from the one in \figurename~\ref{f:01063}. 
The singular points in the $t$-space (forming the Landau set $\cL$) correspond to the 
following topological events:

\begin{itemize}
\item 
One of the triangles $A,B,C,D$ vanishes. This yields  the complex circle
\[
\sigma^t_c: \quad t_1^2 + t^2_2 = 1. 
\]

\item 
One of the two biangles ($A+D$ or $B+C$) formed by $\sigma_1$ and $\sigma_3$ vanishes.  This yields two complex lines
\[
\sigma^t_{1+}: \quad t_1 = 1, \qquad \qquad
\sigma^t_{1-}: \quad t_1 = -1.
\]

\item 
One of the two biangles ($A+B$ or $C+D$) formed by $\sigma_2$ and $\sigma_3$ vanishes.  This yields two complex lines
\[
\sigma^t_{2+}: \quad t_1 = 1, \qquad \qquad
\sigma^t_{2-}: \quad t_1 = -1.
\]

\end{itemize}

\begin{figure}[h]
  \centering{ \includegraphics[width=0.6\textwidth]{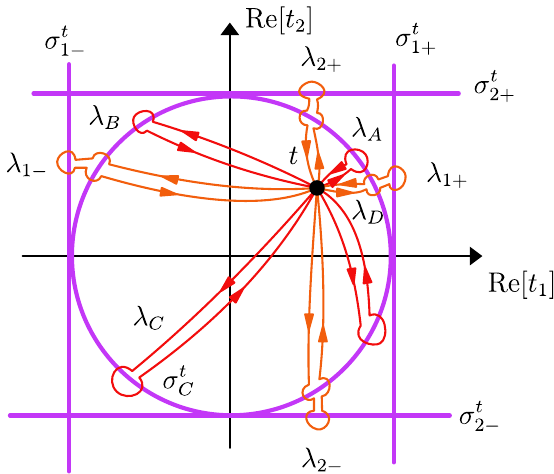}  }
  \caption{Components of $\cL$ and loops in the $t$-space}
  \label{f:01064}
\end{figure}

The set 
\[
\cL = \sigma^t_c \cup \sigma^t_{1+} \cup \sigma^t_{1-} \cup \sigma^t_{2+} \cup \sigma^t_{2-}
\]
is shown in \figurename~\ref{f:01064}.

Introduce 
the loops $\lambda_{1+}$, $\lambda_{1-}$,
$\lambda_{2+}$, $\lambda_{2-}$,
$\lambda_{A}$, $\lambda_{B}$,
$\lambda_{C}$, $\lambda_{D}$
in $\cB = \NR^2 \setminus \cL$
as shown in \figurename~\ref{f:01064}. 
Our aim will be to study the ramification of $H_2(\tilde U, \tilde U' \cup \tpB)$ as $t$
is carried along these loops.

The fundamental group of $\cB$ is of course different from $\PI$.
For example, $\PI$ is Abelian, while 
$\pi_1 (\cB)$ is not.  


\subsubsection{The matrices ${\bf T}_{\alpha}$}

Introduce the basis 
\[
{\bf Q} = \left( \begin{array}{c}
A_e \\
\vdots \\
M_e
\end{array} \right).
\]
Using the methods introduced above, one can compute the matrices 
${\bf T}_{\alpha}$ describing the ramification of the basis
\[
{\bf Q} \stackrel{\lambda_\alpha}{\longrightarrow}
{\bf T}_{\alpha} {\bf Q}.
\]
These matrices are written below. 
We use aliases $a \equiv \gamma_1$, $b \equiv \gamma_2$, $c \equiv \gamma_3$ to simplify notations. 

\begin{equation}
{\bf T}_A = \left( \begin{array}{cccccccc}
a b c	&   0	& 	0	&	0	&	0	& 	0	&	0	&	0	\\ 
e-b c	&   e	& 	0	&	0	&	0	& 	0	&	0	&	0	\\ 
-e+c	&   0	& 	e	&	0	&	0	& 	0	&	0	&	0	\\ 
e-ac	&   0	& 	0	&	e	&	0	& 	0	&	0	&	0	\\ 
c-abc	&   0	& 	0	&	0	&	e	& 	0	&	0	&	0	\\ 
-c +bc	&   0	& 	0	&	0	&	0	& 	e	&	0	&	0	\\ 
	0	&   0	& 	0	&	0	&	0	& 	0	&	e	&	0	\\ 
-c+ac	&   0	& 	0	&	0	&	0	& 	0	&	0	&	e	
\end{array} \right)
\label{e:014016}
\end{equation}

\begin{equation}
{\bf T}_{1+} = \left( \begin{array}{cccccccc}
	0	&   0	& 	0	& -abc	&	0	& 	0	&	0	&	0	\\ 
	e	&   e	& 	0	& bc 	&	0	& 	0	&	0	&	0	\\ 
	c	&   0	& 	e	&	e	&	0	& 	0	&	0	&	0	\\ 
	-ac	&   0	& 	0	&	0	&	0	& 	0	&	0	&	0	\\ 
	c	&   0	& 	0	&	abc	&	e	& 	0	&	0	&	0	\\ 
	-c	&   0	& 	0	&	-bc	&	0	& 	e	&	0	&	0	\\ 
	-c	&   0	& 	0	&	-c	&	0	& 	0	&	e	&	0	\\ 
	ac	&   0	& 	0	&	c	&	0	& 	0	&	0	&	e	
\end{array} \right)
\label{e:014017}
\end{equation}

\begin{equation}
{\bf T}_{2+} = \left( \begin{array}{cccccccc}
	0	&   -abc& 	0	&	0	&	0	& 	0	&	0	&	0	\\ 
	-bc	&   0	& 	0	&	0	&	0	& 	0	&	0	&	0	\\ 
	c	&   e	& 	e	&	0	&	0	& 	0	&	0	&	0	\\ 
	e	&   ac	& 	0	&	e	&	0	& 	0	&	0	&	0	\\ 
	c	&   abc	& 	0	&	0	&	e	& 	0	&	0	&	0	\\ 
	bc	&   c	& 	0	&	0	&	0	& 	e	&	0	&	0	\\ 
	-c	&   -c	& 	0	&	0	&	0	& 	0	&	e	&	0	\\ 
	-c	&   -ac	& 	0	&	0	&	0	& 	0	&	0	&	e	
\end{array} \right)
\label{e:014018}
\end{equation}

\begin{equation}
{\bf T}_{1-} = \left( \begin{array}{cccccccc}
	e	&   a	& 	abc	&	0	&	0	& 	0	&	0	&	0	\\ 
	0	&   0	& 	-abc&	0	&	0	& 	0	&	0	&	0	\\ 
	0	&   -ac	& 	0	&	0	&	0	& 	0	&	0	&	0	\\ 
	0	&   ac	& 	a	&	e	&	0	& 	0	&	0	&	0	\\ 
	0	&   -ac	& 	-abc&	0	&	e	& 	0	&	0	&	0	\\ 
	0	&   c	& 	abc	&	0	&	0	& 	e	&	0	&	0	\\ 
	0	&   ac	& 	c	&	0	&	0	& 	0	&	e	&	0	\\ 
	0	&   -ac	& 	-ac	&	0	&	0	& 	0	&	0	&	e	
\end{array} \right)
\label{e:014019}
\end{equation}

\begin{equation}
{\bf T}_{2-} = \left( \begin{array}{cccccccc}
	e	&   0	& 	abc	&	b	&	0	& 	0	&	0	&	0	\\ 
	0	&   e	& 	b	&	bc	&	0	& 	0	&	0	&	0	\\ 
	0	&   0	& 	0	&	-bc	&	0	& 	0	&	0	&	0	\\ 
	0	&   0	& 	-abc&	0	&	0	& 	0	&	0	&	0	\\ 
	0	&   0	& 	-abc&	-bc	&	e	& 	0	&	0	&	0	\\ 
	0	&   0	& 	-bc	&	-bc	&	0	& 	e	&	0	&	0	\\ 
	0	&   0	& 	c	&	bc	&	0	& 	0	&	e	&	0	\\ 
	0	&   0	& 	abc	&	c	&	0	& 	0	&	0	&	e	
\end{array} \right)
\label{e:014020}
\end{equation}

Using  matrix algebra, we can check some identities expressing non-trivial 
topological properties of~$\cB$. Namely, for example,
\begin{equation}
{\bf T}_{2+} {\bf T}_{A} {\bf T}_{2+}^{-1} {\bf T}_B^{-1} = {\bf I},
\qquad 
{\bf T}_{2+} {\bf T}_{B} {\bf T}_{2+}^{-1} {\bf T}_A^{-1} = {\bf I},
\label{e:014021}
\end{equation}
that are representations of relations (\ref{e:relations_quad}) for a quadratic touch of
$\sigma^t_{2+}$ and~$\sigma^t_c$.

\begin{figure}[h]
  \centering{ \includegraphics[width=0.5\textwidth]{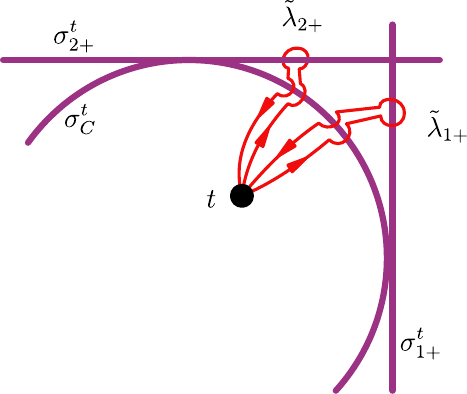}  }
  \caption{Simple loops $\tilde \lambda_{1+}$ and $\tilde \lambda_{2+}$}
  \label{f:tilde_lambda}
\end{figure}

Introduce the bypasses 
\[
\tilde  \lambda_{1+} = \lambda_{1+} \lambda^{-1}_A, 
\qquad
\tilde  \lambda_{2+} = \lambda_{2+} \lambda^{-1}_A, 
\]
that are simple loops about $\sigma^t_{1+}$ and $\sigma^t_{2+}$, 
respectively (see \figurename~\ref{f:tilde_lambda}).
These loops are described by the matrices
\begin{equation}
\tilde {\bf T}_{1+} \equiv {\bf T}_{1+} {\bf T}_A^{-1},
\qquad 
\tilde {\bf T}_{2+} \equiv {\bf T}_{2+} {\bf T}_A^{-1}.
\label{e:014023}
\end{equation}
One can check directly that 
\begin{equation}
\tilde {\bf T}_{1+} \tilde {\bf T}_{2+}
= 
\tilde {\bf T}_{2+} \tilde {\bf T}_{1+}.
\label{e:014024}
\end{equation}
This identity reflects the fact that the corresponding simple loops
commute in~$\pi_1 (\cB)$ since $\sigma^t_{1+}$ and $\sigma^t_{2+}$ intersect transversally.

\subsubsection{The ``additive crossing'' identity and its meaning}

One can check by direct matrix computations in $\Omega$ that 
\begin{equation}
{\bf I} + \tilde {\bf T}_{1+} \tilde {\bf T}_{2+} = 
\tilde {\bf T}_{1+} + \tilde {\bf T}_{2+}.
\label{e:014025}
\end{equation}
This identity is important for application since it is connected with the concept of
{\em additive crossing\/} of the lines $\sigma^t_{1+}$ and $\sigma^t_{2+}$. Let us explain this. 

Consider an integral (\ref{e:i0001}) with $F(z ; t)$ having the singularities (\ref{sing_c2l}), for example, the integral (\ref{e:i0001a}). Let 
\begin{equation}
\Gamma = \cE (w) = \cE ({\bf w} {\bf Q})
\label{e:cont_0}
\end{equation}
for some ${\bf w} \in \cM_8$ chosen such that $\ptl w = 0$.
Fix some reference point $t$, take the value $I(t)$ and denote it by $I_e (t)$. 
Continue $I(t)$ along the closed loop $\tilde \lambda_{1+}$ starting from the reference point, and denote the resulting value of the integral by $I_{\tilde \lambda_{1+}} (t)$. Similarly, introduce $I_{\tilde \lambda_{2+}} (t)$
and  $I_{\tilde \lambda_{1+} \tilde \lambda_{2+}} (t)$.

Let $\lambda$ be a path in $\cB$ going from $t$ to $t$. Indeed, $I(t)$ cannot have singularities other than on $\cL$, so $I(t)$ can be continued along~$\lambda$.
According to the procedure of analytical continuation, the continued value is as follows:
\begin{equation}
I_{\lambda} (t) = \int_{\psi_\lambda (\Gamma)} F(z; t) \, dz_1 \wedge dz_2.
\label{e:cont_1}
\end{equation}
The theorems from this paper yield 
\begin{equation}
\psi_\lambda (\Gamma) = \cE ({\bf w} {\bf T}_{\lambda} {\bf Q}).
\label{e:cont_2}
\end{equation}

Consider the value 
\begin{equation}
I'(t) = I_e(t) + I_{\tilde \lambda_{1+} \tilde \lambda_{2+}}(t) - 
I_{\tilde \lambda_{1+} }(t) - I_{\tilde \lambda_{2+}}(t) .
\label{e:cont_3}
\end{equation}
By linearity, 
\begin{equation}
I'(t) = \int_{\Gamma'} F(z; t) \, dz_1 \wedge dz_2,
\label{e:cont_4}
\end{equation}
where 
\begin{equation}
\Gamma' = \cE \left( {\bf w} ({\bf I} + {\bf T}_{\tilde \lambda_{1+} \tilde \lambda_{2+}}
- {\bf T}_{\tilde \lambda_{1+}}- {\bf T}_{\tilde \lambda_{2+}}) {\bf Q} \right).
\label{e:cont_5a}
\end{equation}
According to (\ref{e:014025}), 
\begin{equation}
\Gamma' = \cE \left( 0 \right) = 0, 
\label{e:cont_5b}
\end{equation}
and thus $I' (t) = 0$. Finally, we can write 
\begin{equation}
I_e(t) + I_{\tilde \lambda_{1+} \tilde \lambda_{2+}}(t) = 
I_{\tilde \lambda_{1+} }(t) + I_{\tilde \lambda_{2+}}(t) .
\label{e:cont_6}
\end{equation}
This identity is valid for $t$ taken in some complex domain, thus it can be continued to the whole domain of analyticity of $I(t)$.

Note that (\ref{e:cont_6}) has a purely topological nature, and it is not based on particular properties of $F(z;t)$ ((\ref{e:i0001a}) is just an example). 

In \cite{Assier2019} the authors referred to functions 
having property (\ref{e:cont_6}) as having {\em additive crossing\/} of singularities 
$\sigma^t_{1+}$ and $\sigma^t_{2+}$.
The reason  for this naming is explained in \cite{Assier2019}: under simple growth conditions, 
(\ref{e:cont_6}) means that at least on a certain sheet of its Riemann surface $I(t)$ can be written near $\sigma^t_{1+} \cap \sigma^t_{2+} = (1,1)$ as
\begin{equation}
I(t) = I_1 (t) + I_2 (t),
\label{e:cont_7}
\end{equation}
where $I_1 (t)$ is only singular at $\sigma^t_{1+}$, and $I_2(t)$ is only singular at $\sigma^t_{2+}$.

The concept of additive crossing is then used in the estimation of Fourier integrals
containing large parameters \cite{Assier2022}. Namely, the authors claim that while, generally, a crossing of two singularities gives a contribution to the asymptotics of an integral, a contribution of an additive crossing is always zero. Indeed, additive crossings of branch singularities play a major role in diffraction theory and has materialized itself in several different cases such as the 3D problem of wave diffraction by a quarter-plane \cite{Assier2019,Assier2024}, the 2D problem of wave diffraction by a penetrable wedge \cite{Kunz2023,Kunz2024}, as well as in the process of analytical continuation of physical wave fields \cite{Assier2021}.

 
\section{Conclusion}

The main results of the paper can be summarized as follows: 

\begin{itemize}

\item 
For the part of complex space $\mathbb{C}^2$ close to the real plane (denoted $\NR^2$) convenient graphic notations are introduced. They allow one to denote paths and loops bypassing the singularities in a clear and unambiguous way. Moreover, these notations enable one to visualize homologies from $H_2 (\tilde U_2 , \tpB )$ (surfaces of integration for instance). 

\item
The theorem of inflation (Theorem~\ref{th:inflattability}) is proven. This allows us to study the ramification of the relative homologies from $H_2 (\tilde U, \tilde U' \cup \tpB)$ instead of 
$H_2(\tilde U_2 , \tpB)$. The group $H_2 (\tilde U, \tilde U' \cup \tpB)$ is an object simpler than $H_2(\tilde U_2 , \tpB)$: 
it is isomorphic to the module $\cM_k$, and its basis is known. Due to linearity, we only need to consider this basis to study the ramification of $H_2 (\tilde U, \tilde U' \cup \tpB)$. The ramification of the basis is described linearly, i.e.\ by  matrix multiplication.

\item
The matrices describing the basis transformations for several elementary cases (or building blocks) are derived and given in (\ref{e:01331}), (\ref{e:01333}), (\ref{e:01334}) and (\ref{e:jump_b}). 
{The matrices can be derived from the Picard--Lefschetz formulae for relative homologies.} 
We claim that together with an auxiliary formula (\ref{e:change_notations}) this set of formulas is enough to describe the ramification along any loop in any reasonable configuration. 
Thus, a description of the ramification can be obtained algorithmically through this matrix formalism. 

\item 
Several examples of computations based on this matrix method are given. Even in the simplest considered cases, the computations demonstrate that the matrices built for the description of the ramification of $H_2 (\tilde U, \tilde U' \cup \tpB)$ contain important topological information. Namely, 
these matrices enable one to compute intersection indices in the $z$-space between the elements 
of $H_2(\tilde U_2 , \tpB)$ and $H_2 (\tilde U, \tilde U' \cup \tpB)$, provide non-trivial relations for the fundamental group of $\cB$ in the $t$-space, and, most surprisingly, provide the additive crossing relations for integrals. 
{As a check of our method, we demonstrate that the matrix computations reproduce the classical Picard--Lefschetz result for absolute homologies. }

\end{itemize}

As directions for future work, we are going to weaken some restrictions imposed on the singularities and apply the result to integrals emerging in physical problems that can be formulated as two-complex-variables Wiener-Hopf equations.

\section*{Acknowledgements}
For A.V.~Shanin, the study was conducted under the state assignment of Lomonosov Moscow State University.
{R.C.~Assier, A.V.~Shanin, and A.I.~Korolkov are grateful to the Isaac Newton Institute (Cambridge, UK) for the retreat program of 2025
that helped the 
team to work together on the subject of the paper. The authors are grateful to D.V.~Artamonov for a helpful discussions.}

\appendix
\appendixpage


\section{Proof of Lemma~\ref{le:locality}}
\label{app:A}

a) Perform a biholomorphic coordinate change $z\to w = (w_1 , w_2)$ such that 
\[
w_1 = g_j(z). 
\]
The singularity $\sigma_j$ becomes given by $w_1 = 0$. It is quite clear that the fundamental group 
of $D \setminus \sigma$ is $\mathbb{Z}$ with a simple loop $\gamma^{el}$ about $w_1 = 0$ as a generator. 

We should prove that there are no ``external'' relations, i.e.\ that there exists no homotopy making $(\gamma^{el})^\nu = e$
in $\cX \setminus \sigma$ outside~$D$. For this, consider the function 
\[
f(z) = \log(g_j(z))
\]
in $\cX \setminus \sigma$. 
By construction, this function is single-valued on the universal Riemann domain. 
The values of this function are different for different $\nu$, so such external homotopy cannot exist.

b) First, let us prove that all paths $\gamma'' \in \Pi_D(z' , z)$ are homotopic to each other. For this, use the coordinates 
$w$ introduced above. Let for simplicity the coordinates $w_2$ of $z$ and $z'$ be both equal~0. 
Let also be ${\rm Arg}[w_1] = 0$ for~$z'$.

Let the path $\gamma''$ be parametrized by a variable $\tau \in [0,1]$, i.e.\ 
\[
\gamma'' : \qquad w_1 = w_1(\tau), \quad w_2 = w_2(\tau).
\]

Consider the following homotopy parametrized by $\alpha \in [0,1]$:
\[
w_1(\tau ; \alpha) = |w_1(\tau)|\exp\{ i (1- \alpha){\rm Arg} [w_1(\tau)] \}  , 
\quad w_2(\tau; \alpha) = (1- \alpha) w_2(\tau).
\]
This homotopy  ``unwinds'' a path $\gamma''$ into a straight segment.

Second, any path $\gamma \in \Pi(z^* , z)$ can be represented as 
\[
\gamma = \gamma' \gamma'' 
\]
for some $\gamma' \in \Pi(z^* , z')$, $\gamma'' \in \Pi(z' , z)$.
Let there be two such representations: 
$\gamma_1 = \gamma_1' \gamma_1''$ and $\gamma_2 = \gamma_2' \gamma_2''$, and let $\gamma_1$ and $\gamma_2$
be connected by a homotopy. We have to show that 
\[
\gamma'_2 = \gamma'_1 \gamma_D,
\qquad 
\gamma_D \in \Pi_D(z',z'). 
\]

Let there be the homotopy variable be $\alpha \in [0,1]$, providing $\gamma(\tau ; \alpha)$
with $\gamma(\tau ; 0) = \gamma_1$, $\gamma(\tau ; 1) = \gamma_2$. By remapping the variable $\tau$ if necessary, we can achieve the following: there will exist some $\tau'$ such that 
\[
\gamma(\tau';0) = z',
\]
\[
\gamma(\tau';1) = z',
\]
\[
\gamma(\tau ; \alpha) \in D \quad \mbox{for} \quad \tau' \le \tau \le 1
\]
Then, $\gamma_D$ is formed by the points $\gamma(\tau', \alpha)$ for $\alpha \in [0,1]$.


\section{ Visual notations for paths and loops}
\label{app:B}

\subsection{Bridge notations for paths} 
 
Consider the domain $\NR^2 \setminus \sigma$ for the variable $z$
with the singularities $\sigma$ having the real property. Here we introduce convenient notations
to describe graphically paths in $\NR^2 \setminus \sigma$. 
We are particularly interested in paths connecting real points of $\NR^2 \setminus \sigma$; 
such points will be denoted by the symbol $x = (x_1 , x_2)$ rather than $z = (z_1 , z_2)$.   

Let the beginning and the end of some path be some points 
$x^b$ and $x^e$
belonging to $\mathbb{R}^2 \setminus \sigma'$.
Our aim is to describe a path $\gamma$ going from 
$x^b$ to $x^e$ in $\NR^2 \setminus \sigma$.

Project the path $\gamma$ onto the plane $\mathbb{R}^2$. The result is the contour~$\gamma'$. 
This contour can cross some real traces $\sigma_j'$. We assume that each such crossing is simple,
i.e.\ $\gamma'$ does not pass through crossings $\sigma'_j \cap \sigma_k'$.

\begin{figure}[h]
  \centering{\includegraphics[width=0.8\textwidth]{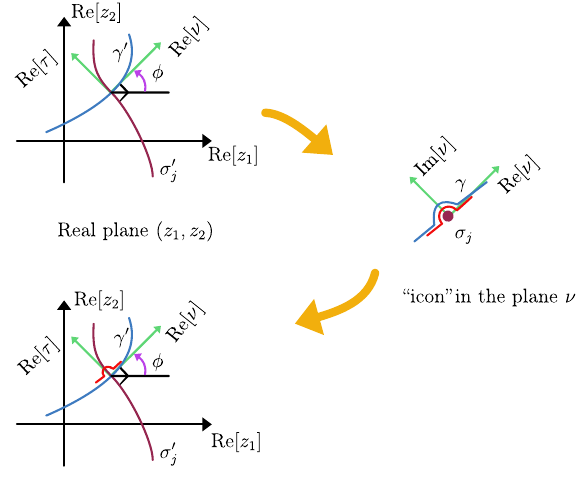}}
  \caption{Icon in the plane $\nu$}
  \label{fig:01030a1}
\end{figure}

At each crossing point $\gamma' \cap \sigma_j'$ we should show how the contour $\gamma$ 
bypasses~$\sigma_j$.
Consider a point $z^* = (z_1^*, z_2^*) \in \sigma_j'$.
Introduce some local variables $(\nu , \tau)$
near $z^*$ by the formulas 
\begin{equation}
\nu = g_j(z_1, z_2),
\qquad 
\tau = -(z_1 - z_1 ^* ) \sin \phi + (z_2 - z_2^* ) \cos \phi, 
\label{e:transformation_1}
\end{equation}
where $\phi$ is an angle between the normal vector to $\sigma_j'$ and the real axis~$z_1$
(see \figurename~\ref{fig:01030a1}, top left).
Note that $\phi$ is a real angle, and $g_j$ has real values for real~$z$, however
we treat (\ref{e:transformation_1}) as a complex transformation.
We say that $\nu$ is the {\em transversal variable}.  
 
Consider (locally) the projection of the contour $\gamma$ onto the complex plane $\nu$ (see \figurename~\ref{fig:01030a1}, right).
The singularity $\sigma_j$ is depicted in the figure by the point $\nu = 0$. The contour $\gamma$ projected
onto the $\nu$-plane coincides with the real axis almost everywhere, and it should bypass the origin 
either above or below. A bypass above or below is drawn in the usual 1D complex analysis by a bridge,  
and we plan to use this bridge as  an ``icon''. 

As shown in \figurename~\ref{fig:01030a1}, the local graph (the right part of the figure) in the $\nu$-plane is drawn in such a way that the 
real axis of $\nu$ is collinear to the real axis of $\nu$-plane in the $({\rm Re}[z_1], {\rm Re}[z_2])$-coordinates (the top left part of the figure). 

Then, we put the local graph atop the graph in the $({\rm Re}[z_1], {\rm Re}[z_2])$-coordinates 
(see \figurename~\ref{fig:01030a1}, bottom left). 
The bridge becomes placed on $\gamma'$, and it shows how $\gamma$ bypasses~$\sigma_j$.


\begin{figure}[h]
  \centering{\includegraphics[width=0.9\textwidth]{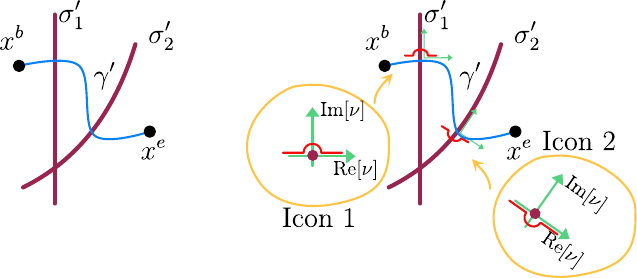}}
  \caption{Projection $\gamma'$ (left); icons at the crossings (right)}
  \label{fig:01030a}
\end{figure} 

The bridge notations for integration surfaces were introduced in \cite{Assier2022}, where one can find some details.

The bridge notations happen to be a flexible and convenient way to display topological objects in~$\NR^2$.
Two bridges related to the same component $\sigma_j$ are equivalent if they can be obtained from each other
either by continuous rotation (such that the ``stems'' of the bridge do not become parallel to $\sigma_j'$) or by parallel sliding along 
$\sigma_j$.
As $\gamma$ moves homotopically in $\NR^2$, a bridge can only be changed to an equivalent one. 

\begin{figure}[h]
  \centering{\includegraphics[width=0.25\textwidth]{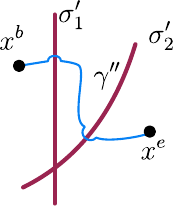}}
  \caption{The resulting contour $\gamma''$}
  \label{fig:01030b}
\end{figure}

Another example of bridges is shown in \figurename~\ref{fig:01030a}.
For display purposes, 
we integrate the bridge notations to the contour $\gamma'$ and obtain 
some resulting contour $\gamma''$ (see \figurename~\ref{fig:01030b}).

\subsection{An addition to bridges: small loops}

\begin{figure}[h]
  \centering{\includegraphics[width=0.9\textwidth]{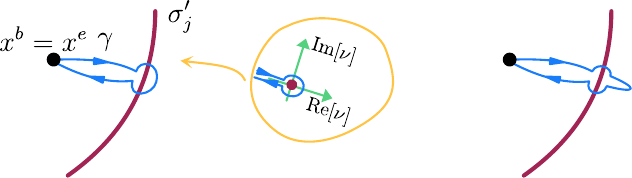}}
  \caption{A small loop on $\gamma$}
  \label{fig:01030c}
\end{figure}

Beside the bridge notations, we also introduce the  small loop notations. Namely, 
a projection of $\gamma$ onto the plane of the transversal 
variable~$\nu$ can make a full circle about $\nu = 0$, i.e.\ about the singularity. 
A corresponding icon is put on $\gamma$ (see \figurename~\ref{fig:01030c}), left.  
Indeed,  a small loop is represented as a combination of two bridges (see the right part of the figure).  

\subsection{Bridge notations for surfaces}
\label{s:vis_surf}

\begin{figure}[h]
  \centering{\includegraphics[width=0.6\textwidth]{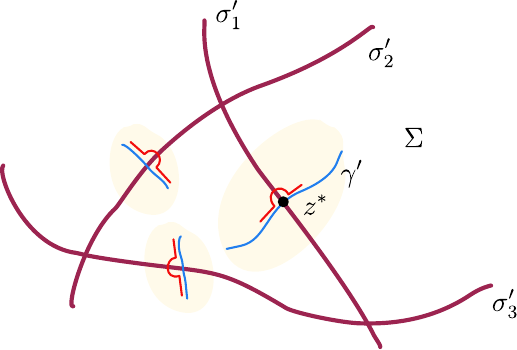}}
  \caption{Indentation of a surface}
  \label{fig:01030d1}
\end{figure}

Let $\Sigma$ be a surface coinciding with the real plane of variables $(z_1, z_2)$ almost 
everywhere and slightly deformed (indented) to avoid some singularity components $\sigma_j$
having the real property. The bridge notation can be used to indicate the indentation of the 
surface. Namely, consider $\Sigma$ near some point $z^*$ belonging to $\sigma_j'$
and not belonging to any other singularity components. 
Take some contour $\gamma \subset \Sigma$, such that the projection of $\gamma$
onto the real plane denoted $\gamma'$ passes through~$z^*$ (see \figurename~\ref{fig:01030d1}).
The same is done for all other irreducible singularity components.
Thus, one obtains a set of bridges showing how $\Sigma$ bypasses  each of the singularity 
components.

\begin{figure}[h]
\centering{\includegraphics[width=0.6\textwidth]{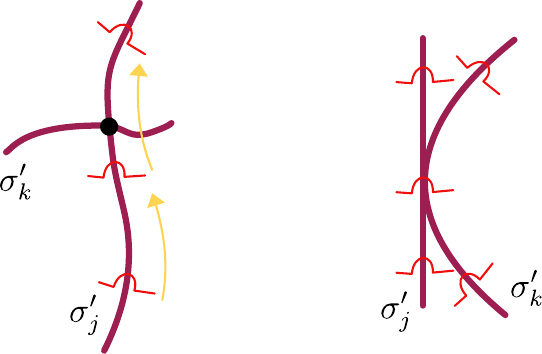}}
  \caption{Rules for bridges indicating indentation of a surface}
  \label{fig:01030d2}
\end{figure}

In \cite{Assier2022}, we formulate some rules of usage of the bridge symbol for surfaces:

\begin{itemize}

\item
The bridge symbol 
can be rotated continuously about $z^*$ unless the ``stems'' become tangent to $\sigma_j'$
at~$z^*$.


\item
The bridge symbol can be carried continuously along a component $\sigma_j'$.
Thus, it is enough to define the bridge symbol at a single point of $\sigma_j'$,
then one can carry the bridge  along $\sigma_j'$ (see \figurename~\ref{fig:01030d2}, left).
Note that, in particular, the bridge symbol can be carried through a crossing point 
of two singularity components. 
 
\item
Let $z^*$ be a point of {\em transversal\/} crossing of two singularity components $\sigma_j'$ and $\sigma_k'$. 
The bridge symbols corresponding to these components can be chosen independently. 

\item
Let $z^*$ be a point of {\em tangential\/} crossing of two singularity components $\sigma_j'$ and $\sigma_k'$. 
The bridge symbols corresponding to these components should match (see \figurename~\ref{fig:01030d2}, right). 

\end{itemize}

%
\FloatBarrier

\section{Proof of the theorem of inflation} 
\label{app:C}

\subsection{A formal proof}

We outline here the proof of Theorem~\ref{th:inflattability} without going too deeply into the details. 

\begin{figure}[h]
  \centering{\includegraphics[width=0.6\textwidth]{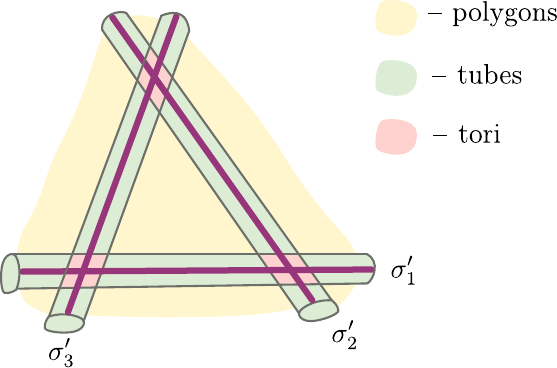}}
  \caption{Different parts of the retract of $H_2(\NR^2 \setminus \sigma)$}
  \label{fig:inf_pr_1}
\end{figure}

\vskip 6pt
\noindent 
{\em Step 1.}
The space $\NR^2 \setminus \sigma$ (assuming transversal crossings and no triple crossings) has a strict deformation retract 
$\check U$
that is an assembly of polygons, thin tubes (built around the singularity components $\sigma_j$), 
and tori (built around the crossings of the singularity components). A sketch of such a retract is shown in \figurename~\ref{fig:inf_pr_1}.

\begin{figure}[h]
  \centering{\includegraphics[width=0.8\textwidth]{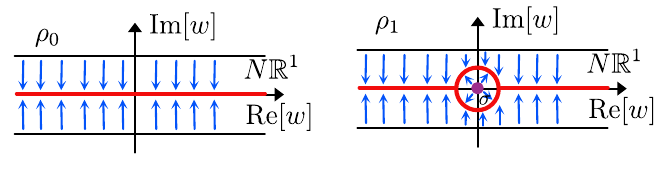}
  }
  \caption{Elementary deformation retractions of $\NR^1$}
  \label{fig:inf_pr_2}
\end{figure}

The structure of this retraction is as follows. Define the retractions $\rho_0$ and $\rho_1$ of $\NR^1$ 
and $\NR^1 \setminus \{ 0\}$ as shown in 
\figurename~\ref{fig:inf_pr_2}. The retract is shown by red curves in both cases.
In a polygon zone (see \figurename~\ref{fig:inf_pr_1}), one can take $z_1$ and $z_2$ as $w$, and build (locally) the retraction as $\rho_0 \times \rho_0$
in these variables. In a tube zone, one takes locally $w_1$ as the variable tangential to the singularity, 
and $w_2$ as the transversal one. Then the retraction is $\rho_0 \times \rho_1$. The product of the line and the circle makes a cylinder. Finally, in a torus zone one takes $w_1$ and $w_2$ as transversal variables to the singularity components, and the retraction is $\rho_1 \times \rho_1$. The product of two circles is a torus. Gluing together the local retractions does not pose any problems in~$\NR^2$.

Denote the retract obtained as the result of this procedure by~$\check U$.

\vskip 6pt
\noindent
{\em Step 2.}
Equip $\check U$ with the structure of the universal Riemann domain, i.e.\ consider the set of points $(p, \gamma)$,
where $p \in \check U$, and $\gamma$ is a path from some reference point $p^*$ to $p$ within~$\check U$.
Denote the set of such pairs by~$\breve U$.
By construction, $\breve U$ is a deformation retract of~$\tilde U_2$.  

\begin{figure}[h]
  \centering{\includegraphics[width=0.4\textwidth]{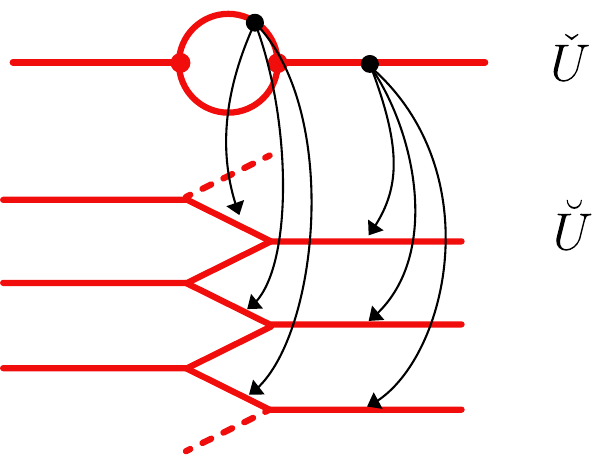}}
  \caption{Transition from $\check U$ to $\breve U$ for the retract shown in \figurename~\ref{fig:inf_pr_2}, right}
  \label{fig:inf_pr_3}
\end{figure}

Let us reveal the local structure of~$\breve U$, i.e.\ study how the procedure of equipping a point with a path-index affects the elements of~$\check U$. Study the 1D factors. A single line (\figurename~\ref{fig:inf_pr_2}, left) 
is transformed in a trivial way, and two lines and a circle (\figurename~\ref{fig:inf_pr_2}, right) are transformed as shown in \figurename~\ref{fig:inf_pr_3}. 

Thus, locally, a neighborhood of a tube is converted into a product of a line and a construction akin to \figurename~\ref{fig:inf_pr_3}, and a vicinity of a torus is converted into a product of two such constructions. 
Again, local structures can be glued together in a simple way. 

\begin{figure}[h]
  \centering{\includegraphics[width=0.6\textwidth]{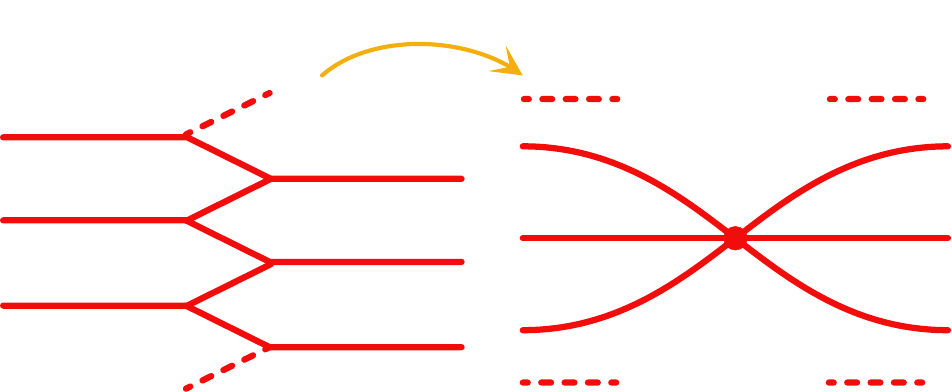}}
  \caption{Contraction of ``ladders''}
  \label{fig:inf_pr_4}
\end{figure}

\vskip 6pt
\noindent
{\em Step 3.}
The construction  of \figurename~\ref{fig:inf_pr_3} consists of ``floors'' and ``ladders''. 
A contraction of all ``ladders'' to a point produces a homotopically equivalent structure 
(see \figurename~\ref{fig:inf_pr_4}). Indeed, the same is valid for products of such structures. 

Finally, we contract all ladders in all factors, and get a topological space homotopically equivalent to $\breve U$. 
Note that this space is the retract~$\cR(\tilde U)$. Thus, we have proven that the retract of $\tilde U_2$
is homotopically equivalent to the retract of~$\tilde U$. Theorem~\ref{th:inflattability} follows directly from this fact.


\subsection{Construction of $\cE$} 

The proof above is rather formal. Here 
we would like to construct explicitly the inflation operator $\cE$, since it may be useful  for the computation of 
intersection indices and for building the integration surfaces. The operator $\cE$ is based on 
\figurename~\ref{fig:inf_pr_1}: an element of $\tilde U$ is first retracted to $\tilde U^*$, then the parts in the polygon zones
are left at their places, and the parts falling in  tube and tori zones are inflated to coverings of tubes and tori (or their parts).
Below we explain how this is done. 

\vskip 6pt
\noindent
{\em Inflating the edge parts.}
Let $w \in H_2 (\tilde U^* , \tpB)$ be the element for which we are building $\cE (w) \in H_2 (\tilde U_2 , \tpB)$. 
Represent $w$ as the sum 
\begin{equation}
w = \sum_m \sum_{j = 1}^k \sum_{\gamma \in \PI} \alpha_{j, \gamma, m} Q^j_{\gamma}
\label{e:inf_pr_1}
\end{equation}
with
$\alpha_{j, \gamma, m} = \pm 1$. The index $m$ is added to avoid integer coefficients $\alpha$ not equal to $\pm 1$ (i.e.\ all terms in (\ref{e:inf_pr_1}) are separate polygons). Each term of the sum belongs to $H_2 (\tilde U , \tilde U' \cup \tpB)$. 

Apply the ``boundary'' homomorphism operator $\ptl$ to (\ref{e:inf_pr_1}). 
On the one hand, on the right, the result is a sum of sides of polygons, all belonging to $H_1 (\tilde U_1, \tilde U_0 \cup \tpB)$.
On the other hand, on the left, the result is 0.  
Let the sides of the polygons cancel each other by pairs. 
Take a pair of such polygons (let them be $\alpha_1 Q^1_{\gamma_1}$
and $\alpha_2 Q^2_{\gamma_2}$) having a common side $a_\gamma \in H_1 (\tilde U_1, \tilde U_0 \cup \tpB)$, 
one with the plus sign and another with the minus sign.  
If $Q^1$ is the same polygon as $Q^2$, the signs $\alpha_1$ and $\alpha_2$ should be opposite, and if $Q^1$ and $Q^2$ are neighboring polygons, then the signs $\alpha_1$ and $\alpha_2$ should be the same.  

\begin{figure}[h]
  \centering{\includegraphics[width=0.8\textwidth]{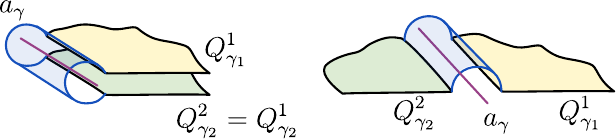}}
  \caption{Inflating the edge parts}
  \label{fig:inf_pr_5}
\end{figure}

According to Lemma~\ref{le:locality}, one can connect the polygons $\alpha_1 Q^1_{\gamma_1}$
and $\alpha_2 Q^2_{\gamma_2}$ by a surface belonging to the cylindrical part of the retract. Such a connection 
for the cases of similar and different $Q^1$ and $Q^2$ is shown in \figurename~\ref{fig:inf_pr_5}. We take this connection (shown in blue)
as the cylindrical part of the inflated homology.
The angles covered by the parts of the cylinder are shown to be $2 \pi$ and $\pi$, but they depend on the indices $\gamma_1, \gamma_2$ and may be different. 

\vskip 6pt
\noindent
{\em Inflating the vertex parts.}
Consider again the sum (\ref{e:inf_pr_1})
and the splitting of the faces into pairs with respect to the boundary elements that has been used above. 
Take some point belonging to $\tilde U_0$ (a vertex) 
and build a sequence of faces (the terms of (\ref{e:inf_pr_1})) 
and edges (fragments of the boundaries of the terms of (\ref{e:inf_pr_1}), i.e.\ elements of $H_1 (\tilde U_1 , \tilde U_0 \cup \tpB)$) 
incident to this vertex.
For this, take some face, say $Q^1_{\gamma_1}$, incident to the selected vertex. There are two edges of this 
face, say $a_{\gamma_2}$ and $b_{\gamma_3}$, meeting at the vertex. Take one of these edges, say $b_{\gamma_3}$. 
Find another face forming a pair with $Q^1_{\gamma_1}$ with respect to 
$b_{\gamma_3}$, let this face be $Q^2_{\gamma_4}$. Find the second edge of $Q^2_{\gamma_4}$ incident to the selected vertex (let it be $c_{\gamma_5}$), etc. 
Thus, get the sequence 
\[
Q^1_{\gamma_1} \to b_{\gamma_3} \to Q^2_{\gamma_4} \to c_{\gamma_5} \dots
\]      
After a finite number of steps, this sequence should come back to~$Q^1_{\gamma_1}$. As a result, we obtain a periodic sequence. 

Starting from another face incident to the same vertex, we get another periodic sequence, and after some iterations, we split all faces and edges incident to the selected vertex into periodic sequences.

\begin{figure}[h]
  \centering{\includegraphics[width=0.9\textwidth]{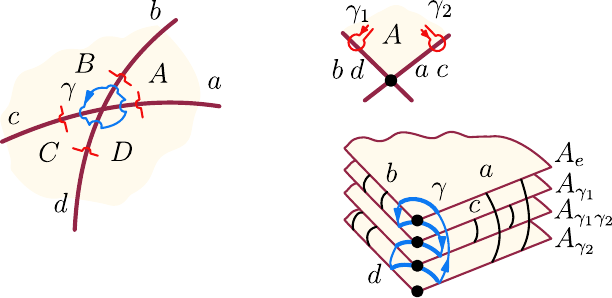}}
  \caption{A periodic sequence of faces and edges at a vertex}
  \label{f:01046a}
\end{figure}

Two examples of such a periodic sequence are shown in \figurename~\ref{f:01046a}. 
In the left part of the figure, $w$ is composed of four different sectors, and in the right 
part, there are four copies of one sector, taken for different paths.  
In both cases, the sequence consists of four faces and four edges. 
For the left part, the sequence is 
\[
A_e \to b \to B_e \to c \to C_e \to d \to D_e \to a \to A_e.
\]
Note that we do not write down the path-indices for the edges since they can be found from the 
path-indices of the faces to which they are incident. 

For the right part, the sequence is 
\[
A_e \to b \to -A_{\gamma_1} \to c \to A_{\gamma_1 \gamma_2} \to d \to -A_{\gamma_2} \to a \to A_e.
\]

\begin{figure}[h]
  \centering{\includegraphics[width=0.8\textwidth]{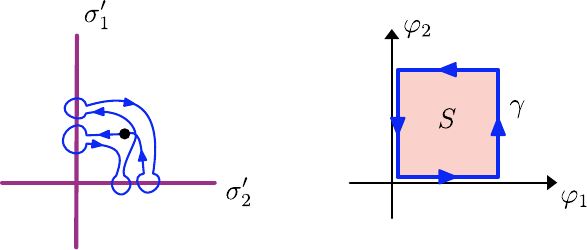}}
  \caption{Contour $\gamma \in H_2(\tilde U_2)$ drawn near a vertex. {Left part is $\gamma$ in the coordinates 
  $(\nu_1, \nu_2)$; right part is $\gamma$ in the coordinates $(\vph_1, \vph_2)$}}
  \label{f:01048}
\end{figure}

For each periodic sequence one can build a contour $\gamma$ near the vertex. Examples of
this contour are drawn in \figurename~\ref{f:01046a} in blue. Assume that this contour lies 
in a small ball $D$ with the center located at the vertex. 
The contour $\gamma$ can be drawn in the bridge/loop style, as above
(see \figurename~\ref{f:01046a}, left, and \figurename~\ref{f:01048}, left).
The contour $\gamma$ can be considered as a boundary of a 
vertex (torus) part of $\cE (w)$. 
 Let us build this part.

Let the vertex be a crossing of 
the singularities $\sigma_1$ and~$\sigma_2$
having defining functions $g_1$ and $g_2$. 
Introduce the local variables 
\[
\nu_1 = g_1(z), \qquad \nu_2 = g_2(z)
\]
and the angles 
\[
\vph_1 = {\rm Arg}[\nu_1], 
\qquad 
\vph_2 = {\rm Arg}[\nu_2]. 
\]
Draw the contour $\gamma$ in the coordinates $(\vph_1 , \vph_2)$ 
(\figurename~\ref{f:01048}, right). 
The result is always a closed polygon since $\gamma$ is closed in $\tilde U_2$.


 
Fill the polygon bounded by $\gamma$, i.e.\ find a chain $S$
in the plane $(\vph_1, \vph_2)$
such that $\ptl S = \gamma$ (here $\ptl$ is just a boundary operator). Indeed, it is possible since the $H_1$ group of the plane is trivial.  
An example of the chain $S$ is shown in \figurename~\ref{f:01048}, right.

The corresponding vertex part of $\cE (w)$ is given by 
\[
\nu_1 = \rho_1(\vph_1, \vph_2) e^{i \vph_1},
\qquad 
\nu_2 = \rho_2(\vph_1, \vph_2) e^{i \vph_2}, 
\]
where $(\vph_1, \vph_2) \in S$. The functions $\rho_1$, $\rho_2$
are real, positive  and continuous, taken in such a way that the boundary of the vertex part can be attached to the cylindrical edge parts constructed above. 

Thus, we described $\cE$ as an assembly of face, edge, and vertex parts that can attached to each other.


\section{Computation of matrices for elementary transformations}
\label{app:D}
\subsection{Proof of Theorem~\ref{the:PLP_triangle}}

We should prove that
\begin{equation}
A_e  \stackrel{\lambda}{\longrightarrow}
A_{\gamma_1 \gamma_2 \gamma_3},
\label{e:01304}
\end{equation}  
\begin{equation}
B_{e } \stackrel{\lambda}{\longrightarrow}
B_{e } 
+
A_{\gamma_2}
-
A_{\gamma_1 \gamma_2 \gamma_3}
,
\label{e:01305}
\end{equation} 
\begin{equation}
C_{e } \stackrel{\lambda}{\longrightarrow}
C_{e } 
+
A_{\gamma_1}
-
A_{\gamma_1 \gamma_2 \gamma_3}
,
\label{e:01306}
\end{equation} 
\begin{equation}
D_{e } \stackrel{\lambda}{\longrightarrow}
D_{e } 
+
A_{\gamma_3 }
-
A_{\gamma_1 \gamma_2 \gamma_3}
,
\label{e:01307a}
\end{equation} 
\begin{equation}
E_{e } \stackrel{\lambda}{\longrightarrow}
E_{e } 
-
A_{\gamma_2 \gamma_3}
+
A_{\gamma_1 \gamma_2 \gamma_3}
,
\label{e:01307}
\end{equation} 
\begin{equation}
F_{e } \stackrel{\lambda}{\longrightarrow}
F_{e } 
-
A_{ \gamma_1 \gamma_2}
+
A_{ \gamma_1 \gamma_2 \gamma_3}
,
\label{e:01308}
\end{equation} 
\begin{equation}
G_{e } \stackrel{\lambda}{\longrightarrow}
G_{e } 
-
A_{ \gamma_1 \gamma_3}
+
A_{ \gamma_1 \gamma_2 \gamma_3}
.
\label{e:01309}
\end{equation} 

As an example, we focus on (\ref{e:01305}), all other formulas are proven in the same way. 
We make three steps.

The computation \eqref{e:01304} is the simplest. Indeed, under monodromy, the projection of this cycle into the base will go into itself, and it is enough to calculate in which sheet the central point $z_A$ of $A$ will be. It is easy to understand that  under the action of the bypass $\lambda$, the point $z_A$ will go along the loop $\gamma_1\gamma_3$, but the bypass $\lambda$ will also act on the bridge via $\sigma_2$ (and on the corresponding path), mapping the sheet  $e$ to the sheet  $\gamma_2$. Thus, the point $z_A$ will be in the sheet $\gamma_1\gamma_2\gamma_3$.

\vskip 6pt
\noindent
{\em Step 1.} 
Let us demonstrate that 
\begin{equation}
B_e \stackrel{\lambda}{\longrightarrow} B_e + \alpha_1 A_{\gamma'} + \alpha_2 A_{\gamma''},
\label{e:shablon}
\end{equation}
where the values $\alpha_1, \alpha_2 = \pm 1$, $\gamma', \gamma'' \in \PI$
are to be determined. In other words, we are proving that ${\rm var}_\lambda (B_e)$ consists of two samples of the triangle $A$ taken on different sheets of $\tilde U$
and possibly oriented in a different way. 

Denote the right-hand side of (\ref{e:shablon}) by $\psi_\lambda(B_e)$.
Note that $\psi_\lambda (B_e)$ (more rigorously, a representative of this class) can be built explicitly. 
Let $\lambda$ be a loop as shown in  \figurename~\ref{f:01061a}. 
In the $t$-plane, the Landau set is defined by $t = t^*$.
Let this loop be parametrized by some real variable $\tau$, i.e.\ 
\[
\lambda: \qquad t = t(\tau), \qquad  \tau \in [0,1].
\]
For each $\tau$ find the crossing points 
\[
M_1(\tau) = \sigma_1 \cap \sigma_2 (t(\tau)),
\qquad 
M_3(\tau) = \sigma_3 \cap \sigma_2 (t(\tau)).
\]
Build a straight segment $[M_1(\tau) \, M_3(\tau)]$ belonging to the complex line $\sigma_2(t(\tau))$. 

As $t$ runs along $\lambda$, a family of such segments is formed. Taking the circular arc in \figurename~\ref{f:01065} vanishingly small (this corresponds to the retraction $\cR$), we find that this family can be deformed into two samples of the triangle $A$ on different sheets. On the other hand, one can see by construction that this family (taken with an appropriate orientation) is ${\rm var}_\lambda (B_e)$. 

\begin{figure}[h]
  \centering{\includegraphics[width=0.4\textwidth]{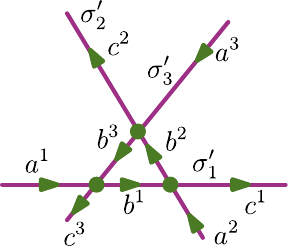}}
  \caption{Boundaries of elements of $H_2(\tilde U, \tilde U')$}
  \label{f:1D_boundary}
\end{figure}

\vskip 6pt
\noindent
{\em Step 2.}
Consider the lines $\sigma'_1$, $\sigma'_2$, $\sigma'_3$. They are split by crossing points into segments and half-lines $a^j$, $b^j$, $c^j$, $j = 1,2,3$ as it is shown in \figurename~\ref{f:1D_boundary}. These elements, oriented as it is shown in the figure, form a basis of $H_1(\sigma , \sigma^{(0)} \cup \tpB)$.  

These elements can be taken with path-indices: e.g. $b^2_\gamma$, where 
$\gamma$ are left cosets in $\PI$ in accordance with Lemma~\ref{le:locality}.
The cosets are labelled by their representatives in~$\PI$. 
Such path-indexed segments form a basis of $H_1(\Tilde U_1 , \tilde U_0 \cup \tpB)$. 
Note that the elements of $H_1(\Tilde U_1 , \tilde U_0 \cup \tpB)$ are parts of boundaries of elements of 
$H_2(\tilde U , \tilde U' \cup \tpB)$.

We will need the following result: 

\begin{proposition}
The elements $a^j_e$, $b^j_e$, $c^j_e$ are transformed under the bypass $\lambda$ (shown in \figurename~\ref{f:01061a}) as follows: 
\begin{equation}
a^1_e 
\stackrel{\lambda}{\longrightarrow}
a^1_e + b^1_{\gamma_3} - b^1_{\gamma_2 \gamma_3},
\qquad 
b^1_e 
\stackrel{\lambda}{\longrightarrow}
b^1_{\gamma_2 \gamma_3},
\qquad 
c^1_e 
\stackrel{\lambda}{\longrightarrow}
c^1_e + b^1_{\gamma_2} - b^1_{\gamma_2 \gamma_3},
\label{e:transformation_boundary_1}
\end{equation}
\begin{equation}
a^2_e 
\stackrel{\lambda}{\longrightarrow}
a^2_e + b^2_{\gamma_1} - b^2_{\gamma_1 \gamma_3},
\qquad 
b^2_e 
\stackrel{\lambda}{\longrightarrow}
b^2_{\gamma_1 \gamma_3},
\qquad 
c^2_e 
\stackrel{\lambda}{\longrightarrow}
c^2_e + b^2_{\gamma_3} - b^2_{\gamma_2 \gamma_3},
\label{e:transformation_boundary_2}
\end{equation}
\begin{equation}
a^3_e 
\stackrel{\lambda}{\longrightarrow}
a^3_e + b^3_{\gamma_2} - b^3_{\gamma_1 \gamma_2},
\qquad 
b^3_e 
\stackrel{\lambda}{\longrightarrow}
b^3_{\gamma_1 \gamma_2},
\qquad 
c^3_e 
\stackrel{\lambda}{\longrightarrow}
c^3_e + b^3_{\gamma_1} - b^3_{\gamma_1 \gamma_2}.
\label{e:transformation_boundary_3}
\end{equation}
\end{proposition}

\begin{figure}[h]
  \centering{\includegraphics[width=0.8\textwidth]{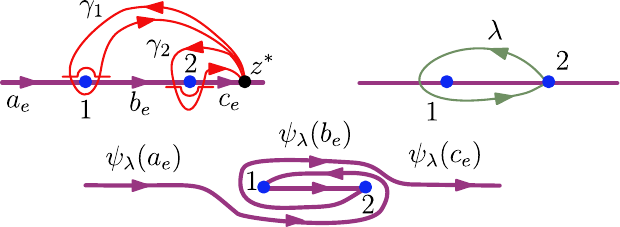}}
  \caption{Boundaries of elements of $H_2(\tilde U, \tilde U' \cup \tpB)$}
  \label{f:1D_ramification}
\end{figure}

To prove this proposition, consider a one-dimensional problem. 
Say, take $\sigma_1$ that is a complex plane~$\mathbb{C}$. There are two singular points in this plane, 1 and 2, shown in \figurename~\ref{f:1D_ramification}, that correspond to $\sigma_1 \cap \sigma_3$ and $\sigma_2\cap\sigma_3$, respectively.
Introduce the universal Riemann domain $\tilde U^{1D}$ in $\NR^1$ with respect to these singularities.

Introduce for the 1D problem the reference point $z^*$, the generators of $\PI$ named $\gamma_1$, $\gamma_2$, and base paths described by bridge notations.  
The bridges in \figurename~\ref{f:1D_ramification} are chosen to match the bridges in \figurename~\ref{f:01061a1}. 

The bypass $\lambda$ for the 2D problem induces a  corresponding bypass for the 1D problem (see \figurename~\ref{f:1D_ramification}). The ramification of the bypasses $a_e, b_e, c_e$ can be computed visually: 
\begin{equation}
a_e \stackrel{\lambda}{\longrightarrow} 
a_{\gamma_1 \gamma_2^{-1} \gamma_1^{-1} \gamma_2} + b_{\gamma_1} - b_{\gamma_1 \gamma_2},
\qquad 
b_e \stackrel{\lambda}{\longrightarrow} 
b_{\gamma_1 \gamma_2},
\qquad 
c_e \stackrel{\lambda}{\longrightarrow} 
c_e  + b_{\gamma_2} - b_{\gamma_1 \gamma_2}.
\label{e:1D_full_ram}
\end{equation}

The formulas (\ref{e:transformation_boundary_1}), (\ref{e:transformation_boundary_2}),
(\ref{e:transformation_boundary_3})  can be obtained from (\ref{e:1D_full_ram}) by substitution of appropriate indices and abelization following from Lemma~\ref{le:commutativity}.

\vskip 6pt
\noindent
{\em Step 3.}
Now we are ready to compute the coefficients in (\ref{e:shablon}) and thus finalize the proof. For this, we consider the transformation of the boundary of $B_e$ as the result of the bypass~$\lambda$. 

According to \figurename~\ref{f:1D_boundary},
\[
\ptl B_e = a^3_e - b^2_e + c^1_e.
\]
By continuity, $\ptl$ and $\lambda$ commute. 
Thus, due to (\ref{e:transformation_boundary_1}), (\ref{e:transformation_boundary_2}),
(\ref{e:transformation_boundary_3}),
\[
\ptl (\psi_{\lambda} (B_e))  = 
\psi_{\lambda} (\ptl B_e) =
a^3_e + b^3_{\gamma_2}  - b^3_{\gamma_1 \gamma_2} - b^2_{\gamma_1 \gamma_3}  - b^1_{\gamma_2 \gamma_3} + b^1_{\gamma_2} + c^1_{e}
\]
and 
\[
\ptl\, {\rm var}_{\lambda}(B_e)  =  b^1_{\gamma_2} +b^2_e+ b^3_{\gamma_2}  - b^1_{\gamma_2 \gamma_3}  - b^2_{\gamma_1 \gamma_3}  - b^3_{\gamma_1 \gamma_2}.
\]
It is easy to prove that the only combination of the form (\ref{e:shablon}) is (\ref{e:01305}). To see this, note that by Lemma~\ref{le:locality},
\[
\ptl A_{\gamma_1^m \gamma_2^n \gamma_3^k} = 
b^1_{\gamma_2^n \gamma_3^k} + b^2_{\gamma_1^m \gamma_3^k}
+ b^3_{\gamma_1^m \gamma_2^n},
\]
and, in particular,
\[
\ptl A_{\gamma_2} = b^1_{\gamma_2} + b^2_e + b^3_{\gamma_2},
\qquad
\ptl A_{\gamma_1 \gamma_2 \gamma_3} = b^1_{\gamma_2 \gamma_3} + b^2_{\gamma_1 \gamma_3} + b^3_{\gamma_1 \gamma_2}.
\]
Moreover, $m,n,k$ may take only values 0 and 1, and all other combinations can be easily analyzed.

\subsection{Proof of Theorem~\ref{the:PLP_biangle}}

We should prove that 
\begin{equation}
A_e \stackrel{\lambda}{\longrightarrow}
- A_{ \gamma_1 \gamma_2},
\label{e:01310}
\end{equation}
\begin{equation}
B_{e } 
\stackrel{\lambda}{\longrightarrow}
B_{e }
+ 
A_{  \gamma_1 }
+
A_{  \gamma_1 \gamma_2},
\label{e:01311}
\end{equation}
\begin{equation}
C_{e } 
\stackrel{\lambda}{\longrightarrow}
C_{e }
+ 
A_{ \gamma_2 }
+
A_{ \gamma_1 \gamma_2},
\label{e:01312}
\end{equation}
\begin{equation}
D_{e } 
\stackrel{\lambda}{\longrightarrow}
D_{e }
-
A_{ \gamma_1 \gamma_2},
\label{e:01313}
\end{equation}
\begin{equation}
E_{e } 
\stackrel{\lambda}{\longrightarrow}
E_{e }
-
A_{ \gamma_1 \gamma_2}.
\label{e:01314}
\end{equation}

The proof basically repeats that of Theorem~~\ref{the:PLP_triangle}.
We focus on some particular line, say (\ref{e:01311})
and perform the same three steps. 
On the first step, we build ${\rm var}_{\lambda} (B_e)$ explicitly and find that
\begin{equation}
B_e \stackrel{\lambda}{\longrightarrow} B_e + \alpha_1 A_{\gamma'} + \alpha_2 A_{\gamma''},
\qquad 
\alpha_1 , \alpha_2 = \pm 1,
\label{e:shablon_2}
\end{equation}
i.e.\ that ${\rm var}_{\lambda} (B_e)$ is comprised of two samples of the biangle~$A$.

\begin{figure}[h]
  \centering{ \includegraphics[width=0.2\textwidth]{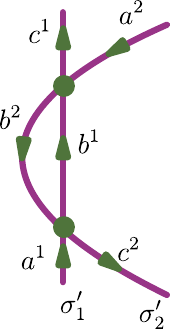}  }
  \caption{Notations for pieces of the boundary of the biangle}
  \label{f:1D_boundary_2}
\end{figure}

On the second step, we study the transformation of the boundaries of the elements of $H_2(\tilde U, \tilde U' \cup \tpB)$. 
Namely, we introduce the notations for segments and half-lines in which $\sigma'$ is split by crossing points 
(see \figurename~\ref{f:1D_boundary_2}).
The fragments of the boundary are transformed as follows: 
\[
a^1_e \stackrel{\lambda}{\longrightarrow} a^1_e + b^1_{\gamma_2} , 
\qquad 
b^1_e    \stackrel{\lambda}{\longrightarrow} - b^1_{\gamma_2},
\qquad 
c^1_e \stackrel{\lambda}{\longrightarrow} c^1_e + b^1_{\gamma_2} , 
\]
\[
a^2_e \stackrel{\lambda}{\longrightarrow} a^2_e + b^2_{\gamma_1} , 
\qquad 
b^2_e    \stackrel{\lambda}{\longrightarrow} - b^2_{\gamma_1},
\qquad 
c^2_e \stackrel{\lambda}{\longrightarrow} c^2_e + b^2_{\gamma_1} .
\]

On the third step, we compute 
\begin{equation}
\ptl(\psi_{\lambda}(B_e))  = 
\psi_{\lambda}(\ptl B_e) = \psi_{\lambda} (a^2_e - b^1_e + c^2_e)  = 
a^2_e + 2 b^2_{\gamma_1} + b^1_{\gamma_2} + c^2_e.
\label{e:bound_biang}
\end{equation}
The only expression of the form (\ref{e:shablon_2}) having the boundary 
(\ref{e:bound_biang}) is (\ref{e:01311}). 

\begin{figure}[h]
  \centering{ \includegraphics[width=0.8\textwidth]{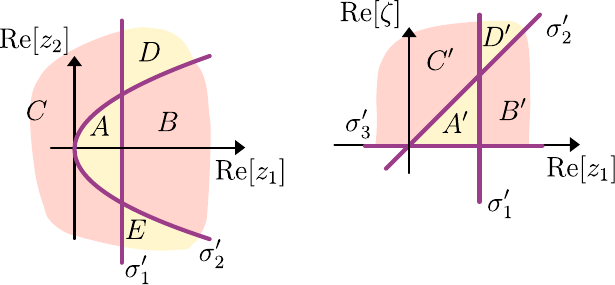}  }
  \caption{``Folding'' of the biangle along the $z_1$-axis}
  \label{f:biangle_folding}
\end{figure}

Let us give an
alternative proof to Theorem~\ref{the:PLP_biangle}. Namely, it can be derived from Theorem~\ref{the:PLP_triangle} using the 
idea proposed in \cite{Pham1965}. Here we put this reasoning in brief. 
Consider the singularities (\ref{e:biangle_sing}). Make the change of variables
\begin{equation}
(z_1 , z_2) \to (z_1 , \zeta) , \qquad \zeta = z_2^2.
\label{e:pham_change}
\end{equation}
The real plane $(z_1, z_2)$ becomes ``folded'' along the axis~$z_1$ as it is shown in \figurename~\ref{f:biangle_folding}.

There are three singularities in the $(z_1, \zeta)$ plane:
\[
\sigma_1: \, \, z_1 = t, \qquad \sigma_2: \, \, z_1 -\zeta =0, \qquad \zeta_3: \, \, \zeta = 0. 
\]
The branch line $\sigma_3$ has order~2. All lines are straight. The bypass $\lambda$ corresponds to the case covered by Theorem~\ref{the:PLP_triangle}.

Introduce polygons $A'$, $B'$, $C'$, $D'$ as it is shown in \figurename~\ref{f:biangle_folding}.
Select a reference point in the polygon~$B'$.
Introduce the bypasses $\gamma_1$, $\gamma_2$, $\gamma_3$ about the singularities $\sigma_1$, $\sigma_2$, $\sigma_3$
in the $(z_1, \zeta)$-plane. There exists a correspondence between the polygons in the $(z_1, z_2)$ and $(z_1, \zeta)$-planes:
\[
A_\gamma \leftrightarrow A'_{\gamma} - A'_{\gamma \gamma_3},
\qquad
B_\gamma \leftrightarrow B'_{\gamma} - B'_{\gamma \gamma_3}, 
\qquad
D_\gamma \leftrightarrow D'_\gamma,
\qquad
E_\gamma \leftrightarrow -D'_{\gamma \gamma_3},
\]
where $\gamma = \gamma_1^m \gamma_2^n$, $m,n \in \mathbb{Z}$.
The polygons $A'$, $B'$, $C'$, $D'$ are transformed according to (\ref{e:01304})--(\ref{e:01307a}): 
\[
A'_e \stackrel{\lambda}{\longrightarrow} A'_{\gamma_1 \gamma_2 \gamma_3}, 
\]
\[
B_e' \stackrel{\lambda}{\longrightarrow} B_e' + A'_{\gamma_1}- A'_{\gamma_1 \gamma_2 \gamma_3}, 
\]
\[
C_e' \stackrel{\lambda}{\longrightarrow} C_e'  +A'_{\gamma_2}-  A'_{\gamma_1 \gamma_2 \gamma_3}, 
\]
\[
D_e' \stackrel{\lambda}{\longrightarrow} D_e'  - A'_{\gamma_1 \gamma_2}+A'_{\gamma_1 \gamma_2 \gamma_3}, 
\]

Then, Lemma~\ref{le:linear_representation} can be applied in the $(z_1, \zeta)$-plane, but the result should be simplified by using the 
relation $\gamma_3^2 = e$: 
\[
A'_{\gamma_3} \stackrel{\lambda}{\longrightarrow} A'_{\gamma_1 \gamma_2}, 
\]
\[
B'_{\gamma_3} \stackrel{\lambda}{\longrightarrow} B_{\gamma_3}' + A'_{\gamma_1 \gamma_3}- A'_{\gamma_1 \gamma_2}, 
\]
\[
C'_{\gamma_3} \stackrel{\lambda}{\longrightarrow} C'_{\gamma_3}  +A'_{\gamma_2 \gamma_3}-  A'_{\gamma_1 \gamma_2}, 
\]
\[
D'_{\gamma_3} \stackrel{\lambda}{\longrightarrow} D'_{\gamma_3}  - A'_{\gamma_1 \gamma_2 \gamma_3}+A'_{\gamma_1 \gamma_2}, 
\]

Finally, combine these relations: 
\[
A_e \leftrightarrow A'_e - A'_{\gamma_3} 
\quad \stackrel{\lambda}{\longrightarrow} \quad 
A'_{\gamma_1 \gamma_2 \gamma_3} - A'_{\gamma_1 \gamma_2} \leftrightarrow - A_{\gamma_1 \gamma_2},
\]

\[
B_e \leftrightarrow B'_e - B'_{\gamma_3} 
\quad \stackrel{\lambda}{\longrightarrow} \quad 
\qquad \qquad \qquad \qquad \qquad \qquad  \qquad \qquad \qquad 
\]
\[
\qquad \qquad \qquad 
B'_{e} + A'_{\gamma_1} - A'_{\gamma_1 \gamma_2 \gamma_3} 
-
B'_{\gamma_3}  - A'_{\gamma_1 \gamma_3} + A'_{\gamma_1 \gamma_2} 
\leftrightarrow
B_e + A_{\gamma_1} + A_{\gamma_1 \gamma_2},
\]

\[
C_e \leftrightarrow C'_e - C'_{\gamma_3} 
\quad \stackrel{\lambda}{\longrightarrow} \quad 
\qquad \qquad \qquad \qquad \qquad \qquad  \qquad \qquad \qquad 
\]
\[
\qquad \qquad \qquad 
C'_{e} + A'_{\gamma_2} - A'_{\gamma_1 \gamma_2 \gamma_3} 
-
C'_{\gamma_3}  - A'_{\gamma_2 \gamma_3} + A'_{\gamma_1 \gamma_2} 
\leftrightarrow
C_e + A_{\gamma_2} + A_{\gamma_1 \gamma_2},
\]

\[
D_e \leftrightarrow D'_e  
\quad \stackrel{\lambda}{\longrightarrow} \quad 
D'_e
-A'_{\gamma_1 \gamma_2} + A'_{\gamma_1 \gamma_2 \gamma_3} \leftrightarrow D_e- A_{\gamma_1 \gamma_2},
\]

\[
E_e \leftrightarrow -D'_{\gamma_3}  
\quad \stackrel{\lambda}{\longrightarrow} \quad 
-D'_{\gamma_3} 
-A'_{\gamma_1 \gamma_2} + A'_{\gamma_1 \gamma_2 \gamma_3} \leftrightarrow E_e- A_{\gamma_1 \gamma_2}.
\]

So, as expected, the relations (\ref{e:01310})--(\ref{e:01314}) are recovered. 

\subsection{Proof of Theorem~\ref{the:PLP_circle}}

\begin{figure}[h]
  \centering{ \includegraphics[width=0.8\textwidth]{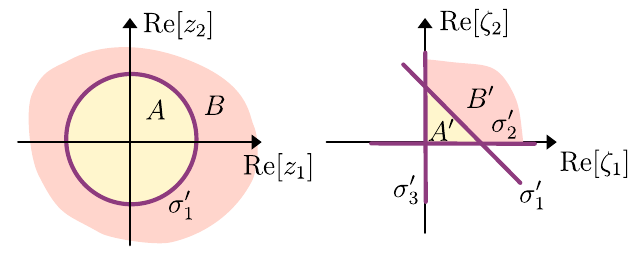}  }
  \caption{``Folding'' of the circle. {Left: the circle in the coordinates $(z_1, z_2)$. Right: 
  the image of the circle in the coordinates $(\zeta_1, \zeta_2)$}}
  \label{f:circle_folding}
\end{figure}

In this appendix we will show how Theorem~\ref{the:PLP_circle} can be reduced to Theorem~\ref{the:PLP_triangle}, in the same spirit of the alternative proof given to Theorem~\ref{the:PLP_biangle} above.
Make the change of variables 
\[
(z_1, z_2) \to (\zeta_1, \zeta_2), \qquad \zeta_1 = z_1^2 , \quad \zeta_2 = z_2^2.
\]
The real plane $(z_1, z_2)$ becomes folded four times in the new variables.
The geometry of the plane $(\zeta_1 , \zeta_2)$ is shown in \figurename~\ref{f:circle_folding}. There are two singularities in this plane: 
\[
\sigma_1: \, \, \zeta_1 + \zeta_2 = t, 
\qquad 
\sigma_2: \, \, \zeta_2 = 0, 
\qquad 
\sigma_3: \, \, \zeta_1 = 0. 
\]
The singularities $\sigma_2$ and $\sigma_3$ are branch lines of order~2.

Denote the polygons $A'$ and $B'$ as shown in \figurename~\ref{f:circle_folding}, right. 
Put the reference point in the polygon~$B'$.
Introduce some simple loops $\gamma_1$, $\gamma_2$, $\gamma_3$ bypassing the singularities $\sigma_1$, $\sigma_2$,
$\sigma_3$, respectively. 

The ``old'' and ``new'' polygons are linked by the following relations: 
\[
A_\gamma \leftrightarrow A'_{\gamma} - A'_{\gamma \gamma_2} - A'_{\gamma \gamma_3} + A'_{\gamma \gamma_2 \gamma_3}, 
\]
\[
B_\gamma \leftrightarrow B'_{\gamma} - B'_{\gamma \gamma_2} - B'_{\gamma \gamma_3} + B'_{\gamma \gamma_2 \gamma_3}, 
\]
The polygons $A'_\gamma$ and $B'_\gamma$ are transformed according to Theorem~\ref{the:PLP_triangle} and Lemma~\ref{le:linear_representation}.
Upon using the additional relations $\gamma_2^2 = e$, $\gamma_3^2 = e$, we obtain
\[
A_e \leftrightarrow 
A'_{e} - A'_{\gamma_2} - A'_{\gamma_3} + A'_{\gamma_2 \gamma_3} 
\quad  \stackrel{\lambda}{\longrightarrow} \quad
A'_{\gamma_1 \gamma_2 \gamma_3} - A'_{\gamma_1  \gamma_3} - A'_{\gamma_1 \gamma_2} + A'_{\gamma_1}  
\leftrightarrow 
A_{\gamma_1},
\]
\[
B_e \leftrightarrow 
B'_{e} - B'_{\gamma_2} - B'_{\gamma_3} + B'_{\gamma_2 \gamma_3} 
\quad  \stackrel{\lambda}{\longrightarrow} \quad 
B'_{e} - B'_{\gamma_2} - B'_{\gamma_3} + B'_{\gamma_2 \gamma_3} 
\leftrightarrow
B_e
\]
This yields (\ref{e:01334}).





\subsection{Proof of Theorem~\ref{the:PLP_jump}}

The proof of this theorem is very similar to that of Theorem~\ref{the:PLP_triangle}, namely, one can construct the surface 
explicitly and then study the boundary to get the proper coefficients. 

The first step of the proof remains the same as for Theorem~\ref{the:PLP_triangle}: one should construct the homologies explicitly, and find that 
\[
A_e \to \alpha_1 A'_{\gamma'_1}
\]
for some $\alpha_1 = \pm 1$, $\gamma'_1 \in \PI$, 
and 
\begin{equation}
Q^n_e \to (Q^n_e)' + \alpha_n A'_{\gamma'_n},
\qquad 
n \ne 1.
\label{e:jump_proof_b}
\end{equation}
Thus it remains only to find the coefficients $\alpha_n$  and path-indices 
$\gamma'_n$.

\begin{figure}[h]
  \centering{ \includegraphics[width=0.4\textwidth]{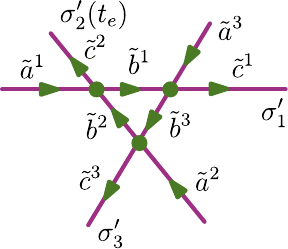}  }
  \caption{Boundaries for \figurename~\ref{f:01061a2a}, right}
  \label{f:jum_proof_1}
\end{figure}

\begin{figure}[h]
  \centering{ \includegraphics[width=0.8\textwidth]{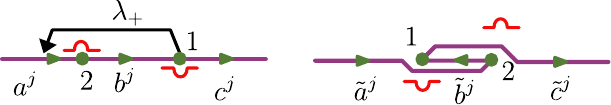}  }
  \caption{Transformation of the boundaries for $\lambda_+$}
  \label{f:jump_proof_2}
\end{figure}

Consider the boundaries of the polygons. Introduce the basis of the boundaries for the configurations shown in \figurename~\ref{f:01061a2a}
as it is shown in \figurename~\ref{f:1D_boundary} and~\ref{f:jum_proof_1}, respectively. Using \figurename~\ref{f:jump_proof_2}, one can 
compute the change of the basis of the boundaries under $\lambda_+$: 
\begin{equation}
a^j_e \stackrel{\lambda_+}{\longrightarrow} \tilde a_e^j + \tilde b_e^j, 
\qquad 
b_e^j \stackrel{\lambda_+}{\longrightarrow} - \tilde b_e^j,
\qquad 
c^j_e \stackrel{\lambda_+}{\longrightarrow} \tilde c_e^j + \tilde b_e^j, 
\qquad 
j = 1,2,3.
\label{e:jump_proof_a}
\end{equation}

Let us find the evolution of $B_e$. 
Note that 
\[
\ptl B_e = a^3_e - b^2_e + c^1_e. 
\]
Then, 
\[
\psi_{\lambda_+} (\ptl B_e) = 
\tilde a^3_e + \tilde b^3_e + \tilde b^2_e + \tilde c^1_e + \tilde b^1_e.
\]
This equation is equal to {$\ptl \psi_{\lambda_+} (B_e)$}. The only combination of the form (\ref{e:jump_proof_b}) having such a boundary is 
\[
\psi_{\lambda_+} (B_e) = B'_e - A'_e, 
\]
i.e.\ the second row of (\ref{e:jump_b1}) is proven. All other rows are proven in the same way.

\section{Computation of $\pi_1 (\cB)$ for a quad\-ra\-tic touch}
\label{app:E}

Consider the Landau set (\ref{e:01353}). Consider the space $\cB = \NR^2 \setminus \cL$ or $\cB = \mathbb{C}^2 \setminus \cL$ (this makes no difference). To compute the fundamental group $\pi_1 (\cB)$, build a strict deformation retract of $\cB$ and study its properties.  

A strict deformation retract of $\cB$ is built in two steps:

\vskip 6pt
\noindent
{\em Step 1. } Represent the variables $t_{1,2}$ in the form
\[
t_{1,2} = \rho_{1,2} e^{i \vph_{1,2}}, 
\]
where $\rho_{1,2}$ are real, and $\vph_{1,2}$ belong to the real circle $\mathcal{S}^1 = [0, 2\pi]$.

Introduce a map 
\begin{equation}
(\rho_1 e^{i \vph_1} , \rho_2 e^{i \vph_2})
\longrightarrow
\left( e^{i \vph_1} , \frac{\rho_2}{\sqrt{\rho_1}}    e^{i \vph_2} \right). 
\label{eq:004}
\end{equation}
This map is a strict deformation retraction both for  $\cB = \NR^2 \setminus \cL$ and for $\cB = \mathbb{C}^2 \setminus \cL$. 

$\cB$ becomes mapped onto a bundle, whose base is $\mathcal{S}^1$
(this corresponds to $t_1 = e^{i \vph_1}$, $\vph_1 \in \mathcal{S}^1$), 
and the fiber is 
a complex plane $\xi$ with two points removed: $\mathbb{C} \setminus \{ e^{i \vph_1 / 2} , -e^{i \vph_1 / 2}  \}$.
The two removed points  belong to $\sigma^t_1$.  

\begin{figure}[h]
  \centering{\includegraphics[width=0.6\textwidth]{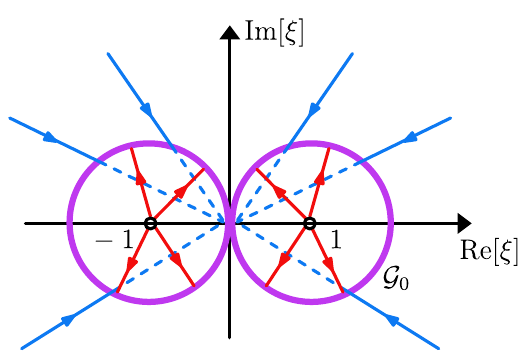}}
  \caption{Projection  $\mathbb{C} \setminus \{1,-1 \} \longrightarrow \mathcal{G}_0$}
\label{fig:eight}
\end{figure}

\vskip 6pt
\noindent
{\em Step 2. } The plane $\xi$ with two  points
removed $\mathbb{C} \setminus \{ e^{i \vph_1 / 2} , -e^{i \vph_1 / 2} \}$
is retracted onto the eight-shaped curve $\mathcal{G}$:
\[
\mathcal{G}_{\vph_1} = \{ \xi \in \mathbb{C} \, : \, |\xi \pm e^{i \vph_1 / 2} | = 1 \}. 
\]
The retraction for the case $\vph_1 =0$ is shown in \figurename~\ref{fig:eight}.
The points outside the eight-shaped curve move along the blue lines until they hit the eight-shaped curve;
the points inside the curve move along the red lines.   

As a result, $\cB$ becomes retracted to the bundle with the base
$\mathcal{S}^1$ and the fiber $\mathcal{G}_{0}$. This bundle is not trivial. 

\begin{figure}[h]  \centering{\includegraphics[width=0.9\textwidth]{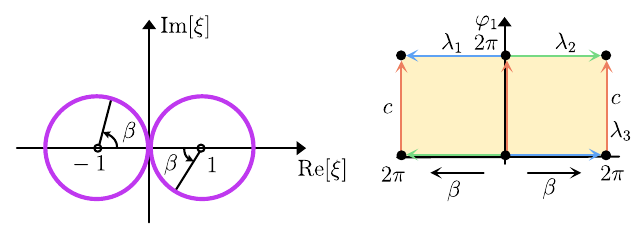}}
  \caption{Coordinate $\beta$ on $\mathcal{G}_0$ (left); the scheme of the retract (right)}
\label{fig:parametrization}
\end{figure}

Let us build the fundamental group for the deformation retract. 
Introduce the angle $\beta$ for the circles forming the eight curve $\mathcal{G}_0$ 
as it is shown in 
\figurename~\ref{fig:parametrization}, left. 
Thus, the retract becomes parametrized by two variables: $\beta, \vph_1 \in \mathcal{S}^1$
(one or two points of the retract correspond to each pair  $(\beta, \vph_1)$).
The scheme of the retract is shown in \figurename~\ref{fig:parametrization}, right. The retract consists of two 
rectangles connected in a sophisticated way. 
The oriented segments marked by the same Latin letters ($a,b,c$) should be attached to each other. 
Note that there are {\em three\/} copies of the segment~$c$.
The reference point (say, $t = (1 , 0.5)$ as it is shown in 
\figurename~\ref{f:01061k})
corresponds to $\beta = \vph_1 = 0$.  
   
The generators for our group are the bypasses  $\lambda_1,\lambda_2,\lambda_3$, 
going along the sides $a,b,c$, respectively (see 
\figurename~\ref{fig:parametrization}, right). 
Explicit representations of the paths $\gamma_1, \gamma_2, \gamma_3$ 
in the space $(t_1, t_2)$
are as follows: 
\[
\lambda_1: \quad (1, 1-e^{i \beta}), \quad \beta \in [0, 2\pi],
\]
\[
\lambda_2: \quad (1, -1+e^{i \beta}), \quad \beta \in [0, 2\pi],
\]
\[
\lambda_3: \quad (e^{i \vph_1}, 0), \quad \vph_1 \in [0, 2\pi].
\]


The relations defining $\pi_1 (\cB)$
are provided by {\rm faces\/} (yellow rectangles) of the graph in \figurename~\ref{fig:parametrization}, right. Namely, bypasses  
along their boundaries yield
\[
\lambda_2 \lambda_3 \lambda_1^{-1} \lambda_3^{-1} = e
,\qquad 
\lambda_1 \lambda_3 \lambda_2^{-1} \lambda_3^{-1} = e. 
\]

\begin{remark}
The group $\pi_1 (\cB)$ can be computed using Zariski's theorem from \cite{Zariski1937,Prasolov2022}. 
The result is a group with two generators $a,b$ and a single relation $abab=baba$.
This group is linked with what we found by the relations $a=\lambda_1,b=\lambda_1^{-1}\lambda_3$.
\end{remark}

\section{Application of formula (\ref{e:proper_PL}) to a triangle}
\label{app:F}
Let us obtain the matrix (\ref{e:01331}) by applying the formula (\ref{e:proper_PL}).
The plan is as follows. 
Consider the notations in \figurename~\ref{f:01061a1}.
Rewrite (\ref{e:proper_PL}) as 
\begin{equation}
{\rm var}_{\lambda} (w) = 
\sum_{\gamma \in \PI} \langle
{\cE(\omega A_e )\, | \, \gamma^{-1} w }
\rangle 
A_\gamma,
\label{e:proper_PL_a}
\end{equation}
\[
\omega = (e - \gamma_1^{-1}) (e - \gamma_2^{-1}) (e - \gamma_3^{-1})
\]
(we take $Q = A$).
Take $A_e, \dots , G_e$ (one by one) as $w$ and obtain (\ref{e:01331a}), (\ref{e:01331}) row by row.

To compute the intersection indices, build a representative $V$ of {$\cE(\omega A_e )$ } (a cycle) that 
is in a general position to $\mathbb{R}^2$, i.e.\ that
intersects 
$\mathbb{R}^2$ in a discrete set of points.
We are building this cycle according to \figurename~\ref{f:01055}, i.e.\ the cycle $V$ consists of 8 triangular face parts, 
12 cylindrical edge parts, and 6 vertex toroidal parts. 
Let us build these parts explicitly, namely, let us build the projection of $V$ onto the affix, i.e.\
$\pp (V)$.
Note that we build $\pp (V)$ in two steps: first we build a topological space that has a simple structure
but that is not in a general position with $\mathbb{R}^2$ (denote it $\pp (V')$), and then we shift it slightly. 

\begin{figure}[h]
  \centering{\includegraphics[width=0.8\textwidth]{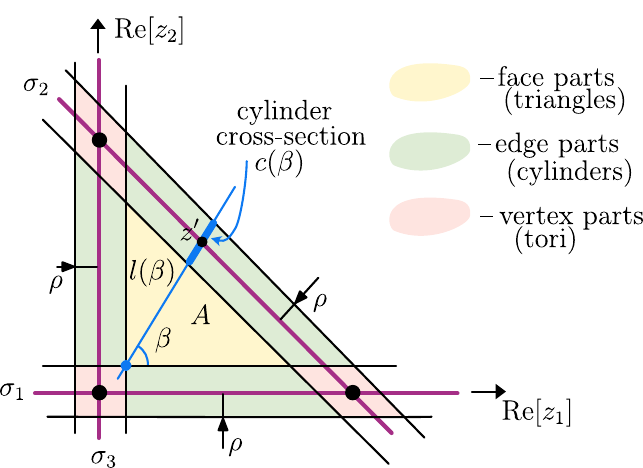}}
  \caption{Building of $V$}
\label{fig:PL_1}
\end{figure}

Without restriction of generality, take 
\[
\sigma_1 : \, \, z_2 = 0, 
\qquad 
\sigma_2 : \, \, z_1 + z_2 = 1, 
\qquad 
\sigma_3 : \, \, z_1 = 0
\]
(see \figurename~\ref{fig:PL_1}).

The traces $\sigma'_1 , \sigma'_2, \sigma'_3$ form a triangle. Take the face parts 
of $\pp(V')$ as parts of $\mathbb{R}^2$ obeying the inequalities 
\[
z_1 > \rho ,
\qquad 
z_2 > \rho ,
\qquad 
z_1 + z_2 < 1 - \sqrt{2} \rho
\]
for some $0 < \rho \ll 1$. This area is shown by a yellow shading in \figurename~\ref{fig:PL_1}. 

Three toroidal parts of $\pp (V')$ are parametrized as follows: 
\[
z_1 = \rho e^{i \alpha_1}, \qquad z_2 = \rho e^{i \alpha_2},
\]
\[
1- z_1 - z_2  = \sqrt{2} \rho e^{i \alpha_1}, \qquad z_2 = \rho e^{i \alpha_2},
\]
\[
1- z_1 - z_2  = \sqrt{2} \rho e^{i \alpha_1}, \qquad z_1 = \rho e^{i \alpha_1} ; 
\]
everywhere $\alpha_1 , \alpha_2 \in [0 , 2\pi]$.

Building the cylindrical parts of $\pp (V')$ is a bit more cumbersome. 
We explain how the cylindrical part around $\sigma'_2$ is built; the two other cylindrical parts 
are built in a similar way. 

The cylindrical part is built as a family of cylindrical cross-sections $c(\beta)$ parametrized by $\beta \in [0 , \pi/2]$,
see \figurename~\ref{fig:PL_1}. Such a cross-section lies in the complex line 
\[
l(\beta): \, \, \frac{z_2 - \rho}{z_1 - \rho} = \tan \beta.
\]
Each cross-section is defined by 
\[
c(\beta) : \, \,
\Delta z_1 = -\rho \frac{e^{i \alpha} \cos \beta}{\cos (\beta - \pi/4)}, 
\qquad 
\Delta z_2 = -\rho \frac{e^{i \alpha} \sin \beta}{\cos (\beta - \pi/4)}, 
\quad 
\alpha \in [0, 2\pi],
\]
\[
\Delta z_1 = z_1 -z_1',
\qquad 
\Delta z_2 = z_2 -z_2',
\]
\[
z' = (z_1' , z_2') = l(\beta) \cap \sigma_2.
\]

One can see that all parts fit each other, such that one can cut cylindrical and toroidal parts along 
the lines $\alpha = 0$, $\alpha_1 = 0$, $\alpha_2 = 0$, take a necessary amount of copies of each part, 
and glue them together. Each corner of the structure will be then assembled according to 
\figurename~\ref{f:01053} (two copies of it). Indeed, this structure can be lifted to 
$\tilde U_2$ as $\cE(\omega A_e)$. 

The construction described here is close to \cite{Vassiliev2002}. 

\begin{figure}[h]
  \centering{\includegraphics[width=0.6\textwidth]{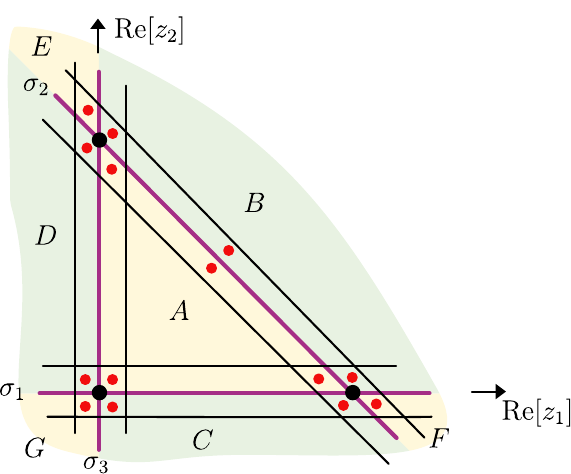}}
  \caption{Intersection $\pp (V) \cap \mathbb{R}^2$}
\label{fig:PL_2}
\end{figure}

To build $V$ that is in a general position with $\mathbb{R}^2$, take $V'$ built above and shift it by 
\[
z_1 \to z_1 - i \epsilon,
\qquad
z_2 \to z_2 - i \epsilon,
\]
where $\epsilon$ is some value $0 < \epsilon \ll \rho$.
One can see that the face parts of $\pp (V)$ will have no intersection with $\mathbb{R}^2$, 
there will be only two points on the cylindrical parts of $\pp (V)$ belonging to $\mathbb{R}^2$
(both belonging to the cylinder about $\sigma_2$), and each of three toroidal parts of $\pp (V)$
will intersect $\mathbb{R}^2$ in four points (see \figurename~\ref{fig:PL_2}, where the crossing points 
are shown by red).
Indeed, each point on the cylindrical part corresponds to four points of~$V$, and each point 
on the toroidal parts corresponds to two points of~$V$,

A careful computation of the sheet and intersection index of each crossing point reveals 
(\ref{e:01331a}), (\ref{e:01331}). 

\section{Homology of local systems and Picard--Lefschetz--Pham formula}
\label{app:G}
If the ramification of a multivalued differential form has only a power {behaviour}, then there exists a \textit{local system} such that the integral of this form over cycles of its homology is well-defined. We describe (following \cite{Vassiliev2002}) this local system and show how formulas \eqref{e:proper_PL} and \eqref{e:013351a} are related with V. Vassiliev's formulas (VI.26), (VI.27) of \cite{Vassiliev2002}, which in turn strengthen F.~Pham's formulas  (PL2), (PL3) \cite[\S4.3]{Pham1965}.
The connection between different types of Picard-Lefschetz formulas is revealed through the geometric meaning of equivariant Poincaré duality.  Finally, we will provide a calculation of certain intersection indices for an example of a singularity of type $P_3$ in~$\mathbb{C}^2$ {(see below)}.



\subsection{Homology of a local system, associated with a branched holomorphic form}

Here we follow  V.~Vassiliev, \cite[VI.1]{Vassiliev2002}.

Let 
\[
\varphi=g_{1}^{\alpha_1}(z)\dots g_{m}^{\alpha_m}(z) \, dz_1\wedge\dots\wedge dz_n,
\qquad z = (z_1 , \dots , z_n),
\qquad 
{\alpha_j \in \mathbb{R},}
\] 
be a branching holomorphic $n$-form in $\mathbb{C}^n$. The singularity (branching) set 
$\sigma$ consists of $m$ irreducible components $\sigma_1,\dots,\sigma_m$ defined by 
\[
g_j(z)=0, \qquad j=1,\dots,m.
\]

An analytic continuation of the form $\varphi$ along a closed path $\rho$ in $\mathbb{C}^n\backslash\sigma$ multiplies $\varphi$ by \begin{equation}
		\mathring\rho=\mathring\gamma_1^{-\text{lk}(\rho,\sigma_1)}\cdot\ldots\cdot\mathring\gamma_m^{-\text{lk}(\rho,\sigma_m)},
\end{equation}
where 
\[
\mathring\gamma_j \equiv e^{-2\pi i\alpha_j}, 
\]
$\text{lk}(\rho,\sigma_j)$ is the \textit{linking number\/}, i.e.\ the intersection index (see p.~\pageref{pr:PL_relative}) of any $2$-chain in $\mathbb{C}^n$ with boundary $\rho$, and  $\sigma_j$.

Introduce a \textit{local system} $L_{\mathring\gamma}$ with a fiber $\mathbb{C}^1$ on $\mathbb{C}^n\backslash\sigma$, associated with a branched holomorphic form $\varphi$. 
According to \cite{Vassiliev2002}, it is a covering $p:L_{\mathring\gamma}\rightarrow\mathbb{C}^n\backslash\sigma$ with a fiber $\mathbb{C}$, 
such that a structure of a vector space is fixed 
in the fibers, and this structure depends continuously on the fiber. Note that $L_{\mathring\gamma}$ is equipped with the topology of the covering space, i.e.\ the full preimage $p^{-1}$ of a small neighborhood $U(a)$ of any point $a\in\mathbb{C}^n\backslash\sigma$ is homeomorphic to the disjoint union of $\mathbb{C}$ copies of $U(a)$. Like any covering, the local system defines a monodromy representation $\pi_1(\mathbb{C}^n\backslash\sigma)\rightarrow\text{Aut}\;\mathbb{C}$: any loop $\rho$  acts by multiplying by $\mathring\rho$ in the fiber over the reference point. Since $\mathbb{C}^{n}\backslash\sigma$ is path-connected, the monodromy representation (considered up to conjugation) uniquely (up to isomorphism) defines the local system $L_{\mathring\gamma}$. 
The set of numbers $\mathring\gamma=(\mathring\gamma_1,\dots,\mathring\gamma_m)$  is called the monodromy coefficients of the local system $L_{\mathring\gamma}$. Each coefficient $\mathring\gamma_j$ corresponds to a simple loop $\gamma_j$ around a smooth piece of the variety~$\sigma_j$.


Let us introduce also a local system $\check L_{\mathring\gamma}$ with fiber $(\mathbb{C})^*\equiv\text{Hom}_{\mathbb C}(\mathbb{C},\mathbb{C})$, dual to the local system $L_{\mathring\gamma}$. The monodromy representations of the local systems $L_{\mathring\gamma}$ and $\check L_{\mathring\gamma}$  are conjugate. In particular, the local system $\check L_{\mathring\gamma}$ has inverse monodromy coefficients $\check{\mathring\gamma}=(\mathring\gamma_1^{-1},\dots,\mathring\gamma_m^{-1})$, and we have an isomorphism $L_{\mathring\gamma^{-1}}\simeq\check L_{\mathring\gamma}$.

Let us define the homology of the space $\mathbb{C}^n\backslash\sigma$ with coefficients in the local system $L_{\mathring\gamma}$. 
These are the homology of the chain complex, the elements of which are formal sums of singular simplices of the space $L_{\mathring\gamma}$, such that: 
\begin{enumerate}
    
\item If simplices
$\bigtriangleup_{\mathring\rho},\bigtriangleup_{\mathring\rho'},\bigtriangleup_{\mathring\rho''}$ 
are located, respectively, on leaves 
$\mathring\rho,\mathring\rho',\mathring\rho''$, 
where 
$\mathring\rho +\mathring\rho'=\mathring\rho''$ 
and 
$p( \bigtriangleup_{\mathring\rho}) =p(\bigtriangleup_{\mathring\rho'})=p(\bigtriangleup_{\mathring\rho''})$, 
then we identify 
$\bigtriangleup_{\mathring\rho}+\bigtriangleup_{\mathring\rho'} \equiv \bigtriangleup_{\mathring\rho''}$.

\item Any simplex from the zero leaf is equal to zero.  

\item The simplex 
$\bigtriangleup_{\mathring\rho}$ taken with coefficient $c\in\mathbb C$ identifies with the simplex $\bigtriangleup_{\mathring\rho'}$ taken with coefficient $c'\in\mathbb C$ if these simplices
are located correspondingly on leaves 
$\mathring\rho,\mathring\rho'$ with $p( \bigtriangleup_{\mathring\rho}) =p(\bigtriangleup_{\mathring\rho'})$ and $c\mathring\rho=c'\mathring\rho'$. 

\end{enumerate}

Now consider the \textit{Riemann covering} (in the usual sense) $\mathfrak{R}$ related to the form $\varphi$. 
In our terms, this corresponds to $\hat U_n$, i.e.\ the main stratum of the Riemann domain.
Let us define the group of chains (with coefficients in $\mathbb{C}$) of $\mathfrak{R}$, factored through the following condition: if $\Delta$ and $\Delta'$ are two singular simplices whose projection to the $\mathbb{C}^n\backslash\sigma$ coincide, $\tilde{\rho}$ is an arbitrary path, connecting the centers of these simplices, and $\rho$ its projection to $\mathbb{C}^n,$ then we identify $\Delta'$ and $\mathring\rho\Delta=\prod_{j=1}^m\mathring\gamma_j^{-\text{lk}(\rho,\sigma_j)}\Delta$. 
These groups of chains are isomorphic to the corresponding groups of chains of the local system $L_{\mathring\gamma}$ by definition.

A branching differential form $\varphi$ can be considered as a single-valued form on $\mathfrak{R}$, and the integrals of the form $\varphi$ along the homology classes of the local system $L_{\mathring{\gamma}}$ are well defined. 

\begin{remark}
     The equivariant homology groups of the space $\mathbb C^n\backslash\sigma$ with the coefficients in the group ring $\Omega\equiv\mathbb{Z}[\pi_1]$ of $\pi_1(\mathbb C^n\backslash\sigma)$ over $\mathbb{Z}$ (which are isomorphic to the integer homology groups of the universal Riemann domain as $\mathbb Z$-modules) are isomorphic to the homology groups of the local system, whose fiber is a free (left) $\Omega$-module of rank $1$ (the structure of the module depends continuously on the fiber), and the element $\rho\in\pi_1$ acts on the fiber by multiplication by~$\rho$.
     \end{remark}

\begin{remark}
We use the terms ``sheet'' and ``sheet with index $\gamma_j$'' when talking about a Riemann covering, and the term ``leaf'' and  ``leaf with index $\mathring {\gamma_j}$'' when talking about a local system.
\end{remark}

\subsection{Equivariant Poincaré duality}

Let $\bB\subset\mathbb{C}^n$ be a sufficiently large open ball centred at the origin, considered throughout the article. Let us denote by $H_*(\bB\backslash\sigma;{L}_{\mathring\gamma})$ and $H^{lf}_*(\bB\backslash\sigma;{L}_{\mathring\gamma})$ the homology groups with coefficients in the local system $L_{\mathring\gamma}$, defined by finite and \textit{locally finite} chains correspondingly. 
The locally finite chains of a local system are defined in \cite{Vassiliev2002} and are used to provide homologies for converging improper integrals.

By the Poincaré duality theorem \cite[VI]{Vassiliev2002} (Theorem 1.1), there is a canonical isomorphism 
\begin{equation}
H^{i}(\bB\backslash\sigma;L_{\mathring\gamma})\simeq H_{2n-i}^{lf}(\bB\backslash\sigma;\check{L}_{\mathring\gamma}).
\end{equation}
In particular, there is a non-degenerate bilinear pairing 
\begin{equation}\label{key}
H_{i}(\bB\backslash\sigma;L_{\mathring\gamma})\otimes H_{2n-i}^{lf}(\bB\backslash\sigma;\check{L}_{\mathring\gamma})\rightarrow\mathbb{C},
\end{equation}
given by the ``twisted'' intersection indices (see Proposition~\ref{twistint}).	

The homology group $H^{lf}_*(\bB\backslash\sigma;{L}_{\mathring\gamma})$ is naturally isomorphic to the relative homology group $H_*(\overline\bB,\sigma\cup\partial \bB;{L}_{\mathring\gamma})$. The last group is defined by finite chains of the branched  covering space $\overline{\mathfrak{R}}$, factorized by chains in the added set (and also by conditions of type $\Delta'=\mathring\rho\Delta$). $\overline{\mathfrak{R}}$ is a usual compactification if the Riemann covering is finite; if the covering is infinite, then 
we are not interested in chains running along an infinite number of sheets (since their projections to the base are not locally finite  
and such chains are not locally finite chains of the local system by definition). In this case, we glue singularities and boundaries similarly to the  universal Riemann domain structure (see Definition~\ref{def:URD}).

The pairing \eqref{key} has the following geometric meaning. We realize the classes $w\in H_i(\overline\bB,\sigma\cup\partial \bB;L_{\mathring\gamma})$ and $\check{\psi}\in H_{2n-i}(\bB\backslash\sigma;\check L_{\mathring\gamma})$ by compact singular chains with smooth simplices,  relative in $\overline{\mathfrak{R}}$ modulo $\tilde U'\cup\tilde{\partial\bB}$ (see \eqref{e:def_Up}, \eqref{e:def_tilde-dB}) and absolute in $\mathfrak R$, respectively. 
Consider the cycle $\psi$,
dual to $\check\psi$ and defined by the set of all points of the local system $L_{\mathring{\gamma}}$ on which the points of the cycle $\check\psi$ take the value~1. Choose an arbitrary chain in $\mathfrak R$, representing this cycle.

\begin{proposition}\label{twistint}
		The ``twisted'' intersection index $\langle\langle w,\check\psi\rangle\rangle$ defining \eqref{key} may be set as follows:
	\begin{equation}\label{intindex}
		\langle\langle w,\check\psi\rangle\rangle
        \equiv
        \sum_{\rho\in\pi_1(\mathbb C^n\backslash\sigma)}\langle\rho^{-1}w|\psi\rangle\mathring\rho,
	\end{equation}	
where $\langle\cdot|\cdot\rangle$ is the standard intersection index (see p.~\pageref{pr:PL_relative}) of chains in $\mathfrak{R}$, defined by the complex orientation.
\end{proposition}

\textbf{Proof.} Let two points $a,b \in \mathfrak{R}$ be connected by a path $\tilde{\rho}$ that 
is projected onto~$\rho$. Consider these points as points of a local system $\check{L}_{\mathring\gamma}$. 
If $a$ takes the value $1$ at the point of the initial leaf of $L_{\mathring\gamma}$, then $b$ takes the value $\mathring\rho^{-1}$ at the same point, and therefore $b$ takes the value $1$ at the point of the leaf with index~$\mathring\rho$. 

 By a small perturbation of the chain representing the cycle $w$, one can always ensure that this chain intersects the chain representing the cycle $\psi$
and all its possible shifts along the sheets of $\mathfrak{R}$ at interior points of the simplices of maximal dimension of both chains, and that all these intersections are transversal  (equivalently, that the projections of chains
into $\mathbb C^n$ intersect transversally). As it is clear from the 
consideration above, the intersection index $\langle w|\psi\rangle$ will participate in the sum with a coefficient equal 
to~$1$. 
Finally, consider the chain $\rho^{-1}w$ and let $a\in\rho^{-1}w\cap\psi$. Let $\tilde{\rho}$ be a path in $\mathfrak{R}$ connecting the points $a$ and $b$ in $w$, which is projected onto~$\rho$. Then the point $a$, considered as a point of $\check{L}_{\mathring\gamma}$, takes the value $\mathring\rho$ at point $b$, considered as a point of $L_{\mathring\gamma}$. Hence the intersection index $\langle \rho^{-1}w|\psi\rangle$ will enter the sum with the coefficient equal to~$\mathring\rho$.

It is easy to check that the obtained number does not depend on the choice of cycles in the homology class (and the chains representing them in $\mathfrak{R}$). Thus, formula \eqref{intindex} is a direct generalization of the intersection index defining classical Poincaré-Lefschetz duality via the realization of homology classes by singular chains. For more details on equivariant Poincaré duality see 
\cite[\S2.1,~2.3]{Hillman2002}. $\Box$

    \begin{remark} 
        Similarly, for $\psi\in H_i(\bB\backslash\sigma;L_{\mathring\gamma})$ and $\check{w}\in H_{2n-i}(\overline\bB,\sigma\cup\partial\bB;\check L_{\mathring\gamma})$ the intersection index $\langle\langle\psi,\check{w}\rangle\rangle$ may be set as 
        \begin{equation}
\langle\langle\psi,\check{w}\rangle\rangle=\sum_{\rho\in\pi_1(\bB\backslash\sigma)}\langle\rho^{-1}\psi|w\rangle\mathring\rho.
        \end{equation}
    \end{remark}
 \begin{proposition}\label{prp}

\begin{equation}\label{7}
	\langle\langle w,\check\psi\rangle\rangle=\sum_{\rho\in\pi_1(\bB\backslash\sigma)}\langle w|\rho\psi\rangle\mathring\rho.
\end{equation}
\end{proposition}
\textbf{Proof.} It follows from \eqref{intindex} and simple geometric considerations. $\Box$

\subsection{Picard-Lefschetz-Pham formula in different interpretations}
	
Let $\sigma$ be a non-degenerate set of singularity 
components $\sigma_j$, close to one of the standard Pham's degenerations (see \cite{Pham1965}, \cite[I.8]{Vassiliev2002}) $P_m$ in $\mathbb{C}^n,$ where $m=1,\dots,n+1$, defined by a set of m equations \begin{equation}\label{PhamD}
    z_1=0,\quad \dots\quad ,z_{m-1}=0, \quad 
    z_1+\dots+z_{m-1}+z_m^2+\dots+z_n^2+t=0.
\end{equation}
Degeneration occurs when the parameter $t$ is equal to zero. Let an open ball $\bB$ contain the central part of the degeneration. 
Since the components of $\sigma$ intersect in general position, the group $\pi_1(\bB\backslash\sigma)$ is commutative. It follows from the local variant of the generalized Zariski theorem \cite{Zariski1937}, see  \cite{Hamm1983}. The singularities of types ``circle'', ``biangle'', ``triangle'', that are considered in this article, are precisely Pham's degenerations of the types $P_1, P_2, P_3$ in $\mathbb{C}^2$.

Let us consider the diagram of homomorphisms (compare with \eqref{cd1}): 

\begin{equation}\label{cd0.5}
	\begin{CD}
		H_{n}(\bB\backslash\sigma;L_{\mathring\gamma}) @>\Theta>> H_{n}(\bB,\sigma;L_{\mathring\gamma})\\
		@VVV @VVV \\
		H_{n}(\overline\bB\backslash\sigma,\partial\bB;L_{\mathring\gamma}) @>>> H_{n}(\overline\bB,\sigma\cup\partial\bB;L_{\mathring\gamma}).\\
	\end{CD}
\end{equation}

The homomorphism of the upper arrow of this diagram is called a \textit{regularizing homomorphism}. It is defined by including finite chains into locally finite ones. We will denote it by~$\Theta$.
By Theorem~5.1 \cite[VI.5]{Vassiliev2002}, if at least one of the monodromy coefficients $\mathring\gamma_1,\dots,\mathring\gamma_m$ is not equal to $1$, then all four groups are isomorphic to $\mathbb{C}^1$ (this is a hard part of this theorem). Moreover, if for all $j$ holds $\mathring\gamma_j\neq 1$ and $\mathring\gamma_1\cdot\ldots\mathring\gamma_m\neq(-1)^{n-m+1}$ then all rows in \eqref{cd0.5} are isomorphisms. 
In this article, an analogue of the isomorphism $\Theta^{-1}$ is called the {\em inflation isomorphism}.

The group $H_n(\bB,\sigma;L_{\mathring\gamma})$ is generated by the class of the \textit{vanishing cell} $\Delta$ (i.e.\ a real domain bounded on all sides by $\sigma$ and equipped with the standard orientation of $\mathbb{R}^n$), lifted to an arbitrary non-zero leaf of our local system (below, without loss of generality, we assume that the cycle $\Delta$ is lifted to the initial sheet of ~$\mathfrak{R}$). The group $H_n(\bB\backslash\sigma;L_{\mathring\gamma})$ is generated by the class of a \textit{double loop} (see \cite[VI.4]{Vassiliev2002} and \figurename~\ref{f:01055} here) $\varkappa\equiv\varkappa(\Delta)$ around $\Delta$. By construction, $\Theta$ maps $\varkappa$ to $\prod_{j=1}^{m}(1-\mathring\gamma_j)\Delta$. 

Consider also a similar diagram for the dual local system $\check{L}_{\mathring\gamma}$. The generators $\check{\Delta}$ and $\check{\varkappa}$ of the same homology groups with coefficients in $\check{L}_{\mathring\gamma}$ can be chosen similarly. Namely, $\check{\Delta}$ is lifted to the leaf of the $\check{L}_{\mathring\gamma}$, the points of which take the value $1$ on the corresponding points of the leaf to which the cycle $\Delta$ is lifted; $\check{\varkappa}$ is a double loop around the cycle $\check{\Delta}$. The homorphism $\check\Theta$ maps $\check\varkappa$ to $\prod_{j=1}^{m}(1-\mathring\gamma_j^{-1})\check\Delta$.

A twisted intersection-pairing \eqref{key} connects any group from diagram \eqref{cd0.5} with the group that stands diagonally from it  in the diagram, 
but taken with coefficients in $\check{L}_{\mathring\gamma}$. 

\begin{remark} Consider the value $\langle\langle w,\check\varkappa\rangle\rangle.$
If the cycle $w$ is located entirely on the initial sheet of $\mathfrak{R}$, the geometric meaning is especially simple: one can take the full preimage of the cycle $w$ in $\mathfrak{R}$ (i.e. the entire orbit under the action of the fundamental group)  and calculate its intersection indices with $\varkappa$ on each sheet, multiplying the obtained values by the inverses of the leaf numbers and adding them together.
\end{remark} 

\vskip 6pt

Denote 
\[
\mathring\Pi\equiv\prod_{j=1}^m\mathring\gamma_j
\]
and 
\[
\check\varkappa^{\circ}\equiv\check\Theta^{-1}\prod_{j=1}^m(1-\mathring\gamma_j)\check\Delta
\] 
(it is a dual double loop in the reverse direction). Note that 
\begin{equation}\label{Pi}
  \check\varkappa=(-1)^{m}\mathring\Pi^{-1}\check\varkappa^{\circ}  
\end{equation}
(this can be checked directly).

\begin{lemma}
    (analogues \S4.3 PL2, PL3 \cite{Pham1965})
Consider one of the standard Pham's degeneration $P_m$ in $\mathbb{C}^n$ and a simple loop $\lambda\in\pi_1(\cB)$ around it.
Suppose that none of the coefficients $\mathring\gamma_1,\dots,\mathring\gamma_m$ is equal to $1$, and their product is not equal to $(-1)^{n-m+1}$. Then for any class $w\in H_n(\overline\bB,\sigma\cup\partial \bB;{L}_{\mathring\gamma})$
\begin{equation}\label{plf3}
  {\rm var_{\lambda}}\;w=(-1)^{\frac{n(n+1)}{2}}(-1)^{n-m+1}\mathring\Pi\langle\langle w,\check\varkappa\rangle\rangle\Delta,   
\end{equation}
or, equivalently, 
\begin{equation}\label{plf2}{\rm var_{\lambda}}\;w=(-1)^{\frac{(n+1)(n+2)}{2}}\langle\langle w,\check\varkappa^{\circ}\rangle\rangle\Delta. 
\end{equation}
   
\end{lemma}
\textbf{Proof.} First, let us note that 
\begin{equation}
\label{varvar}
    \text{var}_{\lambda}\Delta=((-1)^{n-m+1}\mathring\Pi-1)\Delta
\end{equation}
(this doesn't agree with the statement 5 of Theorem 5.1 \cite[VI.5]{Vassiliev2002}).

Indeed, under the action of monodromy the central (and in general any inner) point of
the cycle $\Delta$ will move to the sheet $\gamma_1\cdot\ldots\cdot\gamma_m$, and the calculation of the monodromy of the projection of the cycle $\Delta$ to $\mathbb{C}^n$ is reduced to the usual Picard-Lefschetz formula for the Morse singularity in $\mathbb{C}^{n-m+1}$ (see  \cite{Pham1965}, 
\cite[I.8]{Vassiliev2002}).

From statements 1-3 of Theorem 5.1  \cite[VI.5]{Vassiliev2002} and Poincaré duality, it follows 
\begin{equation}\label{var}
    {\rm var_{\lambda}}\;w=\frac{\langle\langle w,\check\varkappa\rangle\rangle}{\langle\langle\Delta,\check\varkappa\rangle\rangle}{\rm var_{\lambda}}\;\Delta.
\end{equation}
Since in view \eqref{intindex} and the skew-commutativity of the intersection index
\begin{equation}
 \langle\langle\Delta,\check\varkappa\rangle\rangle=\sum_{\rho\in\pi_1(\bB\backslash\sigma)}\langle \rho^{-1}\Delta|\varkappa\rangle\mathring\rho=(-1)^n\sum_{\rho\in\pi_1(\bB\backslash\sigma)}\langle \varkappa|\rho^{-1}\Delta\rangle\mathring\rho,  
\end{equation}
we obtain from statement 4 of the same theorem (see formula (VI.25))
\begin{equation}\label{j}
  \langle\langle\Delta,\check\varkappa\rangle\rangle=(-1)^{\frac{n(n+1)}{2}}(1+(-1)^{n-m}\mathring\Pi^{-1}).  
\end{equation}

Then \eqref{plf3} follows from \eqref{varvar}, \eqref{var} and \eqref{j}, since
$$\frac{(-1)^{n-m+1}\mathring\Pi-1}{1+(-1)^{n-m}\mathring\Pi^{-1}}=(-1)^{n-m+1}\mathring\Pi.$$

Now, by \eqref{Pi} and bilinearity
\begin{equation}
    \langle\langle w,\check\varkappa\rangle\rangle=\langle\langle w,(-1)^{m}\mathring\Pi^{-1}\check\varkappa^{\circ}\rangle\rangle=(-1)^{m}\mathring\Pi^{-1}\langle\langle w,\check\varkappa^{\circ}\rangle\rangle.
\end{equation}

Then we obtain \eqref{plf2} from \eqref{plf3}.
 $\Box$

\begin{corollary}
   For the cases $P_1,P_2,P_3$ in $\mathbb{C}^2$ the formula \eqref{plf3} can be written in the form  \eqref{e:proper_PL}.  
\end{corollary}
\textbf{Proof.} 
It follows from 
substitution $n=2$ to \eqref{plf2} and \eqref{7}. $\Box$

\begin{remark}
Formula \eqref{e:013351a} is obtained from Theorem 5.1 of  
\cite[VI.5]{Vassiliev2002} in a similar way.
\end{remark}

\subsection{Calculation of intersection indices}

\begin{figure}[h]
  \centering{\includegraphics[width=0.5\textwidth]{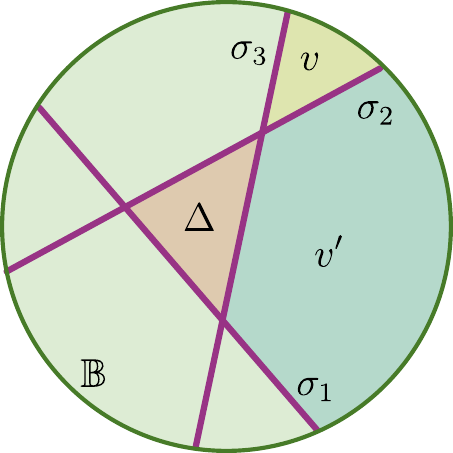}}
  \caption{Domains for calculation of intersection indices}
\label{f:N4}
\end{figure}

Here we will consider the Pham's singularity of type $P_3$ in $\mathbb{C}^2,$ i.e.\ \textit{a triangle} (see \figurename~\ref{f:N4})  and calculate the monodromy of cycles of the relative homology group $H_2(\overline\bB,\sigma\cup\partial\bB;L_{\mathring\gamma}),$ represented by real domains, equipped with the positive orientation $\mathbb R^2$.

According to the formula \eqref{plf3} and \eqref{intindex}
   \begin{equation}\label{P3}
       \text{var}_{\lambda}(w)=-\mathring\Pi\langle\langle w,\check\varkappa\rangle\rangle\Delta=-\mathring\Pi\sum_{\rho\in\pi_1(\bB\backslash\sigma)}\langle\rho^{-1}w|\varkappa\rangle\mathring\rho\Delta.
   \end{equation}

Let us introduce notations for components of a representative of a double loop $ \varkappa$. Note that $ \varkappa$ is the same object that 
{$\cE(\omega A_e)$} in our terms, where $\omega = (e - \gamma_1)(e - \gamma_2)(e - \gamma_3)$.
     
Here we follow the construction  described in \cite{Vassiliev2002}. Double loop is glued from 26 pieces: 8 \textit{triangles} $\bigtriangleup$, 12 \textit{rectangles} $\rectangle$ and 6 \textit{squares} $\Box$ (see \figurename~\ref{f:01055}: triangles are faces, rectangles are cut cylinders, squares are cut tori).  Let the lines $\sigma_j$ be defined by $g_j=0$, respectively, and for any point $a\in\Delta$ we have $g_j(a)<0$. 
Each triangle is identified with a triangle in $\mathbb{R}^2$ formed by a triple of lines $g_j=-\varepsilon$. Let us index these triangles with all possible subsets of the set $\{1,2,3\}$. 
Triangle with the index $\nu$ will be denoted $\bigtriangleup(\nu)$. Rectangles belong to three ``types'', each corresponding to a triangle side~$\sigma_i$. 

Each rectangle of type $i$ will be denoted by $\rectangle_i$ and identified with $I_i\times[-\varepsilon,\varepsilon]$, where $I_i$ is the segment of the line $g_i=-\varepsilon$ located between its intersection points with $g_j=-\varepsilon$ for all $j\neq i$. Let us index the rectangles of type $i$ with all subsets of the set $\{1,2,3\}\backslash \{i\}$. A rectangle of type $i$ with index $\nu$ will be denoted by $\rectangle_i(\nu)$. Finally, the squares are also of three ``types'', each of which corresponds to the intersection point $\sigma_i\cap\sigma_j$. Each square of type $ij$ will be denoted by $\Box_{ij}$ and identified with $[-\varepsilon,\varepsilon]_i\times[-\varepsilon,\varepsilon]_j$. 
Let us index the squares of type $ij$ by subsets of the set $\{1,2,3\}\backslash\{i,j\}$ and denote a square $\Box_{ij}$ with index $\nu$ as $\Box_{ij}(\nu)$.

\begin{figure}[h]
  \centering{\includegraphics[width=0.9\textwidth]{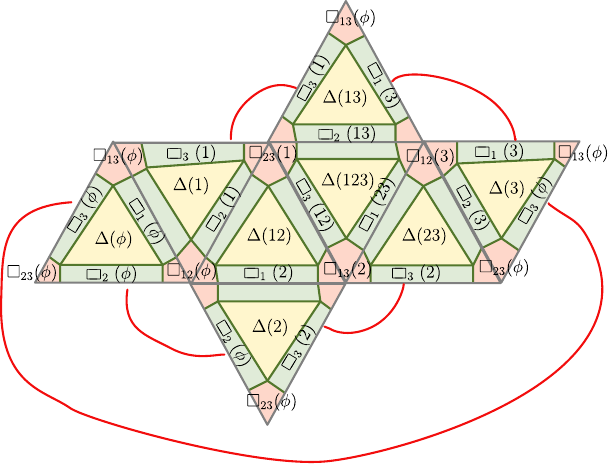}}
  \caption{A double loop $\varkappa$ made of pieces. Edges connected by red lines are glued}
\label{f:N5}
\end{figure}

The representative of $\varkappa$ used for computations is glued as follows. 
Consider $\rectangle_i(\nu)$. Let us 
glue the segment $-\varepsilon\times I_i$ to the corresponding component of the boundary of $\bigtriangleup(\nu)$, 
and  $\varepsilon\times I_i$ to the corresponding component of the boundary of~$\bigtriangleup(\nu\cup i)$. 
In this way, we will glue rectangles to the boundaries of all triangles. 
Then consider~$\Box_{ij}(\nu)$. Let us glue $-\varepsilon_i\times[-\varepsilon,\varepsilon]_j$ to the corresponding component of the boundary of $\rectangle_j(\nu)$ and $\varepsilon_i\times[-\varepsilon,\varepsilon]_j$ to the component of the boundary of $\rectangle_j(\nu\cup i)$. 
Make a similar attachment for $[-\varepsilon,\varepsilon]_i\times\pm\varepsilon_j$. As a result, obtain a  topological space which is homeomorphic to a two-dimensional sphere $S^2$ (see \figurename~\ref{f:N5}).

This corresponds to the terms of the polynomial $\omega$. For example, a triangle $\bigtriangleup (12)$ corresponds to a subset of 
$A{\gamma_1 \gamma_2}$ in our notations, and $\bigtriangleup (\varnothing)$ corresponds to $A_e$. A rectangle $\rectangle_2 (3)$ is a cylinder 
from \figurename~\ref{f:01055} connecting $A_{\gamma_3}$ with $A_{\gamma_2 \gamma_3}$. A square $\Box_{23}(\varnothing)$ is a torus connecting  
the corresponding corners of $A_e$, $A_{\gamma_2}$, $A_{\gamma_3}$, $A_{\gamma_2 \gamma_3}$. 
A scheme of each such corner is shown in \figurename~\ref{f:01053}. 

Indeed, a sphere $S^2$ is a slightly inflated octahedron shown in \figurename~\ref{f:01056}.

Construct an immersion of the obtained space in $\mathbb{C}^2.$
Let us map all triangles identically. Consider a family of real lines $l_{\beta}$ passing through a point $\{g_2=-\varepsilon\}\cap\{g_3=-\varepsilon\}$ and connecting these two lines. On each such line $l_{\beta'}$, 
we choose the origin at the point $l_{\beta'}\cap\sigma_1$ and take $g_1$ as a local coordinate. In this way, we trivialize the  
``trapezoidal'' neighborhood $T(\sigma_1)$ of the singularity component 
$\sigma_1$ which is a polygon bounded by the lines  
$g_1=\pm\varepsilon,g_2=-\varepsilon,g_3=-\varepsilon$ (i.e.\ $T(\sigma_1)\simeq\sigma_1\times[-\varepsilon,\varepsilon]$: the projection of $T(\sigma_1)$ onto the first factor is given by the shift along the pencil $l_\beta$ (see \figurename~\ref{fig:PL_1}), the projection onto the second factor is given by the coordinate $g_1$). This allows us to map the rectangle $\rectangle_1(\nu)$ to $\mathbb{C}^2$ as follows: two components of boundary $I_1\times\pm\varepsilon$ map identically, and all segments $\cdot\times[-\varepsilon,\varepsilon]$ (equipped with the coordinate $\theta$) map by the formula $g_1(\theta)=-\varepsilon e^{\pi i(\theta+\varepsilon)/\varepsilon}$ to the corresponding complexified line $l_{\beta'}$. For $\rectangle_i$ with $i\neq 1$, the immersion is similar. Finally, near each point $\sigma_i\cap\sigma_j$ we take  $g_i,g_j$ as the local coordinates and map $\Box_{ij}$ by the formula $(\theta_i,\theta_j)\rightarrow(-\varepsilon e^{\pi i(\theta_i+\varepsilon)/\varepsilon},-\varepsilon e^{\pi i(\theta_j+\varepsilon)/\varepsilon})$. Here $\theta_i,\theta_j$ are coordinates on the square.

It is reasonable straighforward to see that this immersion can be lifted to an immersion in the space of a Riemannian covering. 
Moreover, the index of each piece of the double loop exactly corresponds to the sheet index of $\mathfrak{R}$, i.e.\ the central point of the piece with index $\nu$ will be lifted to the sheet $\prod_{j\in\nu}\gamma_{j}$. On the image of the $\bigtriangleup(\nu)$  we define the orientation to coincide with the positive orientation of $\mathbb{R}^2$ if the cardinality of the set $\nu$ is even and is the opposite if it is odd. One can check that this orientation can be uniquely extended to an orientation of the immersed double loop. 

The twisted intersection index $\langle\langle\Delta,\check\varkappa\rangle\rangle$ is the most difficult to calculate (see computations in \cite{Vassiliev2002}), but the  monodromy of the cycle $\Delta$ can be calculated directly. 

Let us calculate $\langle\langle v,\check\varkappa\rangle\rangle$. By construction, the immersed double loop intersects $v$ twice in sheets with numbers $1$ and $\gamma_1$, 
and both intersections are transversal. These points are the images of point $(\theta_2,\theta_3)=(0,0)$ 
for the immersion of the pieces~$\Box_{23}$.  
It remains to add the orientations and compare the result with the complex orientation, defined by the frame $(\partial/\partial g_3,i\partial/\partial g_3,\partial/\partial g_2, i\partial/\partial g_2)$. We obtain orienting  frames
\[
(\partial/\partial g_3,\partial/\partial g_2, i\partial/\partial g_3, i\partial/\partial g_2)
\]
at the point on the sheet~1, and 
\[
(\partial/\partial g_3,\partial/\partial g_2, i\partial/\partial g_2, i\partial/\partial g_3)
\]
at the point on the sheet~$\gamma_1$. 
It means that 
\[
\langle\Box_{23}(\varnothing),\varkappa\rangle=-1
\]
and 
\[
\langle\Box_{23}(1),\varkappa\rangle=1.
\]
Therefore 
\[
\langle\langle v,\check\varkappa\rangle\rangle=-1+\mathring\gamma_1^{-1},
\]
and, according to \eqref{P3}, 
\begin{equation}
	\text{var}_{\lambda}(v)=-\mathring\gamma_1\mathring\gamma_2\mathring\gamma_3(-1+\mathring\gamma_1^{-1})\Delta=(\mathring\gamma_1\mathring\gamma_2\mathring\gamma_3-\mathring\gamma_2\mathring\gamma_3)\Delta.
\end{equation}

\begin{figure}[h]
  \centering{\includegraphics[width=0.7\textwidth]{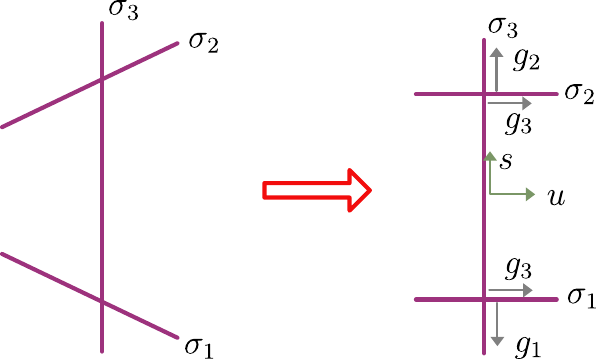}}
  \caption{``Straightening'' diffeomorphism for an edge part}
\label{f:N6}
\end{figure}

Finally, calculate $\langle\langle v',\check\varkappa\rangle\rangle$. 
Apply a ``straightening'' diffeomorphism (see \figurename~\ref{f:N6}) to the $\varepsilon$-neighborhood of $\sigma_3$, i.e.\ diffeomorphism $T_{\varepsilon}(\sigma_3)\simeq\sigma_3\times[-\varepsilon,\varepsilon]$:

1) In the ``trapezoidal'' neighborhood the projection onto the first factor is given by the shift along the pencil $l_\beta$ (see \figurename~\ref{fig:PL_1}); the projection onto the second factor is given by the coordinate $g_3$. 

2) Outside the ``trapezoidal'' neighborhood the projections are given by the coordinates $(g_2,g_3)$ in $T_{\varepsilon}(\sigma_2)\cap T_{\varepsilon}(\sigma_3)$ and $(g_1,g_3)$ in $T_{\varepsilon}(\sigma_1)\cap T_{\varepsilon}(\sigma_3)$.

Now let us introduce a coordinate system $(u,s)$ with the origin in the middle of the stratum $\sigma_3$, such that $u \equiv g_3$ along the direction of the outer normal to the $\Delta$, coordinate $s$ is directed along $\sigma_3$, and this coordinate system orients $\mathbb{R}^2$ positively. Without loss of generality, we will assume that the singularity components $\sigma_1,\sigma_2$ are defined in this coordinate system by the equations $s=\pm1$. The immersed pieces $\Box_{23}$ and $\Box_{13}$ are glued to pieces $\rectangle_3$ along circles of radius $\varepsilon$ about the origin on the complex lines defined by $s=1-\varepsilon$ and $s=-1+\varepsilon$, respectively.

\begin{figure}[h]
  \centering{\includegraphics[width=0.75\textwidth]{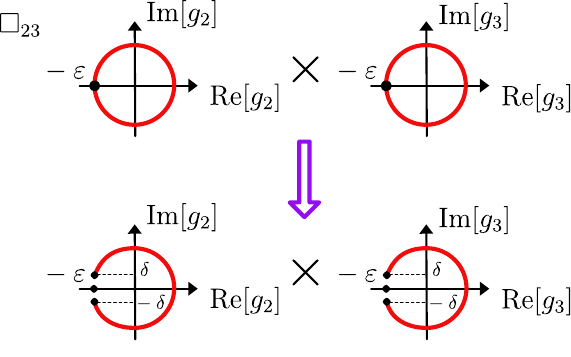}}
  \caption{Perturbation of the edge pieces}
\label{f:N7}
\end{figure}

Let us perturb the immersed pieces $\Box_{23}$ and $\Box_{13}$ as shown in the \figurename~\ref{f:N7} and also in  \figurename~\ref{f:01053}, right. It is important that this deformation preserves the structure of the coordinate-wise direct product of immersion. 

After this, the upper part of the boundary of pieces $\rectangle_3(\varnothing)$ (which is glued with $\Box_{23}(\varnothing)$, see \figurename~\ref{f:N5}) and $\rectangle_3(1)$ (which is glued with $\Box_{23}(2)$, see \figurename~\ref{f:N5}) will be moved to the line $s=1-\varepsilon-i\delta$. 
It follows from the considerations below made for $\square_{23}(\varnothing)$: 

1) According to the gluing rules, the segment $-\varepsilon_2\times[-\varepsilon,\varepsilon]_3$ is glued 
to the upper part of the boundary of $\rectangle_{3}(\varnothing)$.

2) According to the definition of immersion, this segment maps onto circle $\{-\varepsilon e^{\pi i(\theta_i+\varepsilon)/\varepsilon}\}$ on the complex line $g_2=-\varepsilon$ equipped with the coordinate $g_3$.

3) Our perturbation shifts this circle to the line $g_2=-\varepsilon-\i\delta$, breaks
the circle at the point $g_3 = -\varepsilon$ and deforms. 


4) The coordinate systems $(g_3,g_2)$ and $(u,s)$ are identified by the parallel translation: $s\equiv g_2+1$.

Similarly, the lower boundaries of the same pieces will move to the lines $s=-1+\varepsilon+i\delta$ and $s=-1+\varepsilon-i\delta$ correspondingly. The upper part of the boundary of $\rectangle_3(2)$ and $\rectangle_3(12)$ will move to the line $s=1-\varepsilon+i\delta$ and the lower parts will move to the lines $s=-1+\varepsilon-i\delta$ and $s=-1+\varepsilon+i\delta$. 

\begin{figure}[h]
  \centering{\includegraphics[width=0.9\textwidth]{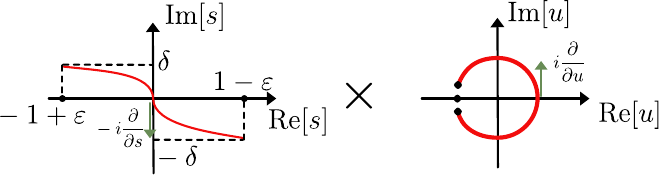}}
  \caption{Computation of the orientation}
\label{f:N8}
\end{figure}

Let us see whether these shifts can be continued inside the cylinders so that there are no intersection points with $\mathbb{R}^2$.
Consider the complex line with coordinate $s$ and note that the points $\pm(1-\varepsilon)-i\delta$ (similarly, $\pm(1-\varepsilon)+i\delta$) can be connected by a straight segment not intersecting the real axis. 
The direct product of this line segment and the perturbed component of the boundary (i.e.\ a cut and deformed circle 
located in the complex line with the coordinate $u$) defines the perturbation of $\rectangle_3(1)$ and $\rectangle_3(2)$ without points of intersection with $\mathbb{R}^2$. Similarly, in the cases $\rectangle_3(\varnothing)$ and $\rectangle_3(12)$, we have a unique intersection point (a pinch with the real axis), coinciding with the central point of the piece. Adding the orientations, we obtain the following orienting frames at the points of intersect: 
\[
(\partial/\partial u,\partial/\partial s, i\partial/\partial u, -i\partial/\partial s)
\]
for $\rectangle_3(\varnothing)$ (see \figurename~\ref{f:N8}) and 
\[
(\partial/\partial u,\partial/\partial s, i\partial/\partial u, i\partial/\partial s)
\]
for $\rectangle_3(12)$. 
Therefore 
\[
\langle\langle v',\check\varkappa\rangle\rangle=1-\mathring\gamma_1^{-1}\mathring\gamma_2^{-1},
\]
and according to \eqref{P3} 
\begin{equation}
	\text{var}_{\lambda}(v')=-\mathring\gamma_1\mathring\gamma_2\mathring\gamma_3(1-\mathring\gamma_1^{-1}\mathring\gamma_2^{-1})\Delta=(-\mathring\gamma_1\mathring\gamma_2\mathring\gamma_3+\mathring\gamma_3)\Delta.
\end{equation}

{Hence we have shown, that using Vassiliev's approach, it is also possible to recover the results presented in the paper, thereby providing an additional and independent validation of our results.}


\bibliographystyle{unsrt}
\bibliography{bibliography}

\begin{thebibliography}{10}

\bibitem{Pham1967IntroductionAL}
F.~Pham.
\newblock {\em Introduction a l'{\'e}tude topologique des singularit{\'e}s de
  Landau}.
\newblock Gauthier-Villars et cie, 1967.

\bibitem{Shabat2}
B.V. Shabat.
\newblock {\em Introduction to Complex Analysis: Functions of Several
  Variables}.
\newblock Translations of Mathematical Monographs. American Mathematical
  Society, 1992.

\bibitem{Pham2011}
F.~Pham.
\newblock {\em Singularities of integrals: Homology, hyperfunctions and
  microlocal analysis}.
\newblock Universitext. Springer London, 2011.

\bibitem{Berghoff2022}
M.~Berghoff and E.~Panzer.
\newblock Hierarchies in relative {Picard-Lefschetz} theory.
\newblock {\em arXiv preprint arXiv:2212.06661}, 2022.

\bibitem{Vassiliev2002}
V.~A. Vassiliev.
\newblock {\em Applied {Picard-Lefschetz} Theory}.
\newblock Mathematical Surveys and Monographs. American Mathematical Society,
  Providence, RI, October 2002.

\bibitem{Kita1994}
M.~Kita and M.~Yoshida.
\newblock Intersection theory for twisted cycles.
\newblock {\em Mathematische Nachrichten}, 166:287–304, 1994.

\bibitem{Assier2019}
R.~C. Assier and A.~V. Shanin.
\newblock Diffraction by a quarter--plane. {A}nalytical continuation of
  spectral functions.
\newblock {\em The Quarterly Journal of Mechanics and Applied Mathematics},
  72(1):51--86, 2019.

\bibitem{Assier2022}
R.~C. Assier, A.~V. Shanin, and A.~I. Korolkov.
\newblock A contribution to the mathematical theory of diffraction: a note on
  double {F}ourier integrals.
\newblock {\em The Quarterly Journal of Mechanics and Applied Mathematics},
  76(1):1--47, 2023.

\bibitem{Assier2024}
R.~C. Assier, A.~V. Shanin, and A.~I. Korolkov.
\newblock A contribution to the mathematical theory of diffraction. part {II}:
  Recovering the far-field asymptotics of the quarter-plane problem.
\newblock {\em Quarterly Journal of Mechanics and Applied Mathematics},
  77(1-2):hbae005, 2024.

\bibitem{Mironov2021}
M.~A. Mironov, A.~V. Shanin, A.~I. Korolkov, and K.~S. Kniazeva.
\newblock Transient processes in a gas/plate structure in the case of light
  loading.
\newblock {\em Proceedings of the Royal Society A: Mathematical, Physical and
  Engineering Sciences}, 477(2253), September 2021.

\bibitem{Shanin2024}
A.~V. Shanin, R.~C. Assier, A.~I. Korolkov, and O.~I. Makarov.
\newblock Double floquet-bloch transforms and the far-field asymptotics of
  green’s functions tailored to periodic structures.
\newblock {\em Physical Review B}, 110(2), July 2024.

\bibitem{Picard1897}
E.~Picard and G.~Simart.
\newblock {\em Th{\'e}orie des fonctions alg{\'e}briques de deux variables
  ind{\'e}pendantes}.
\newblock Gauthier-Villars, 1897.

\bibitem{lefschetz1924analysis}
S.~Lefschetz.
\newblock {\em L'analysis situs et la g{\'e}om{\'e}trie alg{\'e}brique}.
\newblock Collection de monographies sur la th{\'e}orie des fonctions.
  Gauthier-Villars et cie, 1924.

\bibitem{Pham1965}
F.~Pham.
\newblock Formules de {Picard-Lefschetz} g{\'e}n{\'e}ralis{\'e}es et
  ramification des int{\'e}grales.
\newblock {\em Bulletin de la Soci{\'e}t{\'e} Math{\'e}matique de France},
  93:333--367, 1965.

\bibitem{Vassiliev2012-tu}
V.~A. Vassiliev.
\newblock {\em Ramified integrals, singularities and lacunas}.
\newblock Mathematics and Its Applications. Springer, Dordrecht, Netherlands,
  October 2012.

\bibitem{Hwa1966}
R.~C. Hwa and V.~L. Teplitz.
\newblock {\em Homology and Feynman Integrals}.
\newblock Mathematical physics monograph series. W. A. Benjamin, 1966.

\bibitem{Arnold2012-vj}
V.~I. Arnold, S.~M. Gusein-Zade, and A.~N. Varchenko.
\newblock {\em Singularities of differentiable maps, volume 2}.
\newblock Modern Birkh{\"a}user Classics. Birkhauser Boston, Secaucus, NJ, 2012
  edition, May 2012.

\bibitem{Fulton1980}
W.~Fulton.
\newblock On a fundamental group of the complement of a node curve.
\newblock {\em Annals of Mathematics, Second series}, 111(2):407—--409, 1980.

\bibitem{Deligne1981}
P.~Deligne.
\newblock Le groupe fondamental du complément d'une courbe plane n'ayant que
  des points doubles ordinaires est abélien (d'apr\'{e}s {W}. {F}ulton).
\newblock {\em S\'{e}minaire Bourbaki}, 1979/80:1--10, 1981.

\bibitem{Milnor1963}
J.~Milnor.
\newblock {\em Morse Theory}.
\newblock Princeton University Press, 1963.

\bibitem{Kunz2023}
V.~D. Kunz and R.~C. Assier.
\newblock Diffraction by a right-angled no-contrast penetrable wedge:
  Analytical continuation of spectral functions.
\newblock {\em Quarterly Journal of Mechanics and Applied Mathematics},
  76(2):211–241, May 2023.

\bibitem{Kunz2024}
V.~D. Kunz and R.~C. Assier.
\newblock Diffraction by a right-angled no-contrast penetrable wedge: recovery
  of far-field asymptotics.
\newblock {\em IMA Journal of Applied Mathematics}, 89(3):463–497, June 2024.

\bibitem{Assier2021}
R.~C. Assier and A.~V. Shanin.
\newblock Analytical continuation of two-dimensional wave fields.
\newblock {\em Proceedings of the Royal Society A: Mathematical, Physical and
  Engineering Sciences}, 477(2245), January 2021.

\bibitem{Zariski1937}
O.~Zariski.
\newblock A theorem on the {P}oincaré group of an algebraic hypersurface.
\newblock {\em Annals of Mathematics}, 38(1):131--141, 1937.

\bibitem{Prasolov2022}
V.~V. Prasolov.
\newblock {\em Elements of Combinatorial and Differential Topology}.
\newblock Graduate Studies in Mathematics. AMS, March 2022.

\bibitem{Hillman2002}
J.~Hillman.
\newblock {\em Algebraic Invariants of Links}.
\newblock Knots and Everything, Volume 32. World Scientific, March 2002.

\bibitem{Hamm1983}
H.~Hamm.
\newblock Lefschetz theorems for singular varieties.
\newblock {\em Proceedings of symposia in pure mathematics}, 40:547–--557,
  1983.

\end{thebibliography}













\end{document}